\documentclass[a4paper,11pt]{article}
\pdfoutput=1 

\usepackage{jcappub} 

\usepackage[T1]{fontenc} 
\usepackage{natbib}
\usepackage{subfigure}
\usepackage[nottoc]{tocbibind}
\usepackage{booktabs}
\usepackage{multirow}
\usepackage{enumitem}
\usepackage{longtable}
\usepackage{xcolor}
\usepackage{multirow}
\usepackage{soul}
\usepackage{xcolor}
\usepackage{appendix}
\usepackage[normalem]{ulem}
\usepackage{wrapfig}
\usepackage{lscape}
\usepackage{rotating}
\usepackage{epstopdf}
\usepackage{graphicx}   
\usepackage{colortbl}
\usepackage{comment}

\newcommand{\gpcyr}{\,{\rm Gpc}^{-3}{\rm yr}^{-1}}

\title{Signatures of primordial black holes in gravitational wave clustering} 

\author[a,b,c]{Sarah Libanore,}
\author[b,c,d,e]{Michele Liguori,}
\author[b,c,e]{Alvise Raccanelli}

\affiliation[a]{Department of Physics, Ben-Gurion University of the Negev, Be'er Sheva 84105, Israel}
\affiliation[b]{Dipartimento di Fisica e Astronomia G. Galilei, Universit\`{a} degli Studi di Padova, \\ Via Marzolo 8, 35131 Padova, Italy }
\affiliation[c]{INFN, Sezione di Padova, Via Marzolo 8, I$\,-\,$35131, Padova, Italy}
\affiliation[d]{Dipartimento di Fisica, Universit\`a degli Studi di Trento, Via Sommarive 14, \\ I-38123 Povo (TN), Italy}
\affiliation[e]{INAF - Osservatorio Astronomico di Padova, Vicolo dell'Osservatorio 5, \\ I-35122 Padova, Italy}
\emailAdd{libanore@bgu.ac.il}

\abstract{The possible existence of primordial black holes (PBHs) is an open question in modern cosmology. Among the probes to test it, gravitational waves (GW) coming from their mergers constitute a powerful tool. In this work, we study how stellar mass PBH binaries could affect measurements of the clustering of merger events in future GW surveys.
We account for PBH binaries formed both in the early and late Universe and show that the power spectrum modification they introduce can be detected at $\sim 2\sigma-3\sigma$ (depending on some assumptions) whenever PBH mergers make up at least $\sim 60\%$ of the overall number of detected events.
By adding cross-correlations with galaxy surveys, this threshold is lowered to $\sim 40\%$. In the case of a poor redshift determination of GW sources, constraints are degraded by about a factor of 2. Assuming a theoretical model for the PBH merger rate, we can convert our results to constraints on the fraction of dark matter in PBHs, $f_{\rm PBH}$.
Finally, we perform a Bayesian model selection forecast and confirm that the analysis we develop could be able to detect $\sim30\,M_\odot$ PBHs if they account for 
$f_{\rm PBH}\sim 10^{-4}$\,--\,$10^{-3}$, depending on the model uncertainty considered, being thus competitive with other probes.
}

\begin{document}
\maketitle
\flushbottom


\section{Introduction}

We are on the verge of the fourth gravitational wave (GW) observational run by the LIGO-Virgo-Kagra Collaboration (LVK) and the increasing number of GW  detections (e.g.,~\cite{ligo2016,ligo01,ligo03,ligo03A,ligo2021}) foresees the capability of using this observable for statistical studies, in astrophysics and cosmology. 
Many works recently showed that GW provide a valuable tool to study the large scale structures (LSS) of the Universe and their clustering properties, being complementary to galaxy surveys in mapping them~\cite{namikawa_2016,didligo, raccanelli2017, libanore2020,diaz_2021}. 
Next generation GW detectors from the ground, such as the Einstein Telescope (ET)~\cite{Punturo_2010,ET_2012,ET_2020}, Cosmic Explorer (CE)~\cite{CE_2021}, or from space e.g.,~ LISA~\cite{lisa_2020,upcoming_review} and DECIGO~\cite{decigo_2020}, will fully exploit this possibility. 

Previous works~(e.g.,~ \cite{libanore2020,libanore2021,scelfo2018,scelfo2021,raccanelli2017,tamanini_2019,mastrogiovanni_2021,guadalupe_2020,calore_2020,banagiri_2020}) investigated the predicted constraining power of GW surveys alone or through cross-correlations with other LSS surveys, to measure cosmological parameters and the bias of the hosts of binaries that source GWs. Cosmological studies that use GWs, however, have to deal with large uncertainties in the sky localization of merger events, as well as the lack of redshift information, as only the luminosity distance can be directly estimated from the detected waveform. Different solutions exist: either electromagnetic counterparts or host galaxies are observed (see e.g.,~\cite{ligo2021,Fishbach_2019}), or cross-correlations can be used to statistically associate an estimated redshift and position to the GW event~\cite{delpozzo_2012, scelfo2021,muk2021}. A powerful and promising alternative is to directly use luminosity distance from the GW signal as the radial coordinate: by mapping sources in luminosity distance space, one can make use of GW surveys alone, without the need of external datasets or assumptions. The use of luminosity distance space however requires to compute how LSS affect the estimates of the observed position of the sources i.e.,~distortions due to peculiar velocities and relativistic effects~\cite{sasaki_1987,pyne_2004,hui_2006,Bertacca2018,zhang2018,namikawa_2016,Fonseca:2023uay,begnoni2023}. Alternatively, if the goal is not to constrain cosmological parameters, the luminosity distance of the observed GW events can be transformed into redshift by assuming the fiducial $\Lambda$CDM cosmology, provided that large enough bins are used to account for uncertainties in the conversion, in the same fashion as photometric galaxy surveys.

In~\cite{libanore2020,libanore2021} it was shown that the use of luminosity distance space in combination with the large volumes probed and the high luminosity distance resolution of future GW detectors will put tight constraints on the bias parameters of the hosts of GW sources. This will be reached through the tomographic analysis of their angular power spectra even in the case of poor sky localization. 
The analysis of the clustering of GW events and the estimation of the bias of their hosts present a very interesting application, related to the study of the merger progenitors' formation channels. In fact, the different processes through which binaries can form lead to different dependecies of their clustering with respect to the underlying DM field.

Focusing on black hole binary (BBH) mergers, the work of~\cite{raccanelli_2016} firstly investigated the use of the bias as a tool to disentangle between astrophysical black hole (ABH) binaries and 
primordial black hole (PBH) binaries. While the black hole components of the former descend from stellar evolution (see section~\ref{sec:ABH}), the ones of the latter are part of the DM content of the Universe, being formed from perturbations in the very early Universe and bound together across cosmic time (see section~\ref{sec:PBHintro} and~\ref{sec:PBH}). Different models have been proposed for PBH formation, which predict different abundances and spread a wide mass range. Constraints on them, therefore, can be derived through several techniques and observables, such as e.g.,~lensing, CMB distortions, and others (see~\cite{review_carr_2016,review_sasaki_2018,carr2021} and references therein for a review).

GW observations can also be used as a probe for PBHs: being formed in a completely different way with respect to ABHs, the merger of their binaries would trace the LSS in a different way. Even if, considering ABHs and PBHs having comparable masses (i.e.,~$\sim 5-100 M_{\odot}$,) the GW signal produced by their merger would be completely indistinguishable, surveys dominated either by ABHs or PBHs should have different merger rates~\cite{Ng_2022} and show different tomographic angular power spectra. The work of~\cite{raccanelli_2016} and other follow-up analyses~\cite{raccanelli2017, scelfo2018, scelfo2021, bosi2023} showed how the cross-correlation between GW and galaxy surveys changes depending on ABH/PBH relative abundance. 

In this work we refined previous analyses and we propose a complementary way to understand whether future GW surveys will be able to test the existence of PBHs or not. We perform our analysis in redshift space by making use of $5$ redshift bins large enough to take into account the uncertainties in the conversion from observed luminosity distance to redshift under the fiducial {\it Planck 2018} cosmology~\cite{planckcollaboration2019planck}. We consider how the different bias behaviors of ABHs and PBHs will combine in an overall effective bias in future GW datasets. Its functional form depends on the relative abundance of the two, as well as on the presence of early and late binary subpopulations in the PBH component and on their modelling. We adopt a phenomenological prescription to understand how this propagates in the angular power spectra that will be observed by future GW surveys alone or in cross-correlation with galaxy surveys. Depending on the PBH abundance, the estimate of the bias will be suitable to perform model selection analyses aimed at distinguishing the scenario in which only ABHs exist, to the ones where also PBH mergers contribute to the observed GW events. We finally convert our phenomenological results to constraints on the dark matter (DM) fraction possibly made by PBHs. Being based on a statistical analysis, differently from other recent results~\cite{Ng_2022}, our approach does not require precise measurements or assumptions on the very high redshift, neither for the merger rate nor for the luminosity distance of single merger events. Moreover, our method is sensitive to PBHs abundances that leave the overall local merger rate compatible with LVK constraints. 

The paper is organized as follows. In section~\ref{sec:PBHintro} we revise the most important features regarding PBH binary formation and mergers detection, while in section~\ref{sec:sources} we describe the ABH and PBH distributions and biases we adopted. The analysis setup and the statistical tool we adopted are collected in section~\ref{sec:analyses}, while section~\ref{sec:results} presents our results. In particular, in sections~\ref{sec:snr} and~\ref{sec:bias_forecast} we perform SNR and Fisher matrix forecasts for the effective bias detection based on our phenomenological assumption, while in section~\ref{sec:constrain_fpbh} and~\ref{sec:bayes_factor} we move to model-dependent Bayesian model selection forecasts for PBH abundance as DM component. All the analyses are performed with respect to third generation GW surveys, i.e.,~Einstein Telescope, alone or combined with two Cosmic Explorer instruments, and then in cross-correlation with galaxy surveys, for which we assume two example of a wide and deep surveys, modeled roughly with the specifications of SPHEREx\,-\,{\it like}~\cite{Dore:2014cca,Dore:2016tfs,Dore:2018kgp} and ATLAS Explorer\,-\,{\it like}~\cite{Wang_2019,Wang:2019rlo,Spezzati_2023}.
We finally draw our conclusions in section~\ref{sec:concl}.


\section{Primordial black hole binaries}\label{sec:PBHintro}

The hypothesis of the existence of PBH dates back to the '70s~\cite{hawking_1971,carrhawking_1974}: before the matter-radiation equivalence, they can form from high density perturbations which collapse under the effect of gravity, overcoming pressure forces and the cosmic expansion acting at the time~\cite{carrhawking_1974, Musco:2004ak, Musco:2012au, Germani:2018jgr, Kalaja_2019}. Nowadays, different theoretical models exist to explain their possible formation in the early Universe and several probes constrain their abundance (a complete description is beyond the scope of this work; see e.g.,~\cite{Khlopov:2008qy,review_carr_2016, Carr_2010, review_sasaki_2018, carr2021,Carr_2021} for some reviews).
Formation mechanisms are mostly related to the existence of large curvature perturbations, due e.g.,~to blue tilt in the primordial power spectrum 
(e.g.,~\cite{carr_1993, carr_1994, kohri_2007}), inflationary potentials that create peaks on small scales (e.g.,~\cite{yokoyama_1998, lyth_2003, Kalaja_2019}), curvature perturbations (e.g.,~\cite{Polnarev:2006aa, Kalaja_2019}), non-Gaussianity (e.g.,~\cite{Luca_2019,Yoo:2018kvb}) or primordial running spectral index (e.g.,~\cite{leach_2000,kohri_2007}), or processes during inflation (e.g.~\cite{Choudhury:2013woa}). Other explanations for PBH existence involve e.g.,~bubble collisions in phase transitions~\cite{hawking_1982,moss_1994,Belotsky:2014kca}, non linear processes~\cite{Chen:2022usd} or cosmic strings~\cite{hawking_1989,garriga_1993,Jenkins:2020ctp}.

It is customary to define: 
\begin{equation}\label{eq:f_PBH_DM}
    f_{\rm PBH} = \frac{\Omega_{\rm PBH}}{\Omega_{\rm DM}}, 
\end{equation}
as the DM fraction composed by PBH ($\Omega_{\rm PBH,DM}$ are the dimensionless energy densities of PBH and DM, respectively).
The value of $f_{\rm PBH}$ is currently constrained using different techniques and observations for different PBH mass functions (see e.g.,~\cite{review_carr_2016, 
bellomo2018, review_sasaki_2018, carr2021} for a summary): assuming a monochromatic mass distribution, they seem to exclude the possibility of $f_{\rm PBH} = 1$, with the exception of the range $M_{\rm PBH} \in [10^{-16},10^{-10}]\,M_{\odot}$ (asteroid$\,-\,$sublunar masses). 
Lower values of $f_{\rm PBH}$ could still be allowed in other mass ranges, between which of particular interest are the windows~$M_{\rm PBH} \in [10^{-7},10^{-5}]\,M_{\odot}$ and $M_{\rm PBH} \in [5,100]\,M_{\odot}$. By considering a broad mass function, $f_{\rm PBH} = 1$ could be recovered as well.

The $M_{\rm PBH}\in[5,100]\,M_{\odot}$ case is particularly interesting: PBH of such masses could be part of the progenitors of the merger events observed by the LIGO-Virgo-Kagra Collaboration (see e.g.,~\cite{didligo, deluca_2020, Franciolini:2021tla}).
In fact, if such objects exist, they can form binaries and merge, contributing to the signals observed by current and future interferometers. Provided the uncertainty on the mass function, the GW emitted could fall in different frequency ranges. 

To form PBH binaries that merge within a Hubble time, there are two possible formation channels to account for, which we call Early PBH binaries (EPBH) and Late PBH binaries (LPBH), and the two can coexist. Both of them, after formation, evolve via GW emission: the energy released by their quadrupole moment shrinks the orbit until the merger.

Early PBH binaries were firstly theorized by~\cite{nakamura_1997, ioka_1998} as systems that bound together and decouple from the Universe expansion in the radiation dominated era. 
After formation, binaries are affected by tidal forces due to the rest of the DM field, being this made by PBH or other components~\cite{nakamura_1997, ioka_1998, yacine2017, raidal2019}. This enhances the angular momentum of the PBHs in the binary, increasing the time they require to inspiral and retarding the merger, making it observable today. Three\,-\,body interactions can lead to binary disruption as well.
Since early binaries randomly form wherever PBHs are located, they follow closely the DM distribution both at their formation and throughout cosmic time. For this reason, at first approximation, their bias can be modeled as unity, as PBHs in this case would directly trace the DM field:
\begin{equation}
\label{eq:bias_early}
    b^E = 1 \,.
\end{equation}
Late PBH binaries instead form for dynamical capture, when two PBHs approach each other and, after losing energy through GW emission, get bound~\cite{didligo}. Even if single PBHs closely follow the DM distribution, dynamical captures take place when DM halos already formed: therefore, the cross-section of this process depends on the velocity dispersion inside the halos, which in turn depends on the halo mass $M_h$.\footnote{Dynamical capture can also be a formation channel for ABH-PBH binaries. Their  contribution to the overall merger rate depends on the environment where the binary is formed (mainly relative velocity and number densities of the two populations) and is in general subdominant~\cite{dallarmi2022}.} As~\cite{didligo, raccanelli_2016} firstly highlighted, this implies that late PBH binaries are mainly hosted by halos of mass $M_h < 10^6\,M_{\odot}$~\cite{raccanelli_2016}. The bias of late binaries can be approximated through the bias of their hosts, which in this $M_h$ range is~\cite{wernerporciani_2019} 
\begin{equation}\label{eq:bias_late}
    b^L \approx 0.5 \,.
\end{equation}

In this work we focus on future GW experiments, as the number or mergers detected with current and near future detectors (first and second generation) does not allow a powerful enough statistical analysis.
The main proposed future observatories from the ground are the Einstein Telescope (ET, \cite{Punturo_2010,ET_2012,ET_2020}) and Cosmic Explorer (CE, \cite{CE_2021,Srivastava:2022slt}): despite their different shape and technology, these two experiments will have comparable sensitivity on the same frequency range. The number of binary mergers observed and the distance reached will be considerably larger than what will be obtained with second generation detectors. Moreover, combining detections from the two instruments or from a ET2CE configuration will improve the sky localization of GW events (see e.g.,~\cite{Iacovelli:2022bbs} for a recent analysis). 
For this reason, the merger event catalogs they will provide will be ideal to pursue statistical analyses.


\section{Source distribution and bias }
\label{sec:sources}

The progenitors of the BBH mergers observed with GW interferometers are either deriving from stellar evolution and astrophysical processes (which we will call ABH, see e.g.,~\cite{eggleton_2009, ivanova_2013, barack_2019} and references therein) or they date back to primordial origin (PBH, see previous sections and e.g.,~\cite{hawking_1971,carrhawking_1974,review_carr_2016,review_sasaki_2018}).
Assuming that both ABH and PBH masses lie in the range $\sim 5-100 M_{\odot}$, the GW signals produced during the merger of their binaries are in principle indistinguishable.
While specific parameters, e.g.,~redshift range or mass and spin distributions can indicate a preferential astrophysical or primordial origin, in principle any GW survey detects both ABH and PBH mergers as events in luminosity distance space, and is therefore \textit{blind} to their origin. 

A very interesting fact, that is at the base of our work, is that the distribution and clustering properties of ABH binaries are different from the ones of PBH binaries, as they trace the LSS differently. As firstly hinted in~\cite{didligo}, on one hand, ABHs are the endpoint of stellar evolution and for this reason they are found in galaxies, which in the standard hierarchical scenario mainly form within massive DM halos.
On the other hand, PBH binaries, as we saw, can form through more than one channel, each tracing the LSS differently: Poissonian distributed early binaries, in which PBH bound together right after their formation, and late binaries, formed through dynamical capture mainly in small DM halos.
This leads to different redshift distributions and bias properties of mergers in different scenarios.

\subsection{Astrophysical black holes}
\label{sec:ABH}

To model the ABH number distribution, we assume that the merger rate evolution in redshfit is consistent with the star formation rate described by the Madau-Dickinson model~\cite{madau_2014}. Following~\cite{Iacovelli:2022bbs}, we model the ABH normalized merger rate as
\begin{equation}\label{eq:ABHrate}
    \tilde{\mathcal{R}}^A(z) = \frac{(1+z)^{\alpha_1^A}}{1+[(1+z)/(1+\alpha^A_3)]^{\alpha_1^A+\alpha_2^A}}\,,
\end{equation}
where $\alpha_1^A = 2.7,\,\alpha_2^A = 3,\,\alpha_3^A = 2$. 
As for the bias, we use the fit that~\cite{Peron:2023} estimates from clustering measurements on hydrodynamical cosmological simulations combined with population synthesis models~\cite{artale2018, artale2019, artale2019_2}. We therefore parameterize the bias of the ABH population as
\begin{equation}\label{eq:ABHbias}
    b^{A}(z) = A(z+1)^D\,,
\end{equation}
where $A = 1.2,\, D = 0.59$. According to results in~\cite{Peron:2023}, the errorbars on the measurements from which the fit is obtained are $\lesssim 50\%$ for $z\in [1,6]$.


\subsection{Primordial black holes}
\label{sec:PBH}
We parameterize the normalized merger rates of E/L PBH binaries through the power law
\begin{equation}\label{eq:PBHrate}
    \tilde{\mathcal{R}}^{E,L}(z) = (1+z)^{\alpha_{E,L}}\,.
\end{equation}
For early binaries, we interpolate the results of~\cite{raidal2019} and get $\alpha_E = 1.25$, while for late binaries we consider $\alpha_L = 0$, since at first approximation the gravitational capture process in small DM halos (which formed earlier than the $z_{\rm max}$ we adopt) can be considered as redshift independent. 

\subsection{Observed redshift distribution and bias}

In order to account for the different types of progenitors, in this work we consider the total binary black hole merger rate in a GW survey to be
\begin{equation}
\mathcal{R}^{\rm tot}(z) = \mathcal{R}^{A}(z) + \mathcal{R}^{P}(z) = \mathcal{R}^{A}(z) + \mathcal{R}^{E}(z) + \mathcal{R}^{L}(z)\,,
\end{equation}
where $A =$ ABH and $P =$ PBH of $E,L=$ early/late type. We re-express the merger rates of the different binary types by the means of the phenomenological parameters $\{f^E,f^L\}$, defined as the ratio between the early or late PBH and the total local merger rates, namely 
\begin{equation}\label{eq:fEL}
    f^E = \mathcal{R}_0^E/\mathcal{R}_0^{\rm tot}\,,  \quad  f^L = \mathcal{R}_0^L/\mathcal{R}_0^{\rm tot}\,.
\end{equation}
By doing so, the redshift evolution of the total merger rate is written as  
\begin{equation}\label{eq:rtot}
    \mathcal{R}^{\rm tot}(z) = \mathcal{R}^{\rm tot}_0\biggl\{\bigl[1-(f^E+f^L)\bigr]\tilde{\mathcal{R}}^A(z) + f^E\tilde{\mathcal{R}}^E(z) + f^L\tilde{\mathcal{R}}^L(z)\biggr\} \,,
\end{equation}
where the normalized merger rates $\tilde{\mathcal{R}}^{A,E,L}(z)$ are defined in eqs.~\eqref{eq:ABHrate} and~\eqref{eq:PBHrate}; note that the relative ABH, EPBH, LPBH abundances evolve in $z$ depending on the values of $\{f^E,\,f^L\}$. In our setup, the merger rate is model dependent i.e.,~it varies because of the relative abundance not only of ABH and PBH binaries, but also of early and late PBH. Using eq.~\eqref{eq:rtot}, the observed number distribution of events per redshift bin $z$ and solid angle $\Omega$ observed by a GW survey can be estimated as 
\begin{equation}
\label{eq:numberdistr}
\frac{d^2N^{\rm GW}}{dzd\Omega} = \frac{1}{\mathcal{N}}\biggl[\frac{c \,\chi^2(z)}{(1+z)H(z)}T_{\rm obs}\mathcal{R}^{\rm tot}(z)\Theta(z_{\rm max}-z)\biggr] \ ,
\end{equation}
where $c$ is the speed of light, $\chi(z)$ is the comoving distance, $H(z)$ is the Hubble parameter, $T_{\rm obs}$ is the survey observation time, $\Theta$ the Heaviside function and $z_{\rm max}$ the detector horizon (see section~\ref{sec:survey} for further details); the factor $\mathcal{N}$ is defined as 
\begin{equation}
    \mathcal{N} = N^{\rm GW}T_{\rm obs}\biggl[\Omega_{\rm full\,sky}\int_{z_{\rm min}}^{z_{\rm max}}dz\, \frac{d^2N^{\rm GW}}{dzd\Omega}\biggr]^{-1}\,,
\end{equation}
where $\Omega_{\rm full\,sky} \simeq 40000\,{\rm deg}^2$ is the solid angle corresponding to the full sky and $N^{\rm GW}$ the total number of binary black hole mergers the GW survey observes each year. The factor $\mathcal{N}$ is used to rescale the number distribution, so to fix the total number of observed events $N^{\rm GW}$ to a customary value. This is equivalent to define an effective merger rate, which at $z=0$ is anchored\footnote{Substituting eqs.~\eqref{eq:ABHrate},~\eqref{eq:PBHrate} in eq.~\eqref{eq:rtot}, at $z =0$ we obtain
\begin{equation}
    \mathcal{R}^{\rm tot}(z = 0) = \frac{\mathcal{R}_0^{\rm tot} }{(1+\alpha_3^A)^{\alpha_1^A+\alpha^A_2}}+(f^E+f^L)\left(1-\frac{\mathcal{R}_0^{\rm tot} }{(1+\alpha_3^A)^{\alpha_1^A+\alpha^A_2}}\right) \sim \mathcal{R}_0^{\rm tot},
\end{equation}
where the last equality depends on $1-1/(1+\alpha_3^A)^{\alpha_1^A+\alpha^A_2} \sim\mathcal{O}(10^{-3})$. When $\mathcal{N} = 1$, this result is consistent with LVK constraints for any $\{f^E,f^L\}$. Using $\mathcal{N}>0$ allows us to select a fraction of the observed events.
} at $\mathcal{R}_0^{\rm tot}/\mathcal{N}$, while it evolves in $z$ depending on $\{f^E,f^L\}$.

Being the survey {\it blind} to the origin of the merger progenitors, the effective black hole merger bias estimated from it can be computed by weighing the ABH, EPBH, LPBH biases by their relative merger rates at each $z$, namely 
\begin{equation}\label{eq:bias}
\begin{aligned}
b_{\rm eff}(z) &= \frac{\mathcal{R}^A(z)}{\mathcal{R}^{\rm tot}(z)}b^A(z) +  \frac{\mathcal{R}^E(z)}{\mathcal{R}^{\rm tot}(z)}b^E(z) + \frac{\mathcal{R}^L(z)}{\mathcal{R}^{\rm tot}(z)}b^L(z) \\
&= \frac{\mathcal{R}_0^{\rm tot}}{\mathcal{R}^{\rm tot}(z)}\biggl[\bigl[1-(f^E+f_0^L)\bigr]\tilde{\mathcal{R}}^A(z)b^A(z) + f^E\tilde{\mathcal{R}}^E(z)b^E(z) + f^L\tilde{\mathcal{R}}^L(z)b^L(z)\biggr] \,.
\end{aligned}
\end{equation}
The ABH, EPBH, LPBH merger rate functional forms $\tilde{\mathcal{R}}^{A,E,L}(z)$ and biases $b^{A,E,L}(z)$, are computed following the prescriptions in sections~\ref{sec:ABH},~\ref{sec:PBH}. Note that we indicate with $\tilde{\mathcal{R}}(z)$ the redshift evolution of the merger rate normalized by the local value, namely $\mathcal{R}(z) = \mathcal{R}_0^{\rm tot}\tilde{\mathcal{R}}(z)$.
We stress that eq.~\eqref{eq:bias} does not describe the intrinsic bias of one population, but it represents an effective quantity. The slope of its redshift evolution, thus, can also decrease in $z$ since it averages the bias models of ABH, early PBH and late PBH binaries weighted by their relative merger rates; in the rest of the paper we will omit the $_{\rm eff}$ notation for simplicity.

\begin{figure}[ht!]
    \centering
  \includegraphics[width=.9\columnwidth]{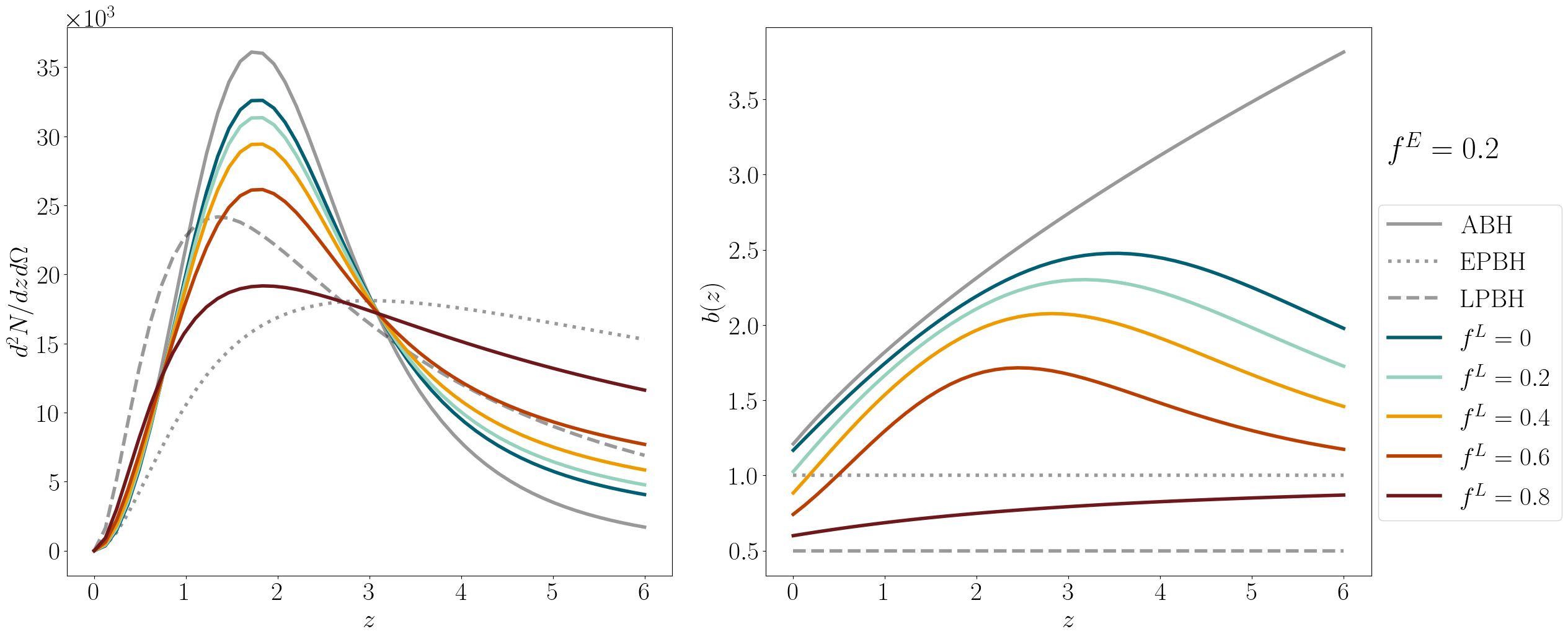}\\
    \caption{Observed number distribution $d^2N^{\rm GW}/dzd\Omega$ (left, note that $\Omega$ is in ${\rm std}$) and effective bias $b(z)$ (right) of the ET2CE survey in the fiducial case $\mathcal{R}^{\rm tot}_0 = 27\gpcyr$, $f^E = 0.2$ and different $f^L$. We consider $N^{\rm GW} = 1.1\times 10^5$ each year and $T_{\rm obs} = 10\,{\rm yr}$. Grey lines illustrate the cases in which only ABH (continuous), only EPBH (dotted) or only LPBH (dashed) are taken into account. }
    \label{fig:model_fid}
\end{figure}

The observed number distribution and effective bias are shown in figure~\ref{fig:model_fid}, where in grey we also show their values for the three populations as if they would comprise the entirety of the observed mergers. 
In our analysis, we explore different $\{f^E,f^L\}$ fractions and detectors setup; for the sake of clarity, we show results for $f^E = 0.2$ and our fiducial ET2CE, in which the local, overall merger rate 
is set to $\mathcal{R}_0^{\rm tot} = 27\gpcyr$, the number of merger events observed each year is $N^{\rm GW} = 1.1\times 10^5$ (following the prescriptions of~\cite{Iacovelli:2022bbs}, where the ET-D design\footnote{ET-D considers an equilateral triangular design, with three 10km-long nested interferometers, each consisting of one instrument optimised for high frequencies and one for low frequencies.} is adopted, and in agreement with the latest LVK~O3 constraints at 95\%\,CL~\cite{ligo2021}; further details in section~\ref{sec:analyses}) and the observation time is $T_{\rm obs} = 10\,{\rm yr}$. We choose the value $f^E = 0.2$ to illustrate our results since it describes an intermediate scenario for the EPBH abundance, 
while at the same time it allows us to span LPBH over a wide range. A further reason why such EPBH fraction is interesting is that~\cite{Franciolini:2021tla} claimed that the presence of $\sim 30\%$ EPBH in the observed events of the second LVK run is statistically favoured by hierarchical Bayesian analysis. Even if such result strongly depends on the assumptions in the astrophysical modelling, it drives the attention to such regime.
In Appendix~\ref{app:all_plots} we show results for other values of $f^E$.

The number distribution in figure~\ref{fig:model_fid} shows all the merger events observed by the detector. As we will discuss in section~\ref{sec:analyses}, only some of these events enter our analysis, depending on the strategy we adopt to treat the poor sky localization for GW events. This does not change the effective bias, since we assume ABH and PBH are similarly affected. Thus, from figure~\ref{fig:model_fid} the goal of our analysis is already evident: if future GW surveys will be able to well constrain the mergers' effective bias, the existence of PBHs will be detected by looking at its flattening or decreasing at high $z$. Our goal is then to understand under which conditions this statistical analysis will have enough constraining power to reliably claim a PBH detection.



\section{Analysis}
\label{sec:analyses}

Here we describe the setup of the analysis, whose results are collected in the next section. We initially rely on the phenomenological parameters $\{f^E,\,f^L\}$ to compute signal-to-noise ratio and Fisher forecasts. To cover the full parameter space, we consider the combination of 6 uniformly spaced values, namely $[0, 0.2, 0.4, 0.6, 0.8, 1]$. For each pair of values, we set the condition $f^E+f^L \leq 1$ (i.e.,~we analyze 21 different models). We then convert our results into constraints on $f_{\rm PBH}$ for the EPBH and LPBH models generally adopted in the literature. Finally, we perform a Bayesian model selection forecast to understand under which conditions future surveys will be able to discriminate between the only-ABH and ABH+PBH scenarios.

A small disclaimer before we start. 
The most natural radial coordinate to map GW surveys is the luminosity distance $D_L$, since it can be directly estimated from data, while the redshift is degenerate with the chirp mass. The observed $D_L$-space can be transformed into $z$-space once that the values of the cosmological parameters is set and the space distortions and general relativistic effects are properly accounted for in the two coordinate systems~\cite{zhang2018, libanore2020, namikawa2021, libanore2021, Fonseca:2023uay}. At the time ET and ET2CE will operate, cosmological parameters will be extremely well constrained with respect to the other quantities in our analysis; moreover, corrections between $D_L$-space and $z$-space contributions to the number counts angular power spectrum have been showed to be small. Given that the measurement of these quantities is not our goal, we can thus safely work in $z$-space by assuming large enough $z$-bins; under this assumption, the GW survey is somehow analogous to a photometric survey. Throughout the analysis, we fix the cosmological parameters to {\it Planck 2018} results~\cite{planck2018}. 

Moreover, depending on parameters such as progenitor masses, sky position, orbit inclination, etc., future gravitational wave detectors will provide different luminosity distance and sky localization uncertainties $\{\Delta D_L /D_L,\,\Delta\Omega\}$ for each of the merger events they will observe. However, since here we are interested in the statistical properties of the overall distribution, we make general assumptions on the measurement uncertainties. 


\subsection{Setup}
\label{sec:survey}

We characterize the GW survey by assuming that it observes a sky fraction $f_{\rm sky} = 1$ for a time $T_{\rm obs} = 10\,{\rm yr}$, up to the detector horizon $z_{\rm max}=6$. To evaluate the statistical luminosity distance uncertainty $\Delta D_L /D_L$ and average sky localization uncertainty $\Delta\Omega$, we consider ET alone and ET2CE.
The sky localization is related to the luminosity distance uncertainty; the analysis in~\cite{Iacovelli:2022bbs} suggests an almost linear dependence between $\log_{10}(\Delta D_L/D_L)\in [-3,\,1]$ and $\log_{10}(\Delta \Omega \,[{\rm deg}^2])\in [-3,\,4]$, with a $\sim \log_{10}[\mathcal{O}(10)]$ dispersion and a decreasing accuracy with increasing redshift. 
We can estimate that $\sim 15\%$ events with $z \leq 6$ will have uncertainties $\Delta D_L /D_L \lesssim 1$ and $\Delta\Omega \lesssim 100\,{\rm deg}^2$ in the single ET case, while $\sim 65\%$ events with $z \leq 6$ will have $\Delta D_L /D_L \lesssim 1$ and $\Delta\Omega \lesssim 10\,{\rm deg}^2$ for ET2CE. The total number of events used in the analysis is then $N^{\rm GW\, obs} = 1.20\times 10^4$/yr for single ET; $N^{\rm GW\, obs} = 7.15\times 10^4$/yr for ET2CE. 
For the purposes of our statistical analysis, we consider only these subsets of events, which we divide into $z_{i,j}$ bins, having amplitude larger than $\Delta D_L /D_L$ if converted to $D_L$-space.

Our estimator is the angular power spectrum $C_\ell(z_i,z_j)$, which we compute in 5 Gaussian tomographic bins, centered in  $z_i \sim [0.4, 0.8, 1.4, 2.4, 4.3]$ with $\Delta z \simeq [0.4, 0.4, 0.6, 1.0, 1.9]$ and r.m.s. either equal to 1/2 or 1/4 of the width. The former case, more conservative, has a reduced constraining power due to the overlap between adjacent bins, which is accounted for in the computation of the covariance matrix (see next section and, e.g.,~\cite{Kovetz_2017}). The latter, instead, relies on a better redshift determination of the observed sources. Depending on the capability next generation GW detectors will reach in determining $z$, results will be somewhere in between these two cases.
In this work, we do not take into account information from EM counterparts, mainly the host galaxy identification for ABH mergers. Since our study is based on a statistical analysis of the overall sample and not on single events, the improvement these could lead depends on how many counterparts can be observed. If their number will be non-negligible, first of all they will increase the number of events observed with a good sky localization $N^{\rm GW\,obs}$, thus decreasing the shot noise in the analysis in section~\ref{sec:analyses}. Moreover, the EM counterparts will improve the redshift measurements, allowing us to perform a better tomography and to use of a larger number of redshift bins, increasing the overall signal. Note that a similar improvement can be achieved by working in luminosity distance space as previously described, once that corrections to redshift space distortions are properly taken into account. Finally, EM counterparts can be used to estimate a prior for the galaxy bias of ABH hosts, which in turn can be used to approximate the prior on the ABH bias itself. 

To estimate the smallest scale accessible by the GW detector due to its lack of sky localization, we further assume to observe mergers in Gaussian beams of average size $\Delta\tilde{\Omega} = 50\,{\rm deg}^2$ for the single ET configuration and $\Delta\tilde{\Omega} = 1\,{\rm deg}^2$ for ET2CE, so to compensate between the constraining power of events with good and bad sensitivities.\footnote{The sky localization we consider for, e.g.,~ET2CE, are distributed between $10{\, \rm deg^2}$ and very good (up to $10^{-3}{\, \rm deg^2}$) values. Thus, by adopting $\Delta\tilde{\Omega}$, we analyze optimistically the events with $1{\, \rm deg^2} \lesssim \Delta\Omega \lesssim 10{\, \rm deg^2}$, while worsening the ones with $10^{-3}{\, \rm deg^2} \lesssim \Delta\Omega \lesssim 1{\, \rm deg^2} $. The smearing we introduce on the angular power spectrum further reduces the constraining power, making our final results conservative.} This allows us to treat the GW maps analogously to e.g.,~CMB and intensity mapping studies, and it implies the angular power spectrum gets smoothed by $\tilde{C}_\ell(z_i,z_j) = B_\ell^2 C_\ell(z^i,z^j)$, where
\begin{equation}\label{eq:bl}
    \tilde{B}_\ell(z_i,z_j) = \exp\biggl[- \frac{\ell(\ell+1)}{2}\frac{\Delta\tilde{\Omega}_{\rm std}}{8\,\log(2)}\biggr]\,,
\end{equation}
with $\Delta\tilde{\Omega}_{\rm std} = \Delta\tilde{\Omega}_{\rm deg^2}(\pi/180)^2$. 
Appendix~\ref{app:cl} recalls the main equations and shows $C_\ell,\,\tilde{C}_\ell$.

\subsection{Cross-correlation with galaxy surveys}
\label{sec:cross_corr}
\begin{figure}
    \centering
    \includegraphics[width=.9\columnwidth]{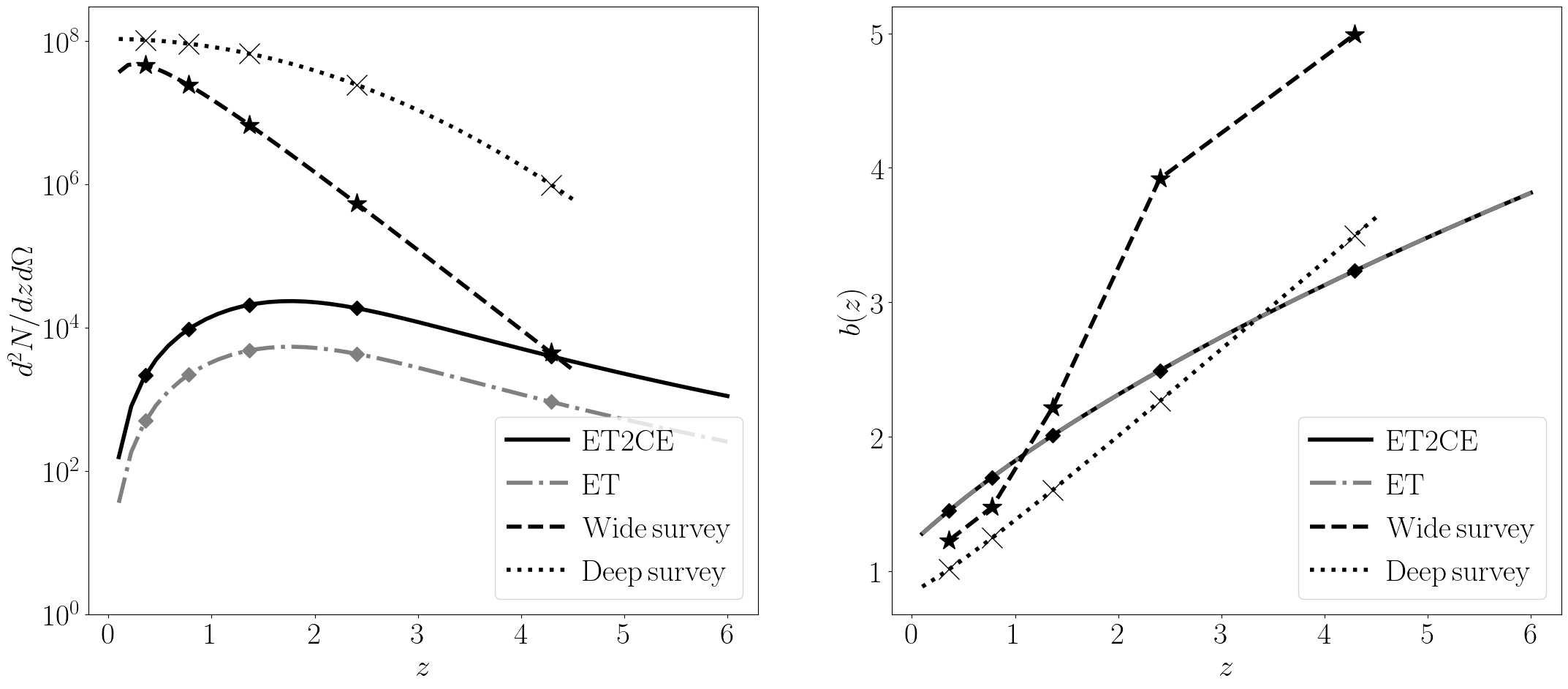}
    \caption{Number of sources per solid angle and redshift bin (left panel) and biases 
    (right panel) of galaxies observed by the wide survey
    (dashed), deep survey 
    (dotted) and for ABH mergers observed by ET2CE (black), ET (gray, same bias). Markers show the mean redshift of the bins we adopted.}
    \label{fig:surveys}
\end{figure}

In this work, we also investigate cross-correlations between future GW and galaxy surveys (see e.g.,~\cite{raccanelli_2016}); to do that, we take as an example two mock surveys, modeled around the forthcoming wide area SPHEREx~\cite{Dore:2014cca, Dore:2016tfs, Dore:2018kgp} and the proposed deep survey ATLAS Explorer~\cite{Wang_2019, Wang:2019rlo,Spezzati_2023}. We do this in order to explore different regimes and scenarios, trying to understand what is the ideal setup for this kind of analyses. A detailed investigation of the optimal galaxy survey strategy and experiment-specific forecasts are beyond the scope of this paper.
Equations describing the cross-angular power spectrum $\tilde{C}_\ell^{XY}(z_i,z_j)$ can be found in appendix~\ref{app:cl}. The surveys are characterized as follows.
\begin{itemize}[label=\raisebox{0.25ex}{\tiny$\bullet$}]
    \item Wide survey, SPHEREx-{\it like}. We consider $f_{\rm sky}^w = 0.75$ and number density and bias from the $\sigma_z /(1+ z) = 0.2$ sample,\footnote{\url{https://github.com/SPHEREx/Public-products}} since the resolution is close to our GW survey. We then compute the source number distribution using bins with $\Delta z = 0.2$, as 
    \begin{equation}\label{eq:spherex}
        \frac{d^2N^{w}}{dzd\Omega} =  \frac{3.26\times 10^8}{4\pi\Delta z}\left(\frac{z}{0.301}\right)^{0.829}\exp\left[-\left(\frac{z}{0.301}\right)^{0.95}\right]\,.
    \end{equation}    
    \item Deep survey, ATLAS Explorer-{\it like}. We consider a configuration covering sky area $f^{d}_{\rm sky}= 0.2$, and we consider
    \begin{equation}
     \frac{d^2N^{d}}{dzd\Omega} = \frac{3.26\times 10^8}{A^{d}_{\rm std}\Delta z}\,0.16\exp(-0.25\,z^2) \; , \quad 
     b^{d}(z) = 0.84\frac{D(z)}{D(z=0)} \; ,
    \end{equation} 
    $D(z)$ being the growth factor and $\Delta z= 0.2$ in this case as well.
    The sky area prevents us observing the largest scales; for this, we set the minimum multipole to $\ell_{\rm min}^{d} \sim {2\pi}/{\theta} \sim 10$, where $\theta = \arccos\left[1-A^{d}_{\rm std}/2\pi\right]$.
\end{itemize}
We show the number density and bias of the galaxies observed in the wide and deep survey 
in figure~\ref{fig:surveys}, where we also compare them to the ABH cases observed by ET2CE and ET. Interestingly, the bias of ABH and of the galaxies observed by the wide survey 
are similar at low redshift, while at high redshift ($z\gtrsim 3$) they deviate from each other, with the galaxy bias being systematically higher than the ABH's host ones. 
This is due to the fact that at high redshift the wide survey 
will mainly observe the brightest galaxies, ultimately hosted by the more massive halos (differently happens for the deep survey, 
sensitive to fainter and smaller galaxies). GW mergers, instead, at first approximation can be observed irrespective from the properties of their host galaxy; therefore, they trace halos of different masses (see~\cite{Peron:2023} for analysis on the relevance of the galaxy and host halo properties in the ABH clustering). 

In the analysis we perform in the following section, we adopt the same $z$ binning both for auto- and cross-spectra, meaning we are dividing the galaxy catalogs in the same bins as the GW events.
When computing galaxy auto-spectra, we consider $\ell_{\rm max} = \chi(z)k_{nl,0}$, where $k_{nl,0} = 0.12\,\rm{Mpc}^{-1}$.
For cross-spectra with GW and GW auto-spectra, instead, the smoothing due to the GW sky localization uncertainty has to be accounted for, see eqs.~\eqref{eq:bl} and~\eqref{eq:cl_cross}.

\section{Results}
\label{sec:results}
Our aim is to understand what confidence level will be reached by future surveys in detecting the presence of a primordial components in the GW catalogs. For both GW auto-spectra and GW$\times$galaxy cross-spectra, we initially perform two different and complementary analyses.
First of all, we compute the signal-to-noise Ratio (SNR) for the presence of PBH mergers in the GW catalogs. Then, we perform a Fisher analysis to constrain the marginalized errors on the mergers' bias. Both the analyses assume as fiducial the case where only ABHs are present. Finally, after converting our constraints to $f_{\rm PBH}$, we use the Bayes factor formalism to forecast model selection between different scenarios.

\subsection{Signal-to-Noise Ratio}
\label{sec:snr}

The first statistical test we perform on our forecast is to understand whether, if PBH contribute to the mergers, the signal observed through the auto- ($X = Y = $ GW) and cross- ($ X = $ GW, $Y = $ galaxies or viceversa) spectra 
$\tilde{C}_{\ell,ij}^{XY{\rm[AP]}}=\tilde{C}_\ell^{XY{\rm [AP]}}(z_i,z_j)$ will be distinguishable from the theoretical assumption in which only ABH exists. This is done by computing:
\begin{equation}\label{eq:snr}
    {\rm SNR}^2  = f_{\rm sky}\sum_\ell\left(\tilde{\mathbf{C}}_{\ell}-\tilde{\mathbf{C}}_{\ell}^{\rm AP}\right)^T\mathcal{M}_\ell^{-1}
\left(\tilde{\mathbf{C}}_{\ell}-\tilde{\mathbf{C}}^{\rm AP}_{\ell}\right)\end{equation}
where we sum all the multipoles 
to estimate the signal-to-noise ratio of the survey. 
In the previous equation, we defined the vector containing the angular power spectra as 
\begin{equation}
    \tilde{\mathbf{C}}_{\ell} = \begin{pmatrix}
    \tilde{C}_{\ell,11}^{XX} & \tilde{C}_{\ell,12}^{XX} & ... & \tilde{C}_{\ell,N_{\rm bin}N_{\rm bin}}^{XX} & \tilde{C}_{\ell,11}^{XY} & ... & \tilde{C}_{\ell,N_{\rm bin}N_{\rm bin}}^{XY} & \tilde{C}_{\ell,11}^{YY} & ... & \tilde{C}_{\ell,N_{\rm bin}N_{\rm bin}}^{YY}  ,
    \end{pmatrix}
\end{equation}
and analogously for the case $\tilde{\mathbf{C}}_\ell^{\rm AP}$ in which PBH are included. The elements of the covariance matrix $\mathcal{M}_\ell$ are defined as~\cite{Bellomo_2020,Scelfo:2020jyw,Cabre_2007,Dio_2014,raccanelli_2016,Kovetz_2017,Sato_Polito_2020}
\begin{equation}
    \mathcal{M}_{\ell,IJ} = \frac{1}{2\ell+1}\biggl[\left(\tilde{C}^{ XX}_{\ell,ip}+\frac{\delta_{ip}}{\bar{n}_{{ X},i}}\right)\left(\tilde{C}^{ YY}_{\ell,jq}+\frac{\delta_{jq}}{\bar{n}_{{ Y},j}}\right)+\left(\tilde{C}^{ XY}_{\ell,iq}+\frac{\delta_{iq}\delta_{XY}}{\bar{n}_{{ X},i}}\right)\left(\tilde{C}^{ XY}_{\ell,jp}+\frac{\delta_{jp}\delta_{XY}}{\bar{n}_{{X},j}}\right)\biggr]
\end{equation}
where $IJ = (X,ij)(Y,pq)$ are the indexes associated with the entries of the $\tilde{\mathbf{C}}_\ell$ vector, each of which refers to a specific source ($X$ or $Y$) and to a specific pair of bins ($ij$ or $pq$). Finally, $\bar{n}_{X,i}^{-1}$ is the shot noise of the $X$ source in the $i$-th bin. 

\begin{figure}[ht!]
    \centering
    \begin{minipage}{0.49\linewidth}
  \includegraphics[width=\columnwidth]{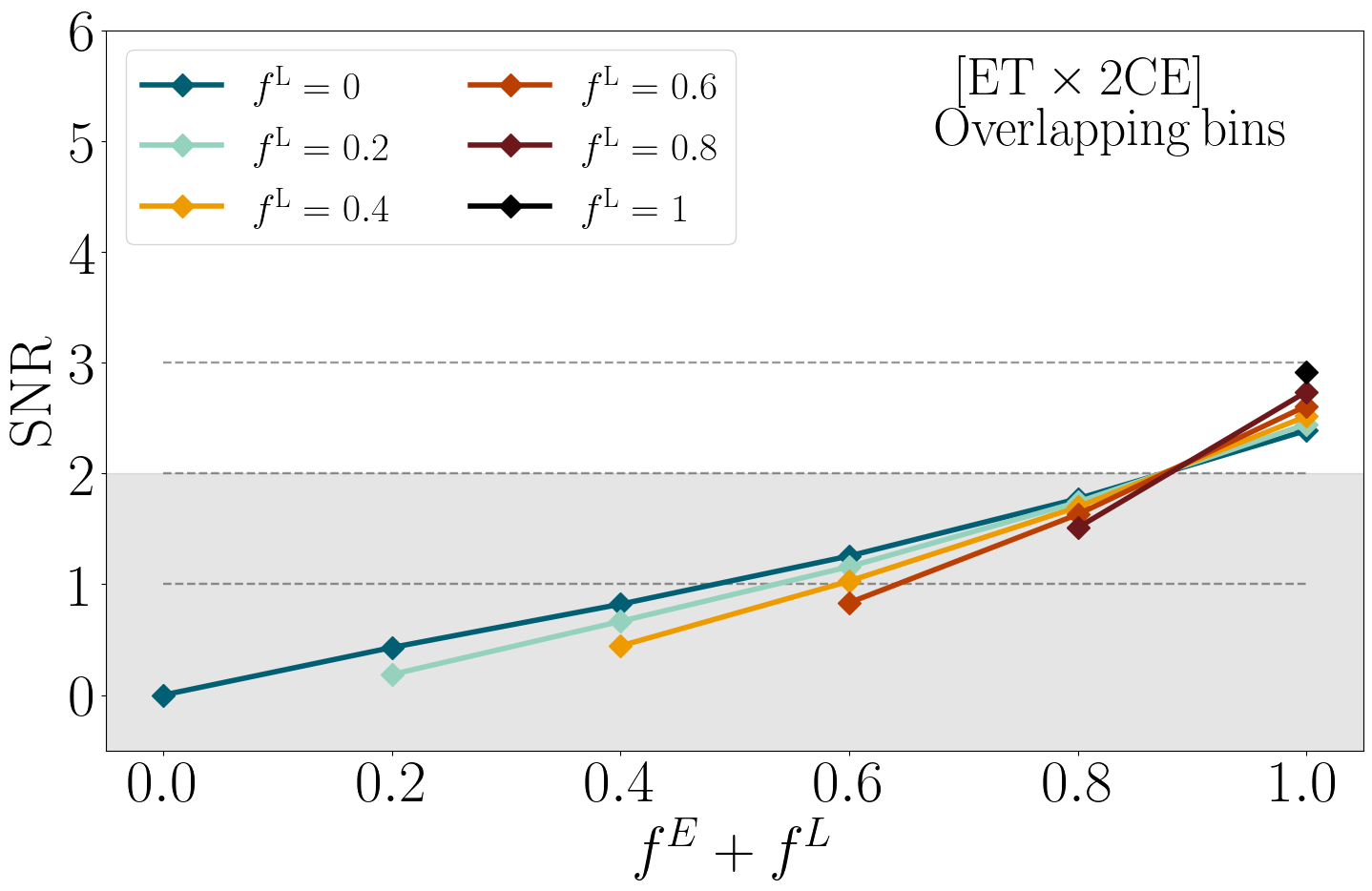}\\
  \includegraphics[width=\columnwidth]{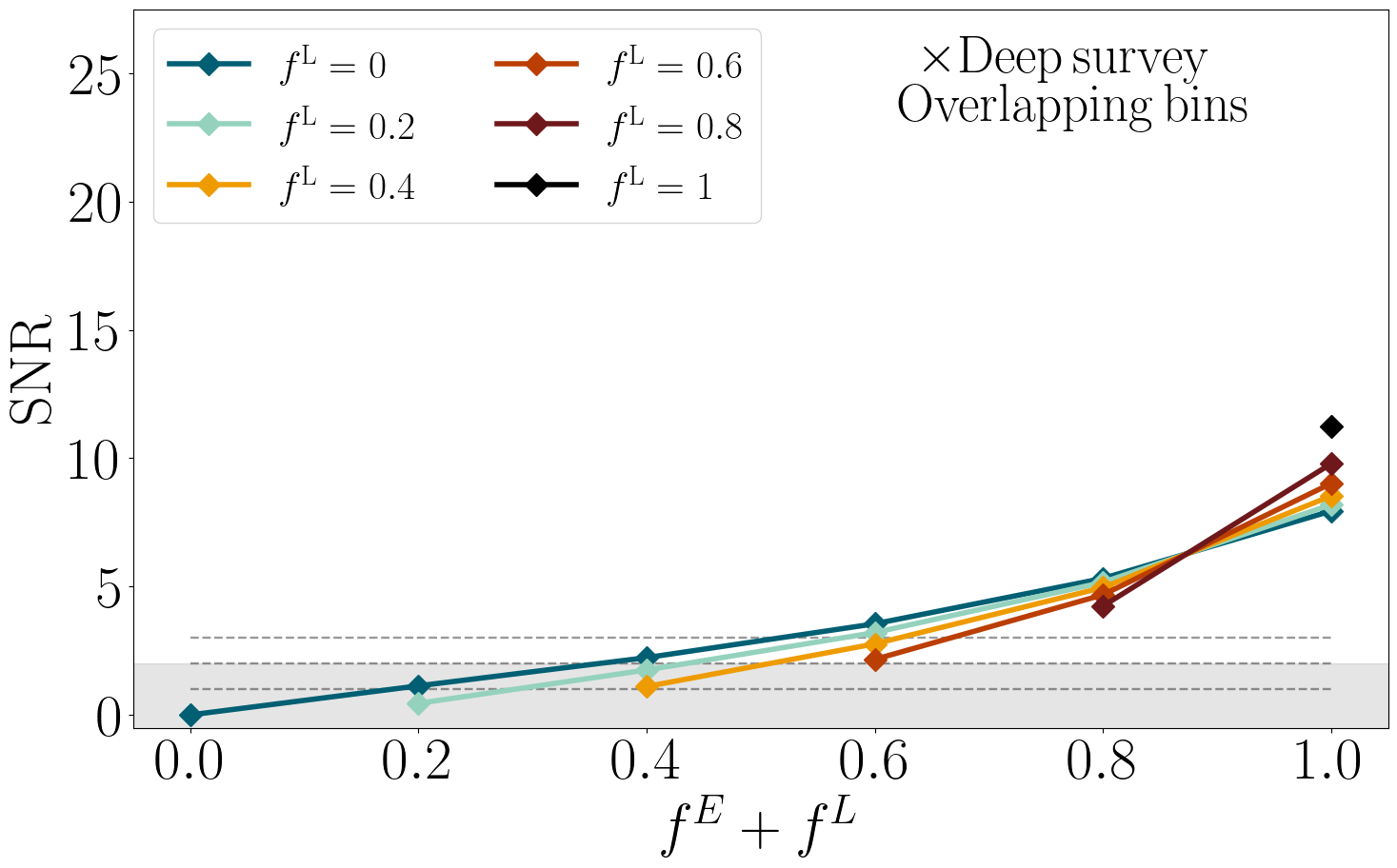}\\
  \includegraphics[width=\columnwidth]{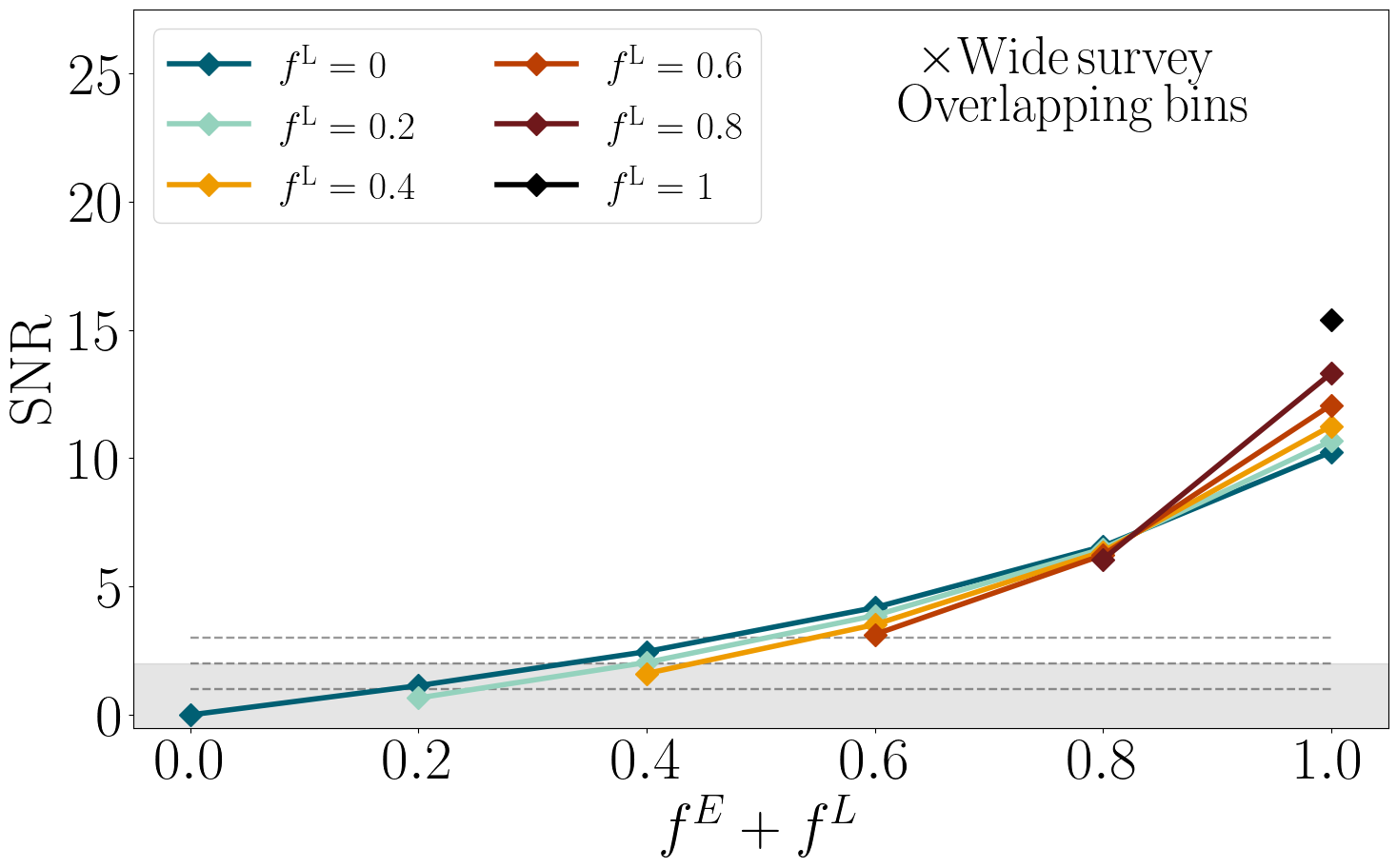}
    \end{minipage}
    \begin{minipage}{0.49\linewidth}
  \includegraphics[width=\columnwidth]{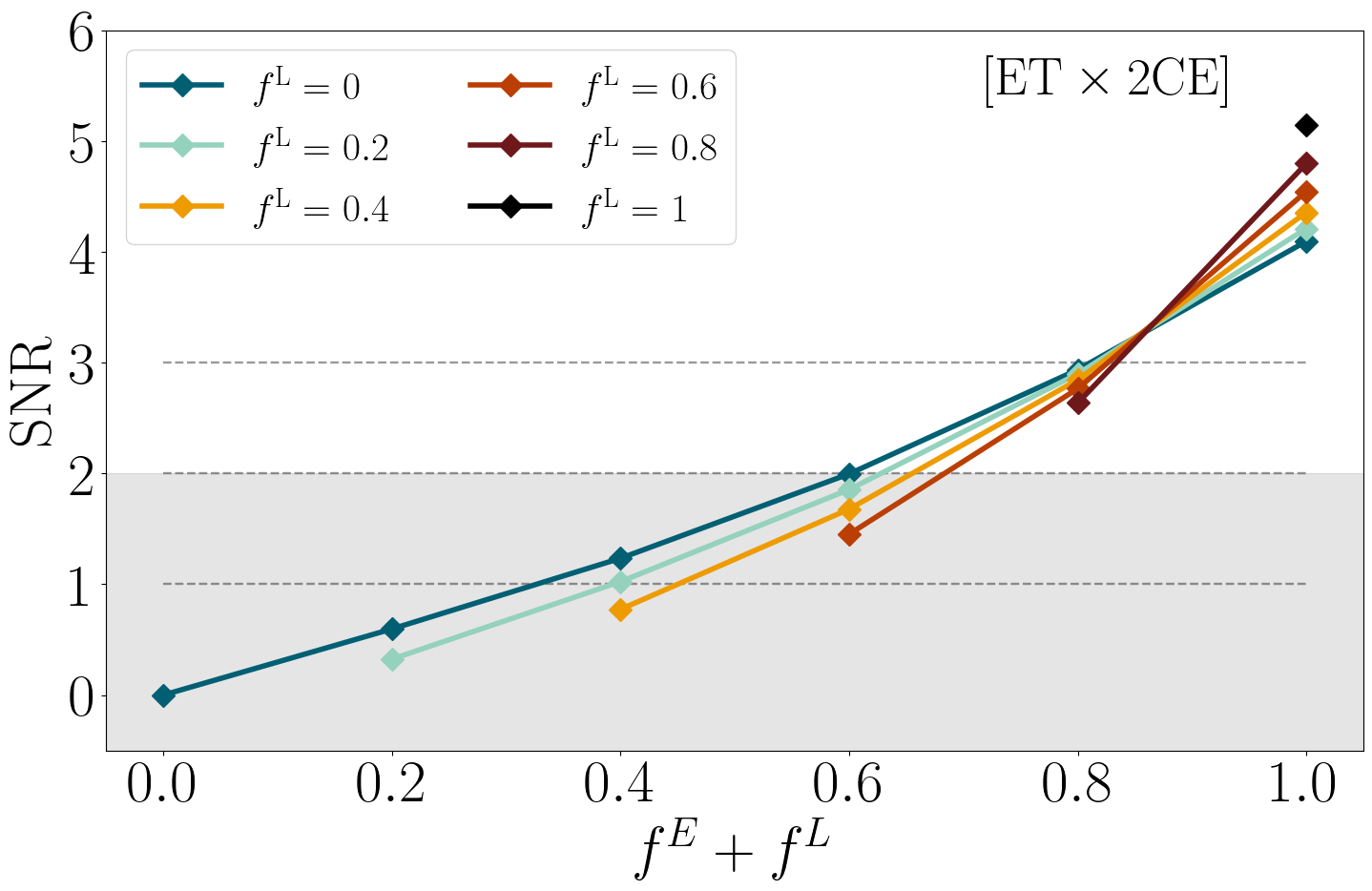}\\
  \includegraphics[width=\columnwidth]{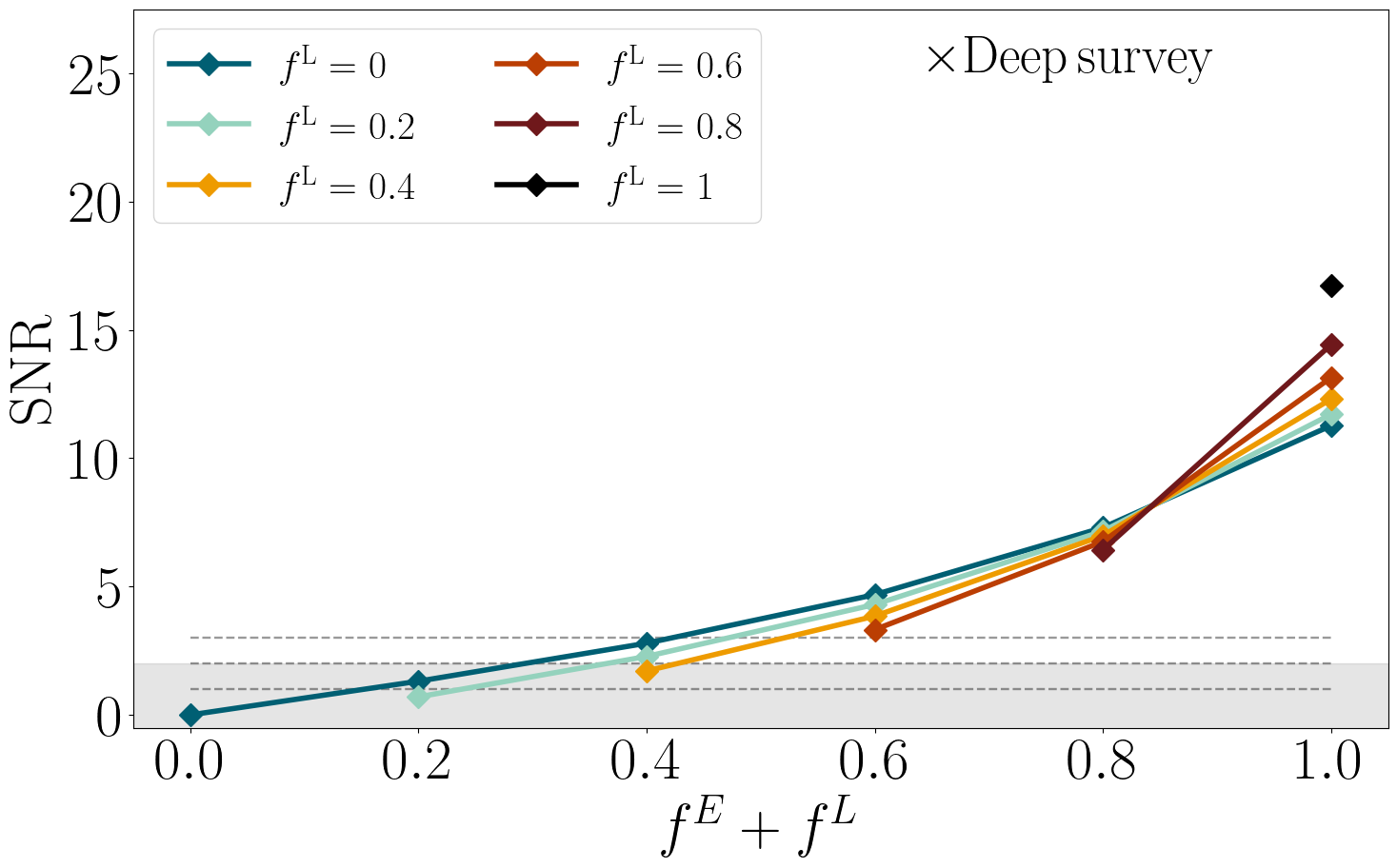}\\
  \includegraphics[width=\columnwidth]{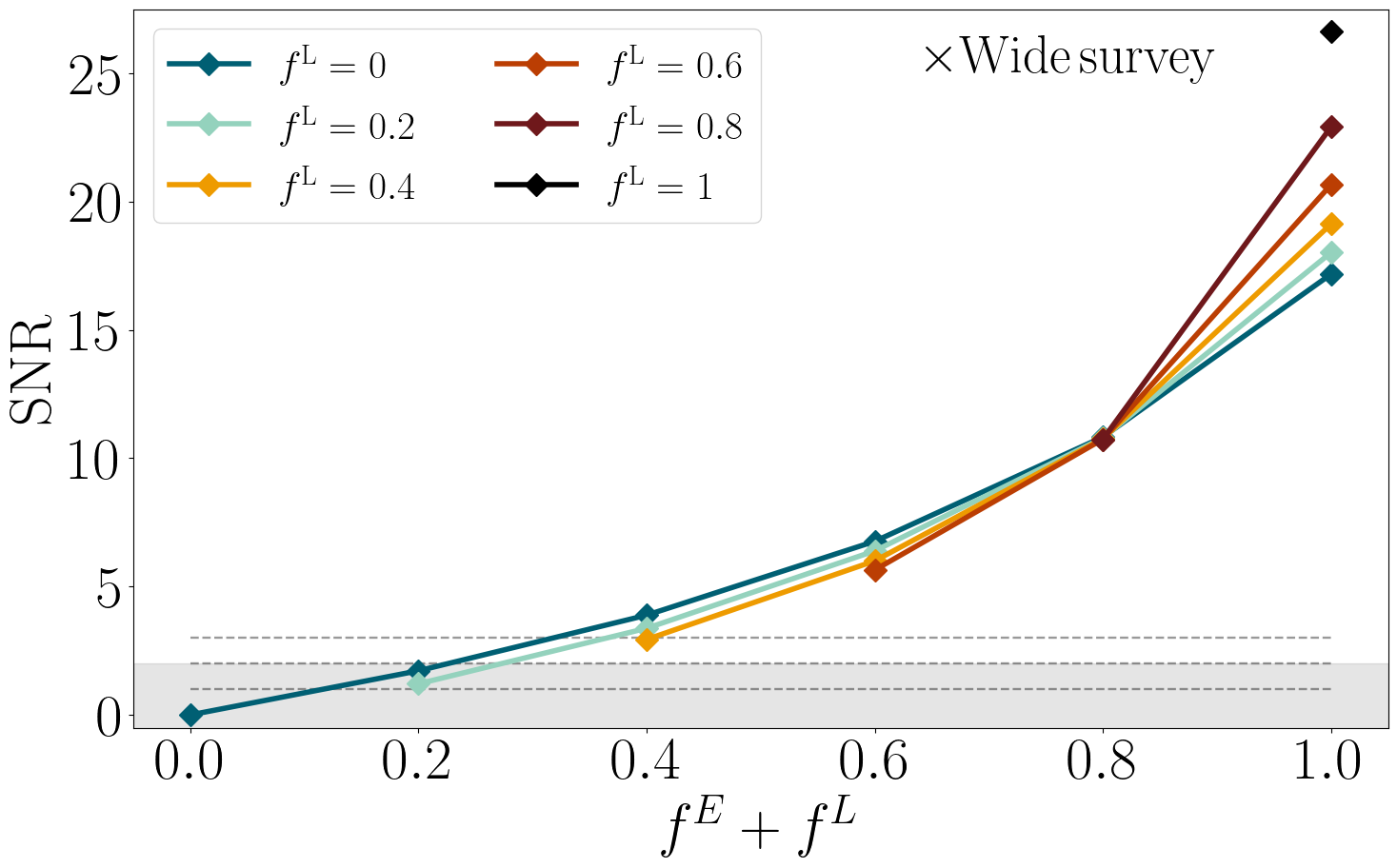}
    \end{minipage}
    \caption{SNR measured using overlapping (left) and non-overlapping (right) bins for the different $\{f^E,f^L\}$ cases, for ET2CE alone and in cross-correlation with a wide and deep survey. The horizontal lines indicates $(1,2,3)\sigma$ detection. }
    \label{fig:snr}
\end{figure}

We compute the $\rm SNR$ as in eq.~\eqref{eq:snr} to study the capability of our GW survey, both alone and in cross-correlation with galaxies, to detect the presence of PBH mergers in the observed catalog. As we show in appendix~\ref{app:cl}, the bulk of the information comes from intermediate redshift $z\sim 2$, making our analysis complementary to the use of high redshift merger rate for PBH detection~\cite{Ng_2022}. 
Figure~\ref{fig:snr} summarizes our results for ET2CE alone or in cross-correlation with galaxy surveys; the SNR values obtained, as well as results for ET, are in appendix~\ref{app:all_plots_snr}.

\begin{figure}[htb!]
    \centering
    \includegraphics[width=\columnwidth]{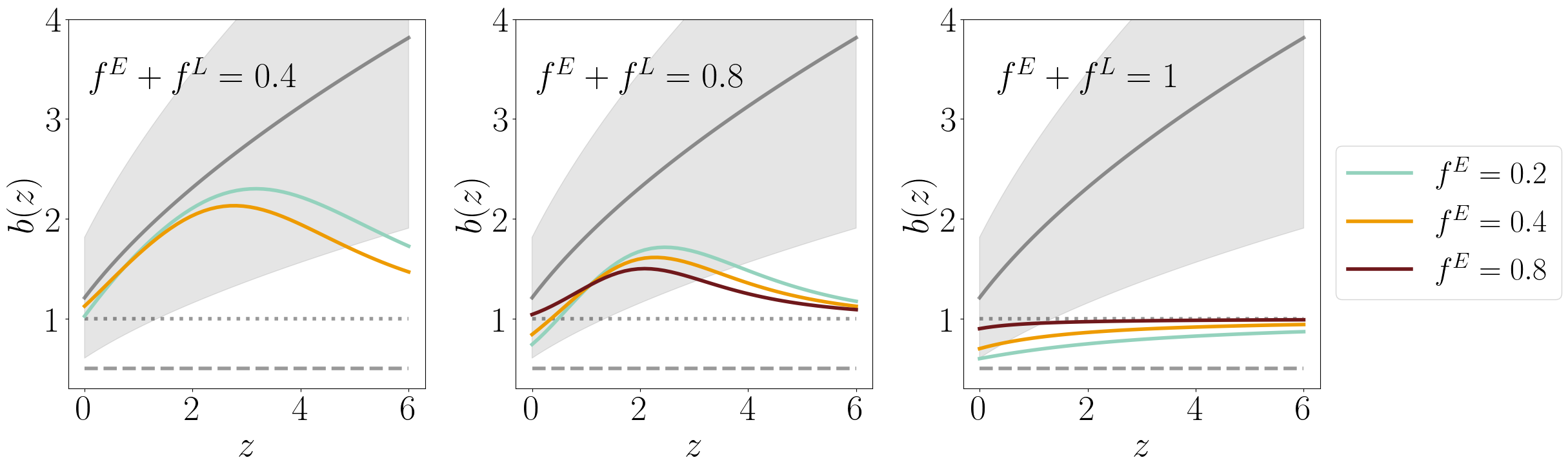}
    \caption{
    ABH bias with $50\%$ prior (gray) compared with the effective bias when $f^E+f^L = 0.4$ (left), $0.8$ (middle) or $1$ (right) with different relative abundances of EPBH anf LPBH. 
    When $f^E+f^L$ is small, a larger number of EPBH ease the detection because of the deviation at high $z$. 
    }
    \label{fig:bias_check}
\end{figure}

The cumulative signal we obtain by summing over $\ell$- and $z$-bin pairs allows us to potentially detect the presence of PBH for some models. The first thing we note is that, given a certain value for $f^E+f^L$, the $\rm SNR$ changes depending on the relative abundance of EPBH and LPBH. When $f^E+f^L$ is low, having a larger number of EPBH leads to a higher $\rm SNR$, while when $f^E+f^L$ is high, the $\rm SNR$ increases when the number of LPBH is larger. When $f^E+f^L\sim 0.8$, the $\rm SNR$ is almost constant independently from the relative abundance of EPBH and LPBH. This may seems surprising at first glance: since $b^L < b^E$, we would expect the same $f^E+f^L$ to have higher $\rm SNR$ with a larger number of LPBH.
However, the effective bias we are using weighs each contribution by its merger rate: as it can be seen in figure~\ref{fig:model_fid}, this is higher for EPBH, particularly at high redshift. 
The balance between these two effects is evident in figure~\ref{fig:bias_check}: a large value of $f^E$ produces a large enough deviation on the effective bias at high $z$ to facilitate the detection when $f^E+f^L$ is small, while a large value of $f^L$ requires an overall higher number of PBH to be effective. 

In the case of only ET2CE, our results show that at least a $2\sigma$ detection could be claimed when PBH binaries constitute more than $\sim 60\%$\,-\,$80\%$ of the observed number of events in the GW catalog, in the case of non-overlapping bins. 
As appendix~\ref{app:all_plots_snr} shows, using a single ET instead leads to no constraining power in the single tracer case, because of the lack of small scales. 
When including cross-correlations with galaxy catalogs, the SNR gets larger even with a smaller amount of PBH binaries in the catalog and we get to $2$\,-\,$3\sigma$ whenever PBH $\gtrsim 40\%$. Cross-correlations between ET and galaxy surveys 
instead reach $1\sigma$ detection only when the number of PBH is very large ($f^E+f^L\gtrsim 80\%$, see appendix~\ref{app:all_plots_snr}). 

When overlapping redshift bins are used, all results degrade by a factor of almost two on average, meaning that similar PBH abundances lead to $1$\,-\,$2\sigma$ detection in the case of ET2CE. Cross-correlations still allow a $3\sigma$ detection when $f^E+f^L \geq 60\%$, reaching $\sim 2\sigma$ for $f^E+f^L \sim 40\%$ instead.

As we will confirm with the Fisher matrix analysis in section~\ref{sec:bias_forecast}, even if both the galaxy surveys lead to an improvement in the capability of distinguish the effective bias from the only-ABH bias, a wide survey allows to better measure the bias. On the other hand, 
even if both the surveys probe high redshifts, the deep survey 
observes a larger number of sources at $z > 2$ (compare with figure~\ref{fig:surveys}) and its cross-power spectra have lower shot noise in the last bin ($z\in[2.4,6.2]$). For the wide survey,  
the larger $f_{\rm sky}$ allows us to access larger $\ell$s, which explains its larger SNR. 
Even if the SNR for the cross-correlation with the wide survey  
is 50\%-100\% times larger than the deep one, the two analyses present a similar trend with respect to the relative abundances of EPBHs and LPBHs. This is because we use only 5 redshift bins and the analysis can not capture the details in the redshift evolution of the source number distribution or bias. Even when the PBH contribution to the mergers can be detected, the relative EPBH and LPBH abundances remain unconstrained with this method.

\subsection{Bias forecasts}
\label{sec:bias_forecast}

The analysis studied in the previous section does not take into account degeneracies between parameters that influence the amplitude of the angular power spectra.  
To strengthen the reliability of our results, we study forecasts constraining power of the different surveys in measuring the effective GW bias parameters. We perform a Fisher matrix analysis
\begin{equation}\label{eq:fisher}
    F_{\alpha\beta} = f_{\rm sky} \sum_{\ell} \frac{2\ell+1}{2} \text{Tr} \left[ \left(\mathcal{C}_{\ell}\right)^{-1} \partial_{\alpha}\mathcal{C}_{\ell} \left(\mathcal{C}_{\ell}\right)^{-1} \partial_{\beta}\mathcal{C}_{\ell} \right],
\end{equation}
with respect to the parameters $\theta_{\alpha,\,\beta} = \{b_1,b_2,b_3,b_4,b_5\}$, each of which represents the effective bias inside one of the redshift bins. Their fiducial values equal the ABH bias from eq.~\eqref{eq:ABHbias} in the central point of the bins.
The covariance matrix $\mathcal{C}_\ell$ (and analogously the derivative matrix $\partial_\alpha \mathcal{C}_\ell$) for the GW survey alone is built as
\begin{equation}
    \mathcal{C}_\ell^{\rm GWGW} = \begin{pmatrix}
    \tilde{C}_{\ell,11}^{\rm GWGW}+\bar{n}_{\rm GW,1}^{-1} & ... & \tilde{C}_{\ell,0N_{\rm bin}}^{\rm GWGW} \\
    ... & ... & ... \\
    \tilde{C}_{\ell,N_{\rm bin}1}^{\rm GWGW} & ... & \tilde{C}_{\ell,N_{\rm bin}N_{\rm bin}}^{\rm GWGW}+\bar{n}_{{\rm GW},N_{\rm bin}}^{-1} \\
    \end{pmatrix}\,,\end{equation}
being $N_{\rm bin}=5$ the overall number of bins used and $\bar{n}_{{\rm GW},i}^{-1}$ the shot noise in the $i$-th bin.
In the multi-tracer scenario instead, the computation of the Fisher matrix requires to update eq.~\eqref{eq:fisher} by using the block-matrix covariance 
\begin{equation}\label{eq:cov_cross}
 \mathcal{C}_\ell^{\rm cross} = \begin{pmatrix}
  \mathcal{C}_\ell^{\rm gg}  &  \mathcal{C}_\ell^{\rm gGW} \\
   \mathcal{C}_\ell^{\rm GWg}  &  \mathcal{C}_\ell^{\rm GWGW} \\
 \end{pmatrix}
\end{equation}
and the derivative matrix defined analogously. 
Moreover, in the multi-tracer case the parameter set used to compute the Fisher matrix also includes the galaxy bias parameters $\{b_{g,1},b_{g,2},b_{g,3},b_{g,4},b_{g,5}\}$, which we marginalize before getting our final results on the GW biases. 
Note that in the previous section the constraints were intergrated over $z$ to obtain the overall $\rm SNR$, while here we separate the information to constrain the bias in different bins.

Table~\ref{tab:bias_forecasts_cross} shows the relative marginalized errors on the bias parameters $\sigma_{b_i}$, both when considering only the GW survey and when including the cross-correlation with galaxies (see appendix~\ref{app:all_plots_snr} for results on ET). Once again, the fiducial for the bias parameters is set to the case where only ABH contributes to the signal.
Our main results assume an uninformative prior on both the merger and galaxy bias; we also consider the possibility of a $50\%$ Gaussian prior on the ABH bias based on the results in~\cite{Peron:2023}. For the cross-correlation analysis, we separately run the Fisher forecast on the only-galaxy surveys, so to estimate their (optimistic) constraining power on the galaxy bias. We checked that using these results as priors for the multi tracer analysis the results do not largely improve, confirming the stability of our analysis. 

\begin{table}[ht!]
    \centering
\renewcommand{\arraystretch}{1.2}
    \begin{tabular}{|c|ccccc|ccccc|}
    \hline
    & \multicolumn{5}{c|}{Overlapping bins}
    & \multicolumn{5}{c|}{Non-overlapping bins}\\
         & $b_1$ & $b_2$ & $b_3$ & $b_4$ & $b_5$& $b_1$ & $b_2$ & $b_3$ & $b_4$ & $b_5$ \\
    \hline
        ET2CE & 85\% & 37\% & 25\% & 31\% & 115\% & 40\% & 19\% & 13\% & 19\% & 87\% \\
        ET2CE, ${b_i}^p$ prior & 43\% & 30\% & 22\% & 26\% & 46\%& 31\% & 18\% & 13\% & 17\% & 43\% \\
        \hline
        $\times$Wide survey & 12\% & 7\% & 6\% & 13\% & 74\%&  9\% & 5\% & 4\% & 8\% & 95\% \\
        \hline
        $\times$Deep survey  & 24\% & 13\% & 10\% & 12\% & 50\%& 17\% & 9\% & 7\% & 9\% & 46\% \\
    \hline
    \end{tabular}
    \caption{Marginalized $1\sigma$ relative errors on ABH bias parameters for the GW survey and the cross-correlation with the different galaxy surveys we take into account. }
    \label{tab:bias_forecasts_cross}
\end{table}

Figure~\ref{fig:bias_forecast} shows predicted errors on the ABH bias for ET2CE in our fiducial case $f^E = 0.2$ (see appendix~\ref{app:all_plots_snr} for the other cases and for ET alone). Coherently with results in the previous section, for non-overlapping bins the observed PBH-induced effective bias can be distinguished from the ABH bias when $f^E+f^L\gtrsim 0.6-0.8$ with GWs only, $\gtrsim 0.4$ with galaxy cross-correlations (i.e.,~$\sim 60\%$\,-\,$80\%$ and $40\%$ of the observed mergers are due to primordial black holes), while not being very sensitive to EPBH and LPBH relative abundances. Marginal errors computed using overlapping bins are $\sim 1.5$\,-\,$2$ times larger than non-overlapping cases. To further improve the bias constraints, higher order statistics e.g.,~the bispectrum can be used~\cite{Matarrese:1997sk}. In the case of GW surveys, the main limit to this approach comes from the poor angular resolution and redshift determination; a possible way out is to use the integrated power spectrum, projected on the sphere~\cite{DiDio:2018unb,Jung:2020zne}. 

Because of the redshift evolution of ABH number density in eq.~\eqref{eq:ABHrate}, the shot noise allows us to have better constraints on ABH bias at intermediate-high redshifts, namely between $z\sim 1$ and $\sim 3$ (consistent with the previous section). This is therefore the range in which our technique is more sensitive to deviations due to PBH contributions and makes our analysis complementary to the use of high $z$ merger detections to test the PBH existence~\cite{Ng_2022}. Moreover, since measurements at higher $z$ are more challenging, intermediate redshift probes can be seen as more reliable. In the case of bias measurements, a strong improvement at high $z$ can be obtained via cross-correlation with the deep survey, thanks to the large number of sources this instrument measures at $z> 3$, as figure~\ref{fig:surveys} shows. On the other side, cross-correlations with the wide survey are more effective to improve constraints at $z\lesssim 3$.

\begin{figure}[ht!]
    \centering
    \begin{minipage}{0.49\linewidth}
  \includegraphics[width=\columnwidth]{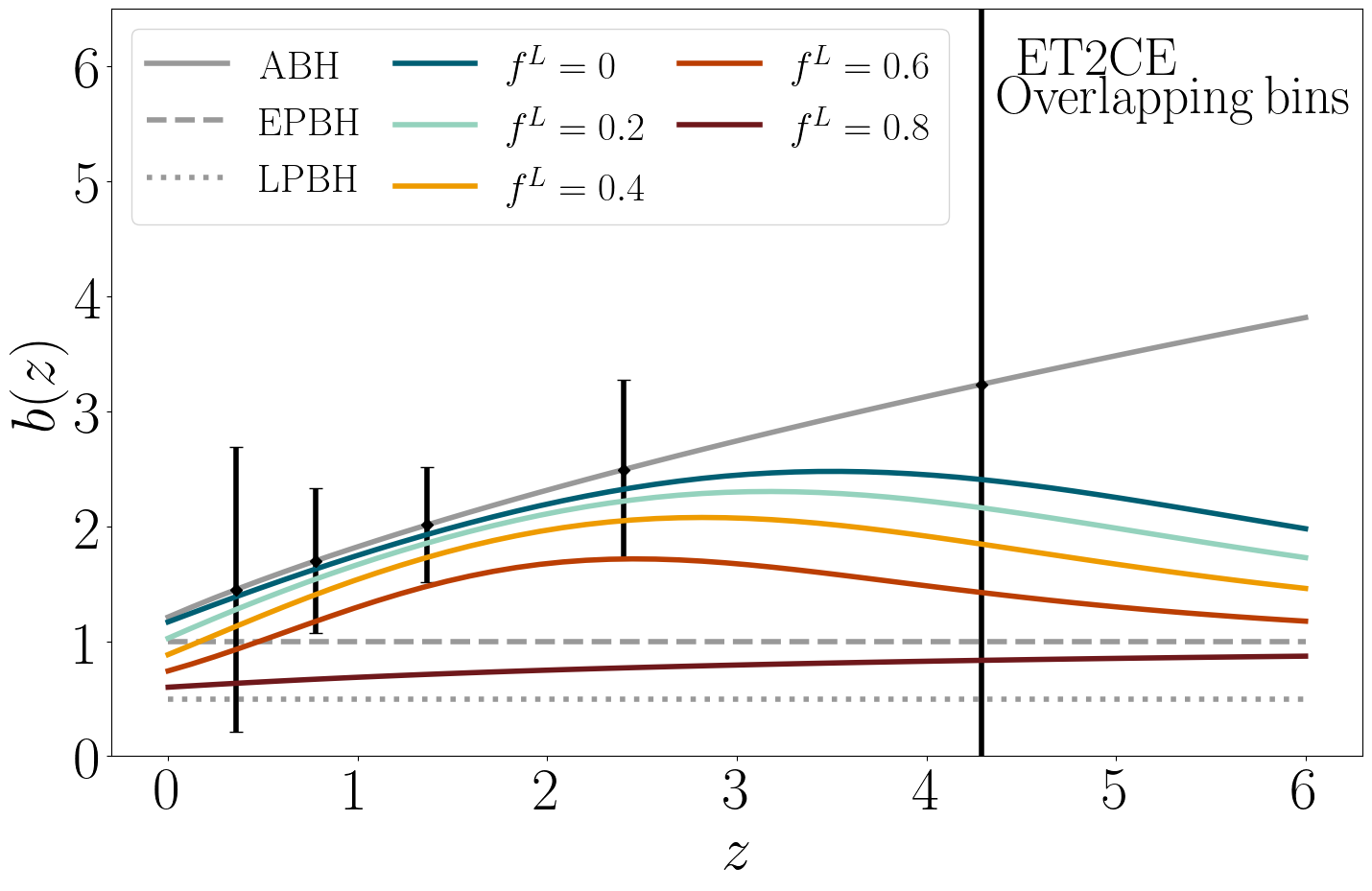}\\
  \includegraphics[width=\columnwidth]{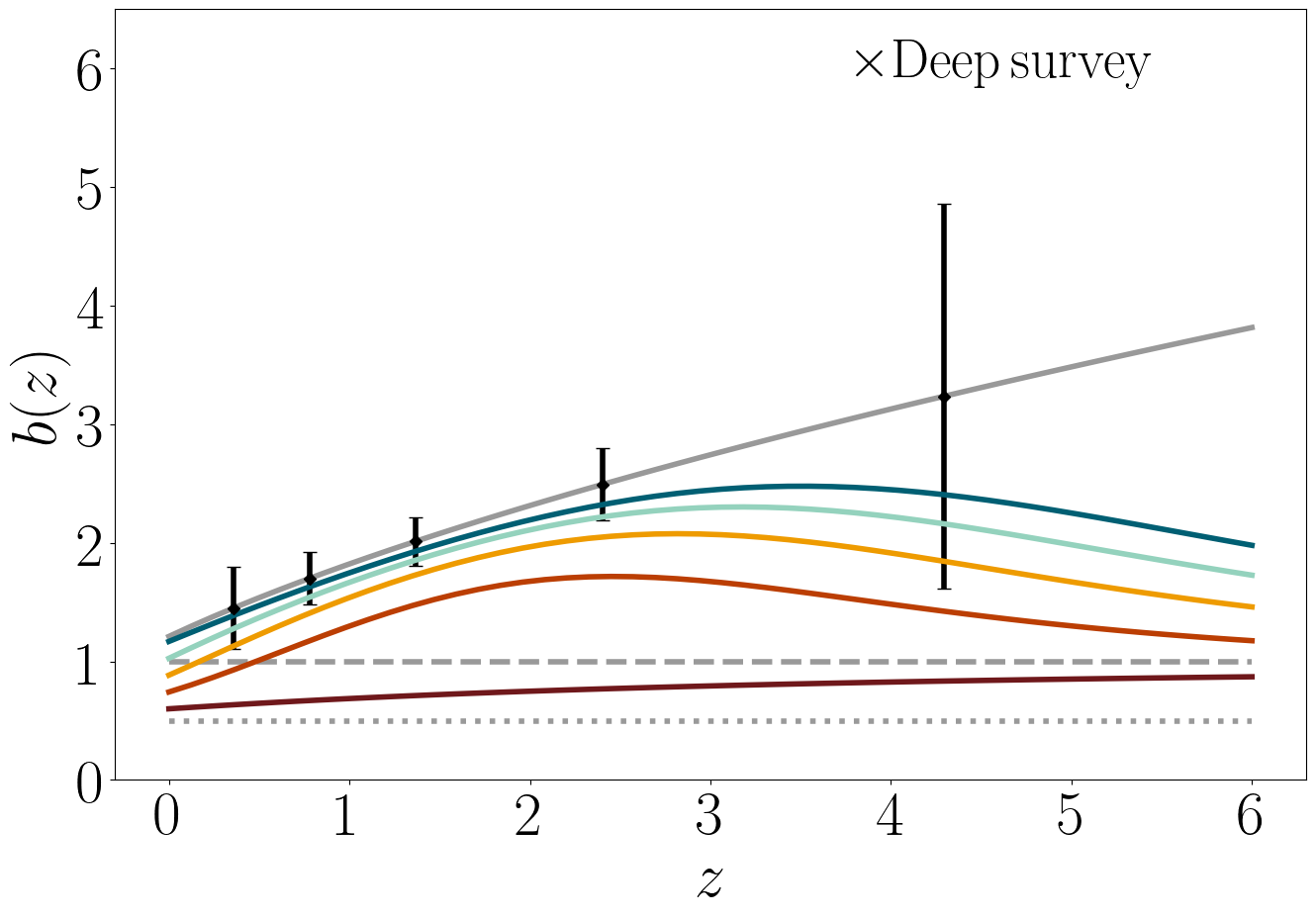}
  \includegraphics[width=\columnwidth]{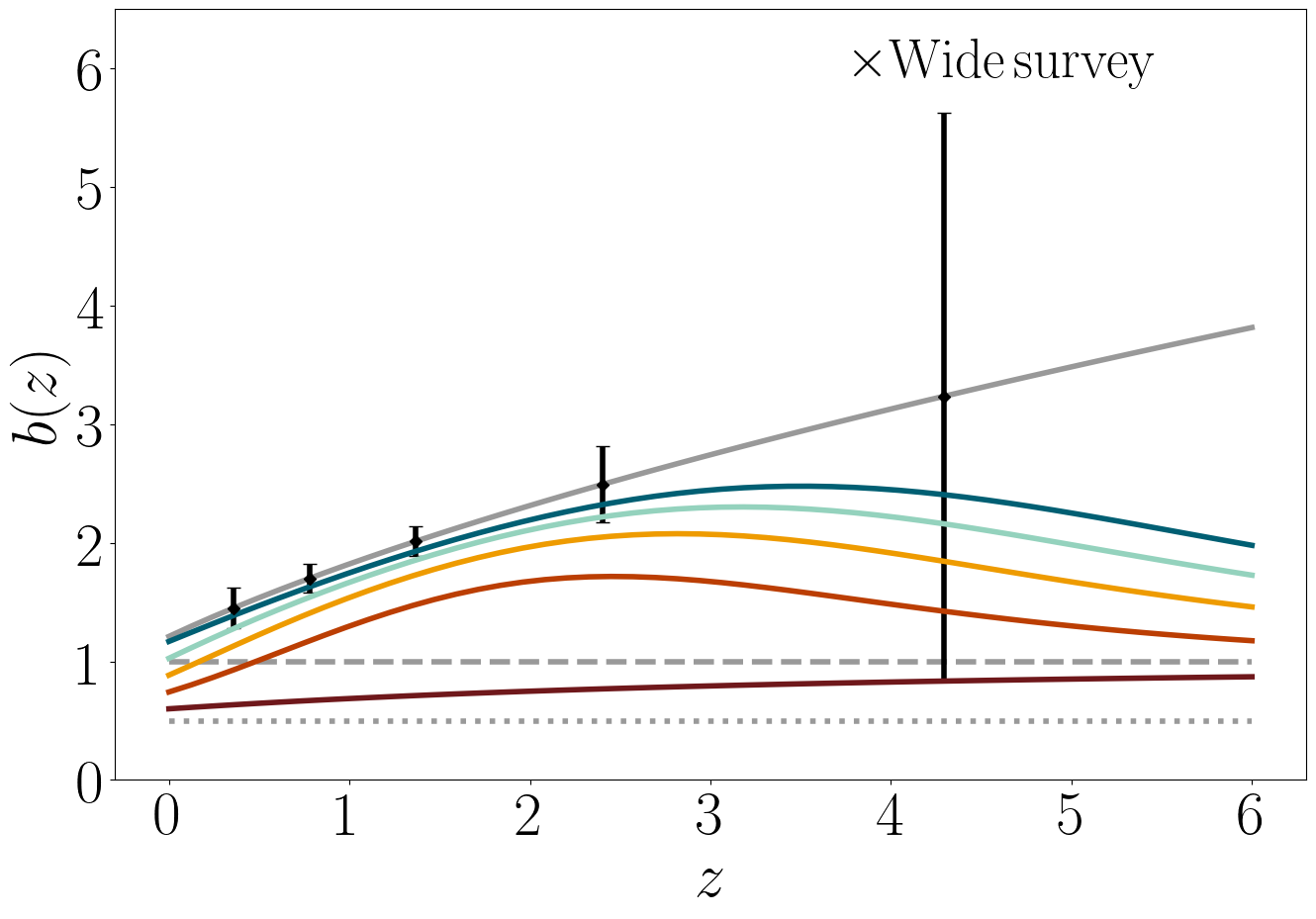}
    \end{minipage}
    \begin{minipage}{0.49\linewidth}
  \includegraphics[width=\columnwidth]{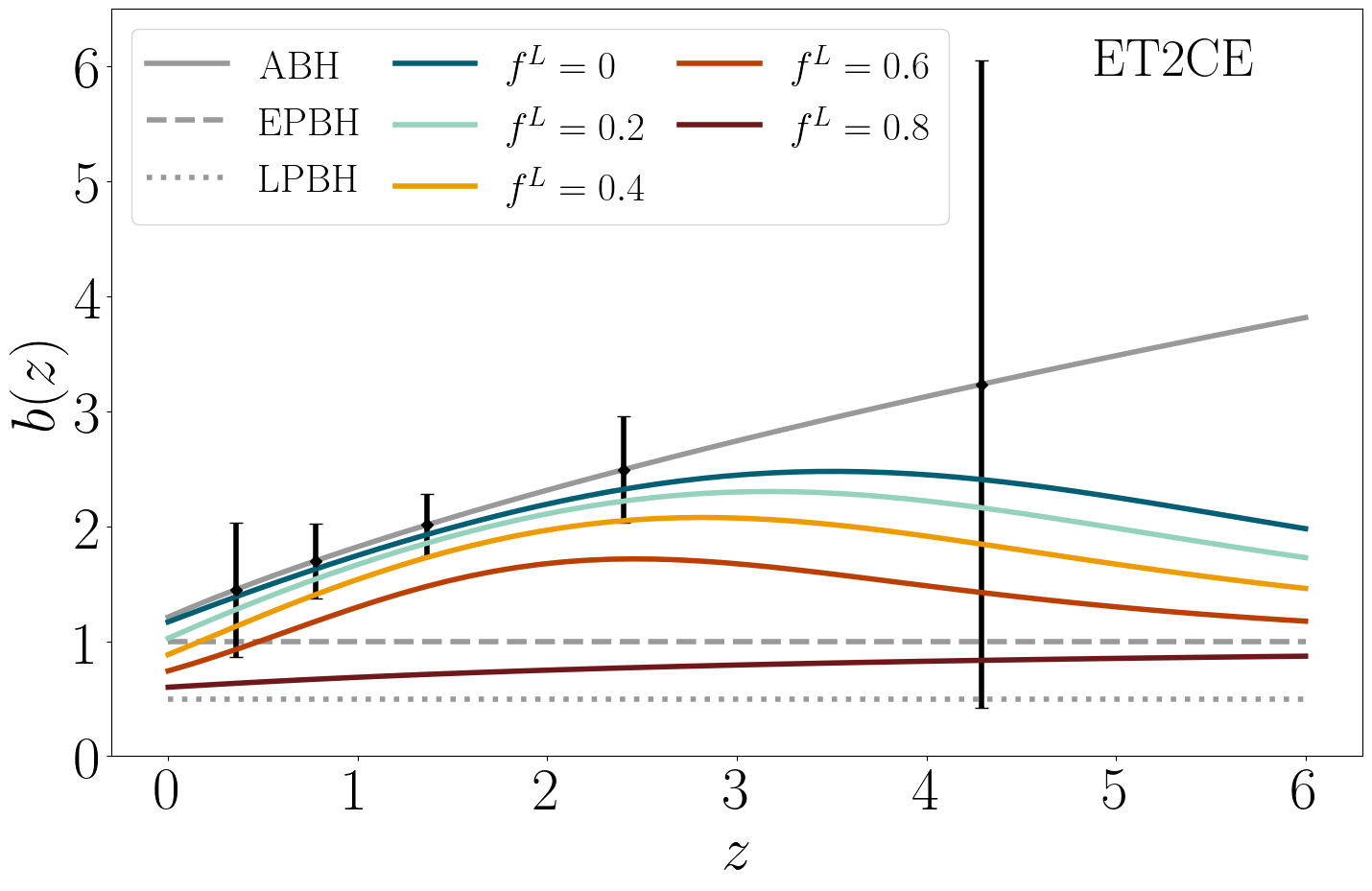}\\
  \includegraphics[width=\columnwidth]{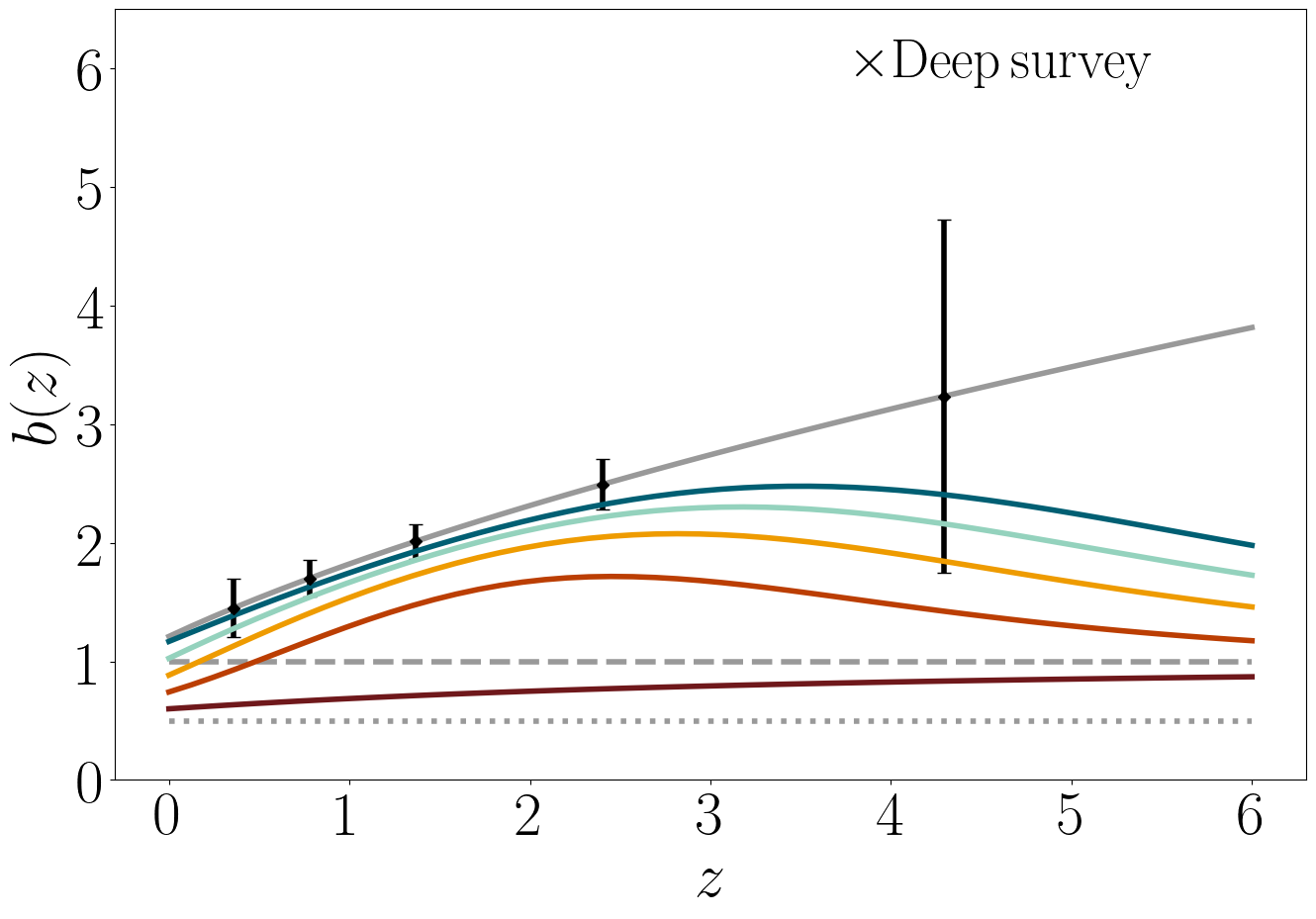}
  \includegraphics[width=\columnwidth]{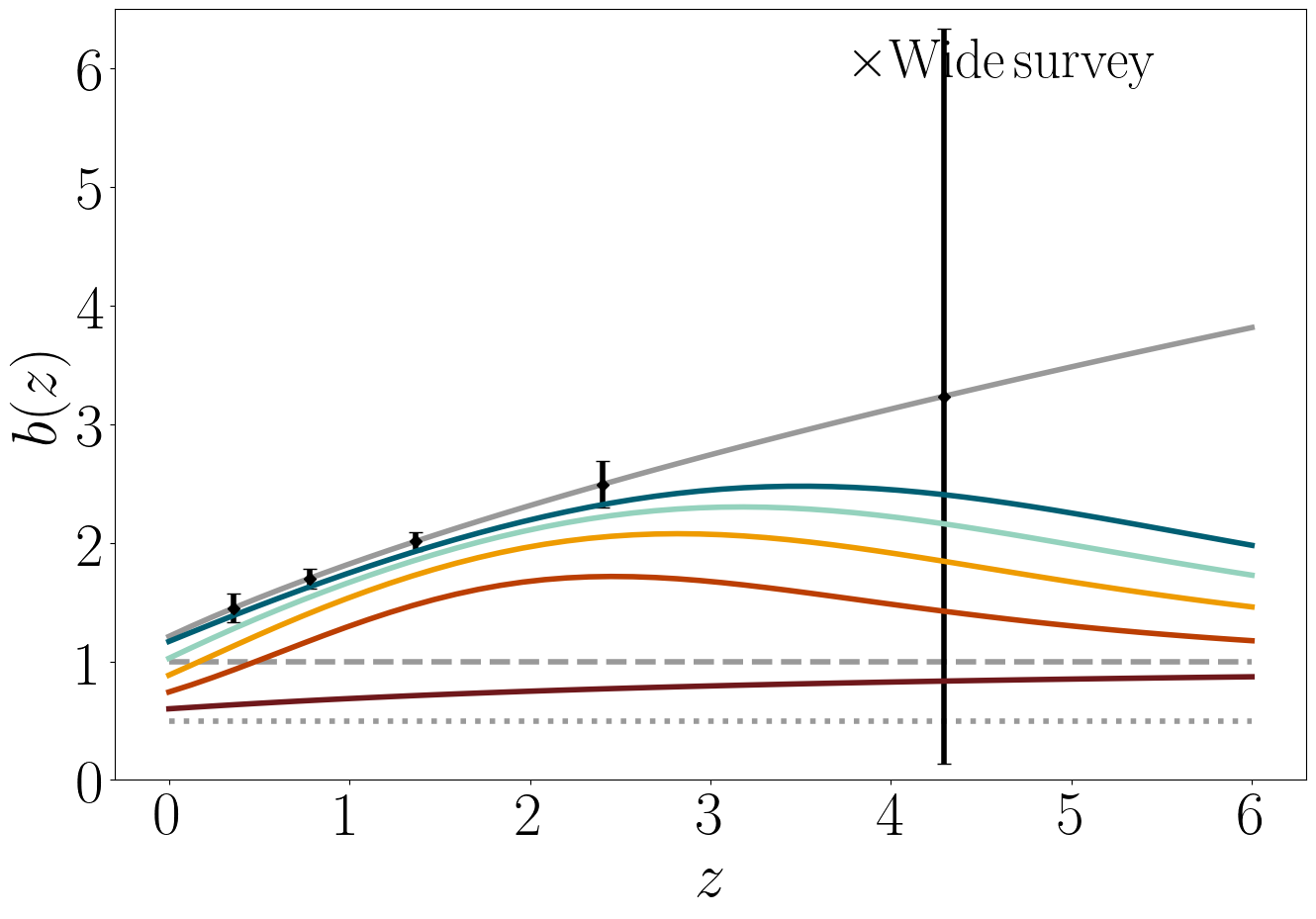}
    \end{minipage}
    \caption{Predicted $1\sigma$ marginalized errors on ABH bias parameters with ET2CE and in cross-correlation with galaxy surveys, using overlapping (left) and non-overlapping (right) bins.
    Errorbars are compared with the effective bias for the models with $f^E = 0.2$.}
    \label{fig:bias_forecast}
\end{figure}

\begin{figure}[ht!]
    \centering
\includegraphics[width=.49\columnwidth]{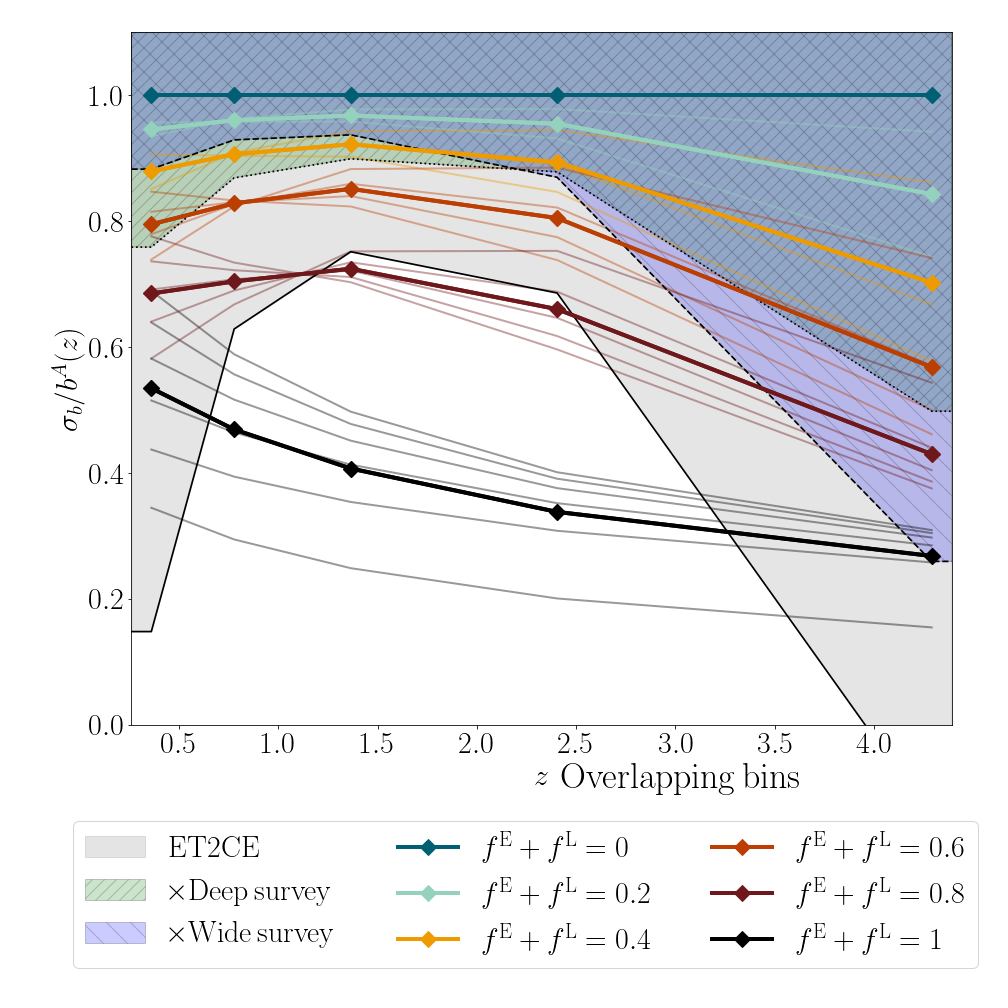}
\includegraphics[width=.49\columnwidth]{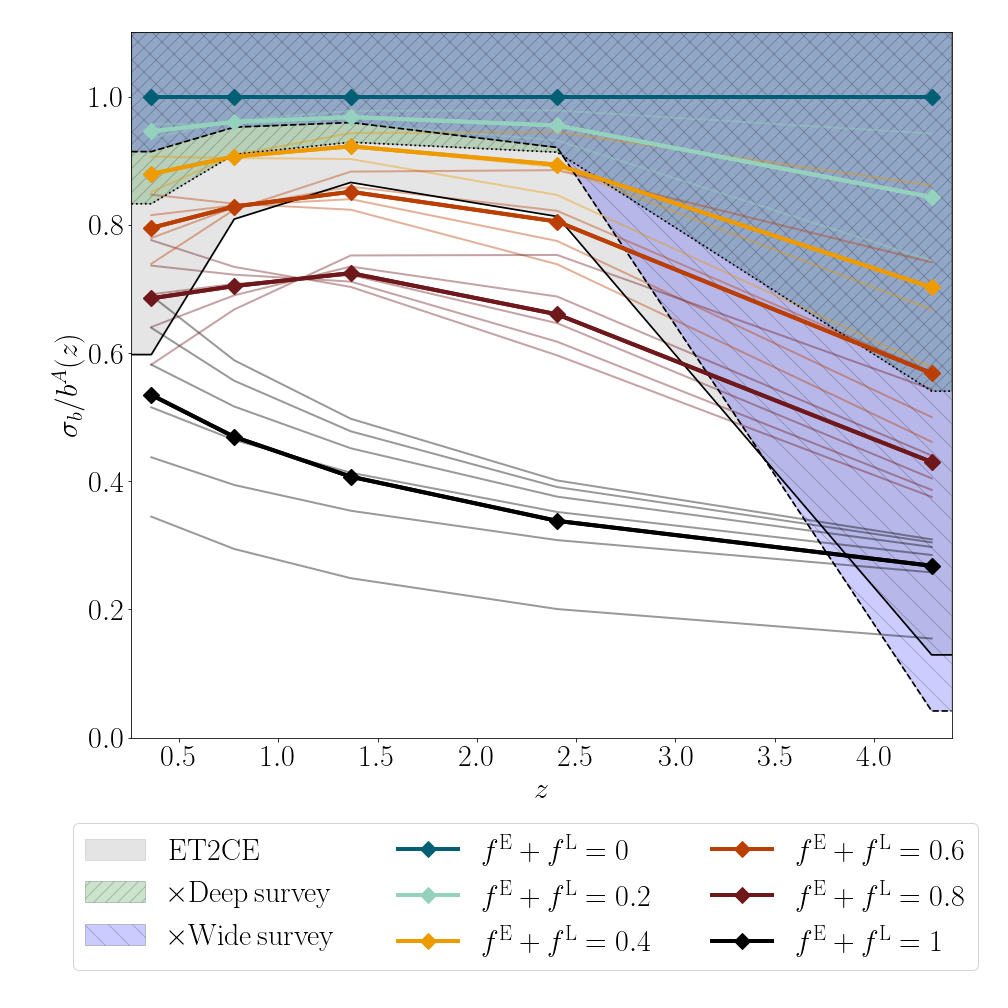}
    \caption{Mean value of the effective merger bias for each $f^E+f^L$ case; lighter lines show different $\{f^E,f^L\}$ combinations,using overlapping (top) and non-overlapping (bottom) bins. Colored areas represent $1\sigma$ confidence intervals obtained through our Fisher analysis for ET2CE, and the cross-correlations with galaxies. 
    Future surveys will be sensitive to PBH abundances providing effective bias values that fall outside the colored areas.
    }
    \label{fig:overall_fisher}
\end{figure}

Since $f^E$ and $f^L$ are almost degenerate in our analysis, we can describe the effective bias $b(z)$ for each value of $f^E+f^L$ using the mean $\bar{b}(z)$ and the standard deviation of the values it assumes in the different $\{f^E,f^L\}$ cases. To visualize the final results of our Fisher analysis, therefore, in figure~\ref{fig:overall_fisher} we compare $\bar{b}(z)/b^A(z)$ for each value of $f^E+f^L$ with the $1\sigma$ confidence level of the fiducial only-ABH scenario. The setup we adopted maximizes the sensitivity to deviations from the only-ABH scenario in the intermediate redshift bins, where the ABH shot noise is smaller. When only ET2CE is adopted, the effective bias deviates $\sim 1\sigma$ from the ABH bias when $f^E+f^L \gtrsim 0.8$, namely when PBH mergers constitute at least $80\%$ of the totality. Cross-correlations with galaxies provide deviation from only-ABH for $f^E+f^L \gtrsim 0.4$, namely when at least 40\% of the overall merger events observed have primordial origin.

A possible way to overcome the degeneracy between $f^E$, $f^L$ in the effective bias is to combine this observable with the merger rate and its redshift dependence. In fact, while the shape of the effective bias is mainly determined by $f^E+f^L$, the values of the two parameters separately have a different impact on the evolution of the merger rate (this can be noted,~comparing figures~\ref{fig:model_fid},~\ref{fig:models_0},~\ref{fig:models_02},~\ref{fig:models_06},~\ref{fig:models_08},~\ref{fig:models_1}). However, since the effective bias and the merger rate estimations would come from the same data-set, to combine the two observables one needs to properly account for the covariance between them, the evaluation of which goes beyond the scope of this work. 

\subsection{Constraints on $f_{\rm PBH}$}
\label{sec:constrain_fpbh}

In the previous sections we showed that GW clustering can be a tool to constrain the existence of PBHs. We parameterized EPBH, LPBH contributions to future GW catalogs with as few model assumptions as possible (maintaining robustness). 
From the cosmological point of view, the quantity of interest is the parameter $f_{\rm PBH}$, introduced in eq.~\eqref{eq:f_PBH_DM}. This describes how much dark matter is comprised of PBHs, either bounded in binaries or isolated. 
To convert the results from sections~\ref{sec:snr},~\ref{sec:bias_forecast} into constraints on $f_{\rm PBH}$, we need to assume a specific model for early and late binary formation, provided the uncertainties that still exist in the literature. 

The goal of this section is to relate the local merger rates $\mathcal{R}_0^{E,L}=f^{E,L}\mathcal{R}_0^{\rm tot}$ to $f_{\rm PBH}$.  We recall that in this work we consider black holes with masses $M_{\rm PBH}\sim 5-100\, M_\odot$, which will be observed by ET and CE~\cite{ligo2021}.
For early binaries, we follow~\cite{yacine2017} and  roughly estimate the local merger rate for a monochromatic PBH mass function as
\begin{equation}\label{eq:fPBH_e}
\begin{aligned}
    \mathcal{R}_0^E =& \frac{1}{2}\frac{f_{\rm PBH}}{0.85}\frac{\rho_m}{M_{\rm PBH}}\frac{dP}{dt}\biggr|_{t_0}\\
    \simeq&\, 
    5.10\times 10^5{\,\rm Gpc^{-3}yr^{-1}}\left(\frac{f_{\rm PBH}}{0.85}\right)^2\left(\frac{M_{\rm PBH}}{30\,M_\odot}\right)^{-32/37}\left[\left(\frac{f_{\rm PBH}}{0.85}\right)^2+\sigma_{\rm eq}^2\right]^{-21/74},
\end{aligned}
\end{equation}
where $\rho_m$ is the matter density at present time $t_0$ and $dP/dt$ the probability distribution of the time of merger, which can be computed as a function of $M_{\rm PBH}$, $f_{\rm PBH}$ and the variance of DM density perturbations due to non-PBH components at matter-radiation equality, $\sigma_{\rm eq}\sim 0.005$.
In our work, $\mathcal{R}_0^E = f^E\mathcal{R}_0^{\rm tot}$, therefore we can numerically solve the previous equation to get $f_{(\rm PBH|E)}$, namely the overall DM fraction in the form of EPBH. 
For late binaries, instead, we adopt the formalism in~\cite{didligo}, which provides
\begin{equation}\label{eq:fpbh_l}
\begin{aligned}
    \mathcal{R}_0^{L} =& \,(f_{\rm PBH})^{53/21} \int dM_h\frac{dn}{dM_h}\mathcal{R}_h(M_h) \\
    \simeq& \,\mathcal{R}_0^{\rm tot} (f_{\rm PBH})^{53/21}\left(\frac{M_h}{400\,M_\odot}\right)^{-11/21}\,,
\end{aligned}
\end{equation}
where $\mathcal{R}_h(M_h)$ is the merger rate in an halo of mass $M_h$ and $dn/dM_h$ the halo mass function. The mass dependence is negligible and we normalize the result to the merger rate observed by the detector. 
Using $\mathcal{R}^L_0 = f^L\mathcal{R}_0^{\rm tot}$, we invert the equation to get $f_{(\rm PBH|L)} = \left({f^L}\right)^{21/53}$.

Clearly, the constraints on $f_{\rm PBH}$ obtained through this procedure will be model dependent; in particular, for late binaries they will be affected by assumptions on the host halo properties, while uncertainties for early binaries arise from effects that could alter the binary formation process and merger time (primordial non-Gaussianity~\cite{Young_2015,Young_2015_1,Young_2020}, non-linearities~\cite{Young_2019,deluca_2020}, nearest and next-to-nearest neighbor interactions~\cite{Ballesteros_2018,Atal_2020}, etc.). Since model uncertainties are larger for early binary scenarios, we collect all the uncertainties in the parameter $\mathcal{U} = \mathcal{R}_{0}^{E}/\mathcal{R}_{0}^{E,nom}$ where $\mathcal{R}_{0}^{E,nom}$ is computed via eq.~\eqref{eq:fPBH_e} in the {\it nominal} scenario, i.e.,~relying on assumptions in~\cite{yacine2017}. Constraints on $f_{\rm PBH}$ are then computed as a function of $\mathcal{U}$, which can be seen as the early mergers' reduction factor due to, e.g.,~third-body interactions, or any other effect that could disrupt early binaries or delay mergers. It is important to note that different effects that could impact the EPBH merger rate may become relevant in different parts of the parameter space. For example, there are indications that the nearest and next-to nearest neighbor interactions reduce the merger rate more for larger values of $f_{\rm PBH}$, so that the final merger rate corrections could be a mix of different values of $\mathcal{U}$.

\begin{figure}[ht!]
    \centering
    \includegraphics[width=.8\columnwidth]{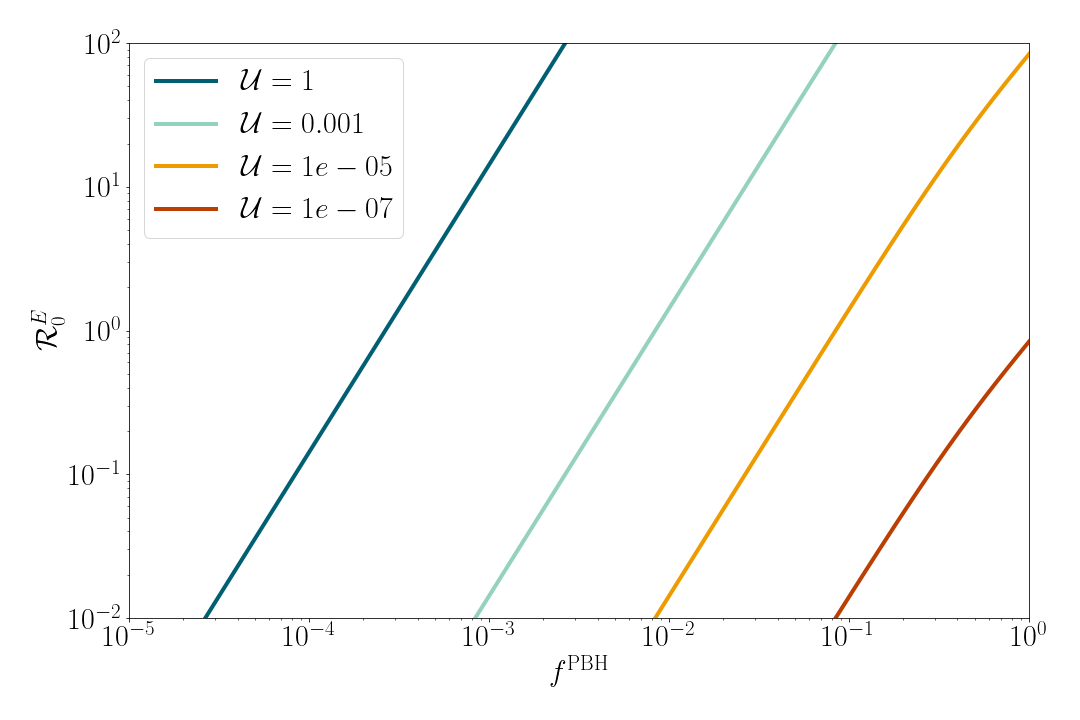}
    \caption{Local merger rate for EPBH for different values of the merger reduction factor $\mathcal{U}$. We here consider $M_{\rm PBH}= 30\,M_\odot$.}
    \label{fig:fpbh_fefl}
\end{figure}

From eqs.~\eqref{eq:fPBH_e} and~\eqref{eq:fpbh_l}, choosing a value for $f_{\rm PBH}$ sets both $\{f^E,\,f^L\}$ and determines the observed local merger rate. We are interested in values comparable with LVK constraints, namely $\mathcal{R}_0^{E,L}\in[0.01,100]\,{\rm Gpc^{-3}yr^{-1}}$: as figure~\ref{fig:fpbh_fefl} shows, this implies that in the {\it nominal} case we would be probing values of $f_{\rm PBH}$ of the order of $\sim \mathcal{O}(10^{-4})$ for EPBH, since larger values of $f_{\rm PBH}$ in the {\it nominal} scenario would lead to values of $\mathcal{R}_0^E$ that are too large to be accepted. In this range, the number of LPBH is negligible.
When $\mathcal{U}$ decreases (i.e.,~the EPBH merger rate is smaller than what~\cite{yacine2017} estimates), larger values of $f_{\rm PBH}$ can be taken into account; when $\mathcal{U} \sim \mathcal{O}(10^{-7})$, EPBH contributions to the overall number of merger events can be neglected.  

\begin{figure}[ht!]
    \centering
    \begin{minipage}{0.49\linewidth}
  \includegraphics[width=\columnwidth]{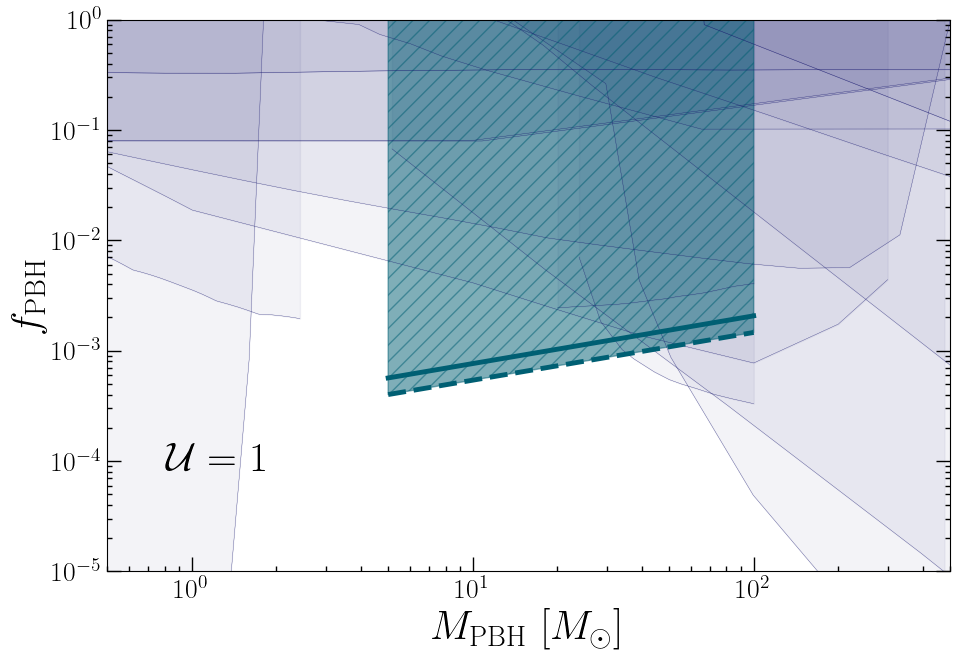}
    \end{minipage}
    \begin{minipage}{0.49\linewidth}
  \includegraphics[width=\columnwidth]{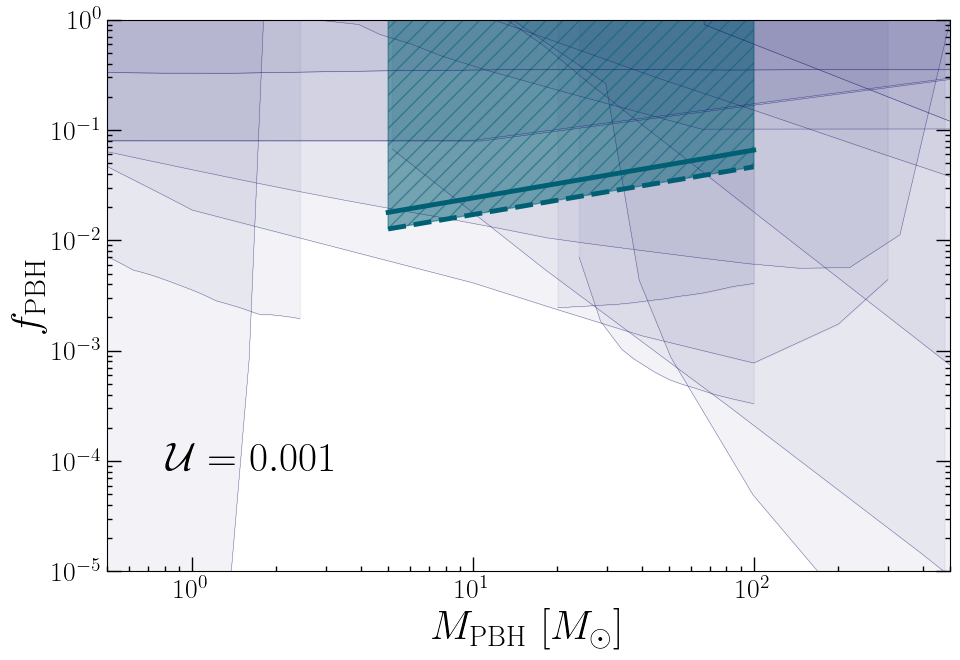}
    \end{minipage}
    \begin{minipage}{0.49\linewidth}
  \includegraphics[width=\columnwidth]{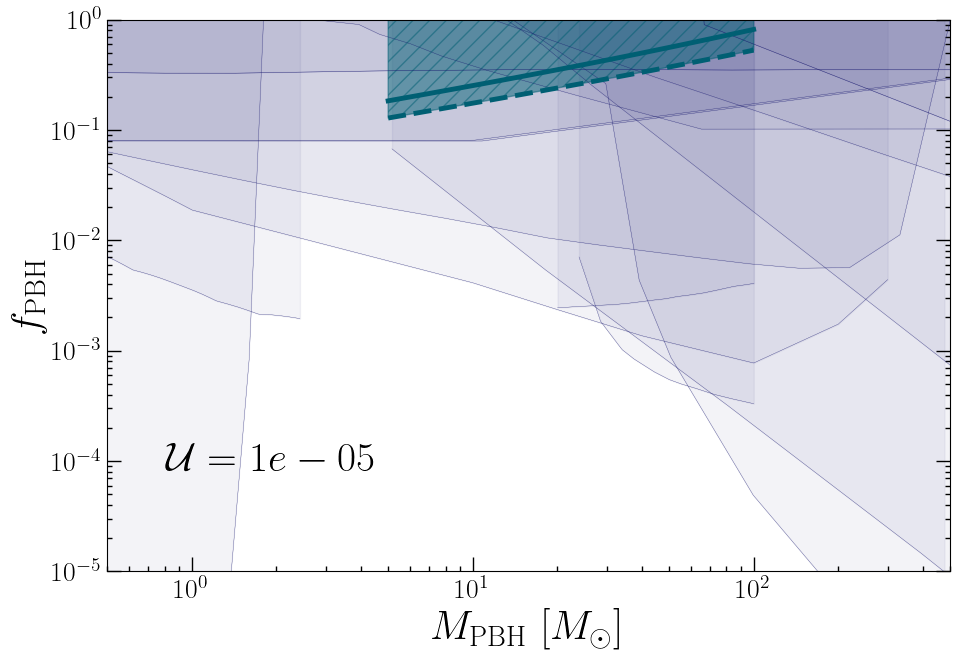}
    \end{minipage}
    \begin{minipage}{0.49\linewidth}
  \includegraphics[width=\columnwidth]{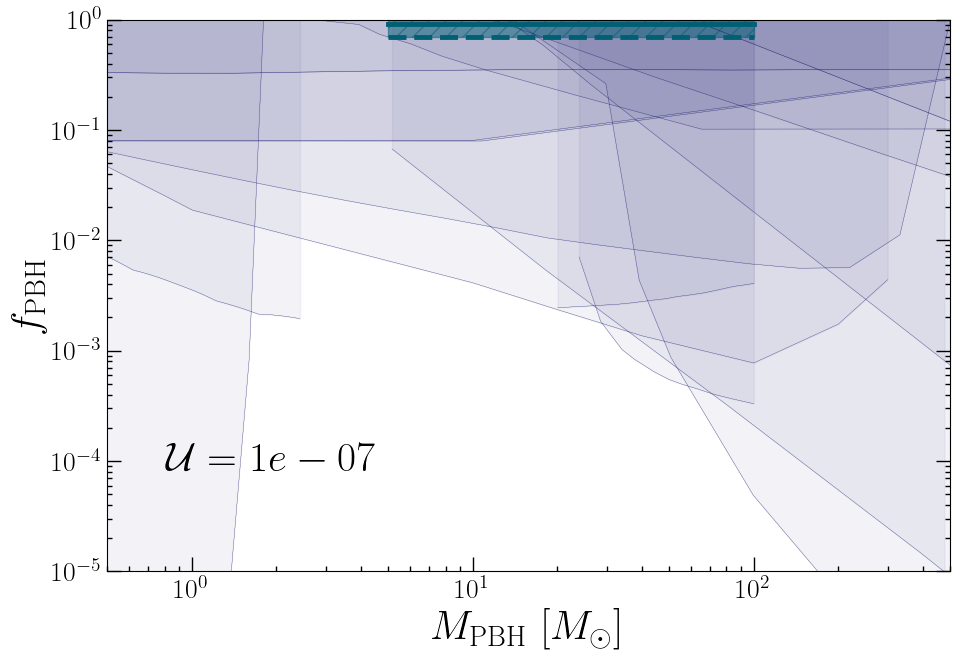}
    \end{minipage}
    \caption{
    Constraints on $f_{\rm PBH}$ from our clustering technique for ET2CE alone (thick, continuous teal lines) or in cross-correlation with 
    galaxy surveys (thick, teal dashed lines) for $\mathcal{U} = 1$ (top left), $10^{-3}$ (top right), $10^{-5}$ (bottom left) and $10^{-7}$ (bottom right). In the first three cases we consider $f^L \sim 0$, while in the latter we adopt $f^E\sim 0$: since $\mathcal{R}_0^L$ in eq.~\eqref{eq:fpbh_l} is mass-independent, $\mathcal{U} = 10^{-7}$ is the only case with flat constraints in $M_{\rm PBH}$. All cases are computed with non-overlapping bins.
    We compare our results with current constraints from other observables (blue, see main text for details).}
    \label{fig:constraints_fPBH}
\end{figure}

To cover all the parameter space, we consider the cases $\mathcal{U} = \{1,10^{-3},\,10^{-5},\,10^{-7}\}$ and we convert constraints (with non-overlapping bins) from section~\ref{sec:bias_forecast} into constraints for $f_{\rm PBH}$. 
We showed in section~\ref{sec:snr} and~\ref{sec:bias_forecast} that the effective bias can be distinguished from the ABH-only case if $f^E+f^L \gtrsim 0.6$ for ET2CE and $\gtrsim 0.4$ when cross-correlating with galaxy surveys. When $\mathcal{U} = 1$, this implies that LPBH are negligible and our technique can detect (or rule out) $f^E\simeq 0.4-0.6$, which corresponds to $f_{\rm PBH}\simeq 1\times 10^{-3}$; when $\mathcal{U} = 10^{-7}$, $f^E\sim 0$ instead and constraints on $f^L$ can be used to constrain $f_{\rm PBH}\sim 0.7-0.8$. Figure~\ref{fig:constraints_fPBH} summarizes our results, for ET2CE alone and in cross correlation with galaxy surveys. Our constraints are compared to current, tentative constraints coming from microlensing~\cite{Alcock_2001,Oguri_2018,Tisserand_2007}, accretion contributions to Galactic radio and X-ray emissions~\cite{Manshanden_2019}, CMB spectral distortions~\cite{Serpico_2020,yacine2017}, 21cm signal~\cite{Hektor_2018}; dynamical constraints from dwarf galaxies~\cite{Lu_2019,Brandt_2016}, wide binaries~\cite{Monroy_Rodr_guez_2014};  NANOGRAV~\cite{Chen_2020}; supernova lensing~\cite{Zumalacarregui:2017qqd}, Lyman-$\alpha$~\cite{Murgia:2019duy}, and LIGO GW data~\cite{Kavanagh_2018,Abbott_2019,Nitz_2022}. Other techniques have been used to forecast PBH constraints in a variety of mass ranges: e.g.,~\cite{_nal_2021} considers the combination of CMB distortions and the pulsar timing array measurements,~\cite{bosi2023} studies how the merger rate changes depending on the early and late abundances and gravity and dark energy models, while in~\cite{Ng_2022} the high redshift merger rate is considered.
Note that all these constraints are model-dependent as well; the ones based on GW measurements from binary mergers, in particular, share the same uncertainties our model has on EPBH and LPBH formation processes.

In the {\it nominal} case $\mathcal{U} = 1$, studying the clustering allows us to tighten the constraints on PBH with respect to currently available ones. For smaller values of  $\mathcal{U}$, our technique is still competitive and can be used as an independent tool to confirm other probes, which is extremely important since PBH constraints usually suffer from strong model-dependency. In particular, our analysis provides a complementary constraint to the one obtained from merger detections at high redshift~\cite{Ng_2022}. Both techniques suffer of some limitations: to constrain PBHs via clustering measurements we need a large number of events with a good sky localization, while the use of the merger rate will require to observe events at high redshifts with a small distance uncertainty. A clear detection is challenging in both cases; for this reason, using both approaches will be fundamental in order to obtain solid limits or detections. Moreover, as stressed above, the conversion from observed (that being low-bias or high-z) primordial mergers to $f_{\rm PBH}$ is model dependent and subject to several assumptions and uncertainties, making it even more useful to use both techniques on the same catalog.

\subsection{Model selection}\label{sec:bayes_factor}

As a final remark for our work, we  perform a Bayesian model selection forecast on the capability future GW surveys will have to distinguish scenarios with and without PBHs. 
It is well known that a Bayesian analysis of this kind is based on the prior choice for the underlying model; therefore, results of this section have to rely on one specific choice for the models described in the previous section. We rely on the {\it nominal} case $\mathcal{U} = 1$, i.e.,~we assume the models in~\cite{yacine2017} and~\cite{didligo} to be the correct ones for early and late binaries respectively, thus we can safely set $f^L\sim 0$. The same analysis can be repeated for other early merger models.

In our Bayesian model selection forecast, we compare the following two models:
\begin{itemize}[label=\raisebox{0.25ex}{\tiny$\bullet$}]
    \item {\it ABH-only} model ($A$): In the GW survey there are only ABH mergers. The parameter set that characterizes this model is $\theta = \{A\}$, which describes the redshift evolution of the ABH bias (see eq.~\eqref{eq:ABHbias}).
    We will also refer to this case as the ``{\it simple} model''.
    \item {\it ABH-PBH} model ($AP$): The GW survey contains both mergers from ABH and PBH early binaries. The parameter set to take into account is now $ \theta = \{A,\,f^E\}$. We will refer to this as the ``{\it complex} model''.
\end{itemize}

Since this is a nested model scenario, we can apply the model selection forecasting procedure described in~\cite{heavens2007}. In this approach, we work in Laplace approximation i.e.,~we assume that the expected likelihoods are multivariate Gaussians. Moreover, we model the precision matrices using the Fisher matrices $F_{A},\,F_{AP}$.\footnote{Note that $F_A,F_{AP}$ in this section differ from the Fisher analysis in the previous section, because of the different parameter choice. Results can be transformed through change of coordinates.} This leads to the following expression for the ensemble average of the Bayes factor, in the assumption that the likelihoods are narrowly peaked around the fiducial values:

\begin{equation}\label{eq:Bfactuse}
    \bigl<B\bigr> = \frac{1}{(2\pi)^{(n_{AP}-n_{A})/2}}\frac{\sqrt{\det F_{AP}}}{\sqrt{\det F_{A}}}\exp\biggl[-\frac{1}{2}\delta\theta_2^{\alpha}F_{AP,\alpha\beta}\,\delta\theta_2^{\beta}\biggr]\prod_{q=1}^{n_{AP}-n_{A}} \Delta\theta_{AP}^{n_{AP}-n_{A}+q} \ .
\end{equation}
In the previous formula, $n_{AP}$ is the number of parameters in the {\it complex} model whereas $n_{A}$ is the number of parameters in the {\it simple} model. Therefore, $n_{AP}-n_{A}$ is the number of extra-parameters in the {\it complex} scenarios, while $\Delta\theta_{AP}^{n_{A}+1,...\ n_{AP}}$ are their prior ranges. 

Note that, in the {\it ABH-PBH} model, the shot noise will include PBH contributions, thus depending on the parameter $f^E$. 
If we consider the {\it ABH-only} model, the associated Fisher matrix is just the submatrix of $F_{AP}$, obtained by fixing the extra parameter to $f^E = 0$; in other words, we are conditioning the value of $f^E$ in the multivariate Gaussian likelihood describing the {\it ABH-PBH} model, in order to reproduce the {\it simple ABH-only} scenario.
This conditioning procedure also leads to shifts in the best-fitting values of all other parameters, according to the following formula (see~\cite{heavens2007} for a detailed explanation) 
\begin{equation}\label{eq:shifts}
\delta\theta^{\alpha} = 
\begin{cases} 
& \theta_{A}^{\alpha}-\bar{\theta}_{AP}^{\alpha} \quad \text{if } \alpha > n_{A} \ (\text{i.e.,~extra-parameters}) \quad \\
& -(F_{A}^{-1})^{\alpha\gamma}\,G^{\gamma\zeta}\,\delta\theta_{AP}^{\zeta} \quad \text{if } \alpha < n_{A} \ (\text{i.e.,~common parameters}) \\ 
\end{cases} 
\end{equation}
In the above, 
$G^{\gamma\zeta}$ is the $n_{A} \times (n_{AP}-n_{A})$ subset of $F_{AP}$, obtained by considering the $\gamma = 1,...\ n_{A}$ rows related to the common parameters and the $\zeta = n_{A}+\,...\ (n_{AP}-n_{A})$ columns related to the extra parameters in the {\it complex} model. 

To sum up, in our case:
\begin{itemize}[label=\raisebox{0.25ex}{\tiny$\bullet$}]
    \item In the {\it ABH-PBH} model, we have $n_{AP} = 2$ parameters, namely $\{A,\,f^E\}$. Since $f^E$ spans from 0 to 1 depending on the specific model adopted, we can safely assume a uniform prior on the extra parameter $\Delta\theta_{AP}^{extra} = \Delta f^E = 1$.
    \item In the {\it ABH-only} model we have $n_{A} = 1$ parameter, namely the slope of the ABH bias $A$. The extra-parameter $f^E$ is set to 0 to recover this model. In this case, the Fisher matrix collapses to a single element, namely $F_{A} = f_{\rm sky}\sum_\ell (2\ell+1)[(\tilde{C}_\ell)^{-1}\partial_A\tilde{C}_\ell(\tilde{C}_\ell)^{-1}\partial_A\tilde{C}_\ell]/2$, where the power spectra are computed in the {\it ABH-only} condition.
\end{itemize}
Under these conditions, we apply eq.~\eqref{eq:shifts} to compute the shifts in the fiducial parameters and then we insert them in eq.~\eqref{eq:Bfactuse} to estimate the Bayes factor. 

In the multi tracer case, the only difference is that we build up the Fisher matrix considering the galaxy-galaxy and cross-spectra as well; note that the galaxy power spectra are independent on both the $A$ and $f^E$ parameters, therefore the only extra signal we gain comes from cross-correlations. On the other side, adding the galaxy bias parameters $b_g$ to the analysis can introduce extra degeneracies, for which the Laplace approximation cannot be trusted: this happens e.g.,~when considering ET alone, even in correlation with galaxy surveys. 
Moreover, the cases $f^E = \{0, 1\}$ can not be analysed with this method since they present singular Fisher matrices: in the former case, $f^E$ can not be constrained since there are no PBH in the computation ($\partial_{f^E}\tilde{C}_\ell = 0$ in $F_{AP}$ and the {\it complex} model collapses on the {\it simple} one), while in the latter the same happens with $A$ since there are no ABH ($F_{AP}^{AA} = 0$).

\begin{figure}[ht!]
    \centering
   \includegraphics[width=.8\columnwidth]{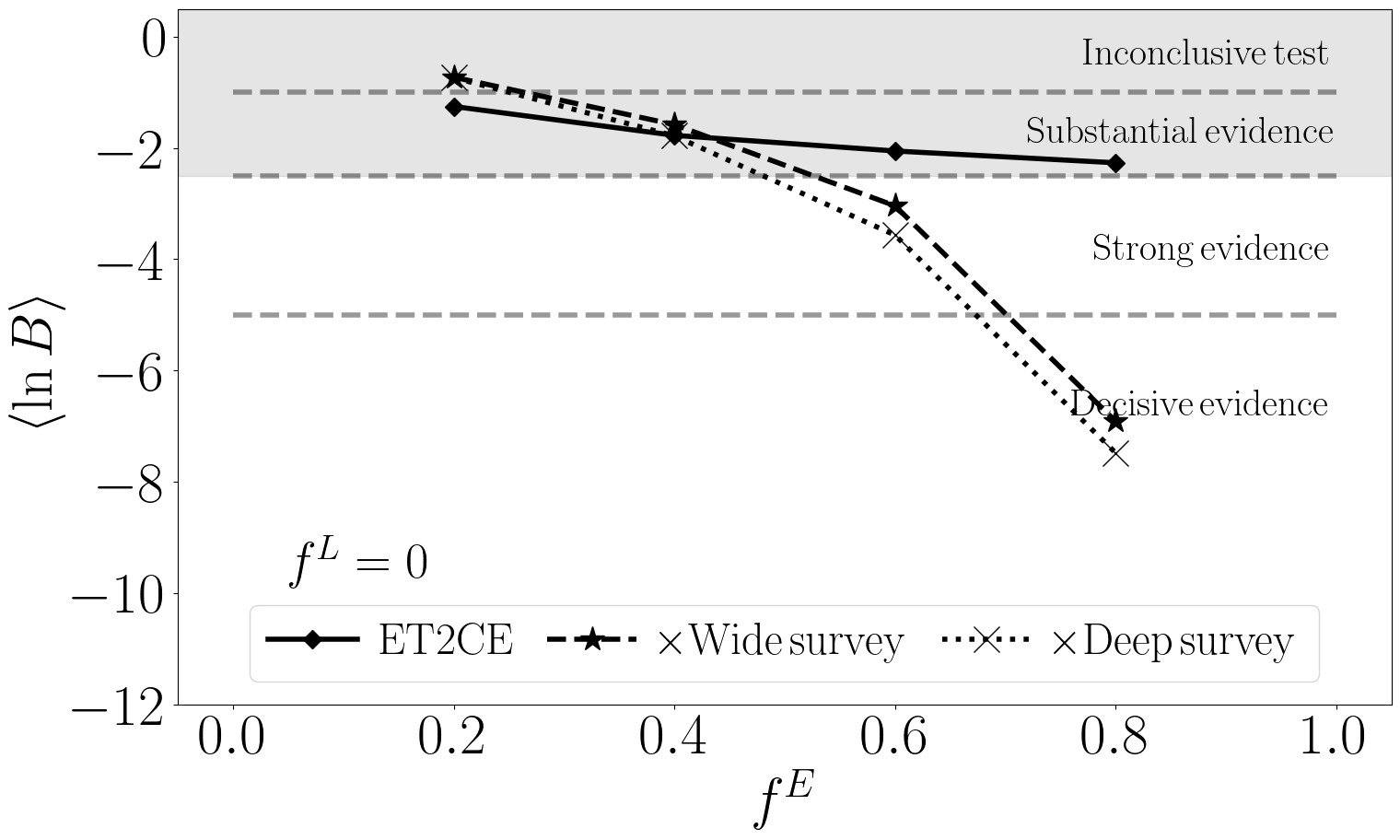}\vspace{-.3cm}
    \caption{Bayes factor $\langle \ln B\rangle$ depending on $f^E$ using non-overlapping redshift bins. We account for auto- and cross-correlations of ET2CE with galaxy surveys. The horizontal lines show the Jefferey's values for evidence in favor of the {\it complex ABH-PBH} model. \vspace{-.3cm}}
    \label{fig:bayes_fefl}
\end{figure}

Figure~\ref{fig:bayes_fefl} shows our Bayes factor results in terms of $f^E$, compared with the Jeffrey's values for detection~\cite{Jeffreys61} as described by~\cite{heavens2007}:
\begin{itemize}[label=\raisebox{0.25ex}{\tiny$\bullet$}]
    \item $-1 \leq \langle \ln B\rangle < 1$ : inconclusive test;
    \item $-2.5 \leq \langle \ln B\rangle < -1$ : substantial evidence for {\it complex} model {\it AP} ($\sim 1\sigma$ detection);
    \item $-5 \leq \langle \ln B\rangle < -2.5$ : strong evidence for {\it complex} model {\it AP} ($\sim 2\sigma$ detection);
    \item $\langle \ln B\rangle < -5$ : decisive evidence for {\it complex} model {\it AP} ($\sim 3\sigma$ detection).
\end{itemize}
We only show results for ET2CE and its cross-correlations with the wide and deep surveys,
since the cases related with a single ET have large shot noise and break the Laplace approximation: this is consistent with the large uncertainties obtained from the Fisher matrix analysis.
Our results for the Bayes factor are consistent with those shown in the previous sections (for non-overlapping bins): evidence for the {\it ABH-PBH} model is reached when $f^E \gtrsim 40\%$ of the mergers observed by future GW detectors, when the cross-correlation with galaxies is taken into account. Above this value, we obtain a strong or decisive evidence, namely a $\gtrsim 2\sigma$ detection of early PBH binary contribution to the effective bias. Using eq.~\eqref{eq:fPBH_e}, this translates into evidence for $f_{\rm PBH}\gtrsim 9\times 10^{-4}$ for monochromatic PBH mass distribution $M_{\rm PBH}\sim 30\,M_\odot$.

Our result is stable if we increase the parameter space,
even if adopt a minimal parameter choice, which only includes the ABH bias parameter $A$ in the {\it single} model, and $f^E$ in the {\it complex} model. This choice is motivated by the fact that these parameters are the most relevant in setting the amplitude and shape of the effective bias: $A$ determines both the amplitude at $z=0$ of the ABH bias and the slope of its redshift evolution in the linear case, while in the nominal $\mathcal{U} = 1$ case when early binaries dominate the PBH contribution, under the assumption that they Poisson sample of the underlying DM field, their intrinsic bias is well described by $b^E = 1$ (see section~\ref{sec:PBH} for detail).

Other parameters entering the effective bias computation are fixed throughout the analysis. The parameter $D$ characterizes the slope of the ABH bias evolution: if we leave it completely free to vary, in combination with $A$ it could lead the ABH bias to mimic some of the ABH+PBH effective bias shapes. However, we can set well motivated constrains on it: since $D$ it describes the intrinsic bias of a population, its value must be either constant or monotonically increasing at high redshift. Moreover, since ABH trace the host galaxy population, in first approximation we can set a $D$ value which is based on galaxy bias measurements. Under these prior conditions, $D$ can not be used to mimic the bias behaviour {\it complex} model, when we are considering the {\it simple} model. 

As for late binary contributions, based on previous studies~\cite{didligo,scelfo2018}, we set their bias at the constant value $b^L = 0.5$. To go beyond this approximation, its value could be computed by marginalizing the bias of the host DM halos over their masses. Since the halo bias grows with $z$, a larger contribution of late binaries and a redshift dependent model for $b^L$ could in principle mimic the ABH bias and the {\it simple} model in the context of the {\it complex} one. Also in this case, however, this possibility is tuled out by a reasonable choice of prior, based on physical considerations: late binaries are mainly hosted by small DM halos, that have bias $< 1$ at low $z$ and reach values slightly higher than $1$ at high $z$ (see e.g.~\cite{Tinker:2010my,wernerporciani_2019}). Therefore, also in this case, it is very difficult to mimic the {\it simple}, only-ABH scenario, in the context of the {\it complex} scenario. The LPBH merger rate $\mathcal{R}_0^L$, as well as the ABH and EPBH ones, mainly affect the constraints as contributions to the shot noise.

Last but not least, a final crucial reason why we decided to rely on a minimal parameter choice is due to the use of the Fisher matrix in our analysis. In fact, every small degeneracy between parameters propagates into the value of the determinants in eq.~\eqref{eq:Bfactuse} and breaks the Laplace approximation under which their computation is performed. Our Fisher matrix analysis, unfortunately, is sensitive to this problem; an actual analysis on real data or realistic mock data-sets, would instead allow us to take into account all the possible degeneracies, by considering the full posterior and not relying on any Gaussian approximation. As we previously motivated, we expect the results to be consistent with our forecast analysis.

\section{Conclusions}
\label{sec:concl}

The physics acting during the early stages of the Universe is still largely unknown; in particular, there are no constraints on the primordial power spectrum on scales smaller than what is probed by the CMB.
Regarding the physics relevant for this work, this leaves the window open to the possible existence of large curvature perturbations, which might be responsible for the formation of primordial black holes. Independently on their formation model, if PBHs exist, they could be (or become) bound in binaries and eventually merge, producing gravitational waves. If their masses are comparable to black holes produced via stellar evolution, namely $\mathcal{O}(5-100)\,M_\odot$, the GW signals produced by their mergers would be in principle indistinguishable from the signal produced by astrophysical binary mergers. 

Among the probes to distinguish between different BBH origin, in this work we chose to focus on the study of the clustering of BBHs mergers, making predictions for future GW surveys. Previous studies already demonstrated the capability of this tool, which comes from the fact that PBH and ABH binaries trace the underlying matter distribution differently. Gravitational wave surveys are blind to the origin of the mergers' progenitors, thus the angular power spectrum estimated from their catalogs will be characterized by an effective bias due to the weighted average contribution of ABHs and PBHs. If the latter can be formed both in radiation dominated era and via dynamical captures in the late Universe, the relative abundances of the two sub-populations is needed to determine the deviation of the effective bias from the standard bias of astrophysical mergers.

In this paper, we investigated for the first time the possibility to use GW clustering alone to search for signatures of PBH mergers in catalogs that should be observed by future experiments such as the Einstein Telescope and Cosmic Explorer, alone or in combination. The measured effective bias of the mergers' hosts will be sensitive to the presence of a primordial component, in ways that depend on the formation channel of primordial binaries.
Our results showed that interferometers of this type could detect (or rule out) the presence of primordial mergers if those make up at least 60\%-80\% of the detected mergers, when sources are divided into non-overlapping redshift bins.
We then used updated limits on merger rates for the different BBH populations to produce new limits from the cross-correlation between GW and galaxy catalogs, investigating wide and deep future surveys. We found that the fraction of primordial mergers these cross-correlations could be sensitive to lowers to $\sim 40\%$. A poor redshift determination leads to overlapping redshift bins, with a consequent degradation of our results of about a factor of two. A good sky localization is crucial as well to gain constraining power, particularly when relying on GW auto-power spectra. Our analysis shows that ET alone can hardly constrain the bias of the astrophysical and primordial black hole mergers: improvements come either from a better sky localization or from cross correlation with galaxy surveys. The former is reached combining more GW detectors: in this work, we relied on the ET2CE configuration, but intermediate results can possibly be reached triangulating ET with the (advanced version of) second generation GW detectors. The main limitation in this case comes from the smaller detector horizon, which allows to reach a lower $z_{\rm max}$.  

For the first time in this type of analyses, we connected constraints on the fraction of primordial mergers in the observed catalogs to the fraction of dark matter in PBHs. This connection is model-dependent, as the predicted merger rate for early PBH mergers is still very uncertain and depends on a variety of parameters and assumptions.
For this reason, we computed our constraints as a function of a fudge factor that encapsulates such uncertainties, allowing the merger rate of early PBH mergers to vary by several orders of magnitude. We found that, due to the large uncertainty in the modeling of early PBH mergers, constraints on $f_{\rm PBH}$ from this observable could range from some $10^{-4}$ to $\sim 85\%$ of dark matter in PBHs. Clearly, this methodology could become potentially one of the best ones to constrain PBH abundance in the LVK mass range, provided the physics of early binary formation and mergers gets clarified in future work, possibly using simulations accounting for a variety of effects.

Finally, we performed a new type of analysis for the GW clustering observable, in the form of a forecast Bayesian model selection technique aimed at understanding in what cases we could be able to claim evidence for the presence of PBH mergers in the catalogs. The analysis compared the model in which a parameter accounting for the PBH presence in the effective bias model is included, from the one in which only ABH are taken into account.
GW clustering alone can reach substantial evidence, and most likely go beyond that in the case when the vast majority of detected mergers is of primordial origin.
When looking at cross-correlations with galaxy catalogs, strong and even decisive evidence can be reached for certain values of PBH mergers. Results from all methodologies are consistent with each other.

In parallel to the development of this work, another article investigated the potential of the cross-correlation between gravitational waves from binary black hole mergers and galaxy maps, in order to constrain modified gravity and dark energy models, and withing those theories, PBH abundance~\cite{bosi2023}. In that work, the authors focus on slightly different situations and use GW catalogs specifically generated; in the limits where the two analyses can overlap and be compared, the results agree and present a complementary point of view.

The bulk of constraining power for these clustering analyses comes from intermediate redshifts $z\sim [1,2.5]$, showing that our tool is complementary to other tests based on high-$z$ merger rate. Given the relevance of the topic and the numerous uncertainties arising when trying to rule out (or detect) PBHs in the stellar mass range, we advocate for a combination of different techniques and observables, which could potentially provide robust results.


\section*{Acknowledgements}

\noindent
We especially thank Nicola~Bellomo, Michele~Bosi and Andrea~Ravenni for insightful comments on the manuscript. The authors also thank Jos\'e~L.~Bernal, Ely~D.~Kovetz, Sabino~Matarrese, Caner~\"Unal, Lorenzo~Valbusa Dall'Armi, Eleonora Vanzan, Yun Wang and the Padova Cosmology group for interesting and fruitful discussions. SL~acknowledges funds from the Fondazione Ing.\ Aldo Gini scholarship and thanks the Azrieli Foundation for the support. 
We acknowledge support by the project ``Combining Cosmic Microwave Background and Large Scale Structure data: an Integrated Approach for Addressing Fundamental Questions in Cosmology", funded by the MIUR Progetti di Ricerca di Rilevante Interesse Nazionale (PRIN) Bando 2017 - grant 2017YJYZAH.
AR acknowledges funding from the Italian Ministry of University and Research (MIUR) through the ``Dipartimenti di eccellenza'' project ``Science of the Universe''.
We acknowledge the use of the public library~\url{https://zenodo.org/record/3538999} to realize the plots in figure~\ref{fig:constraints_fPBH}.

\appendix
\section{Model dependence on subpopulation abundances}\label{app:all_plots}

\begin{figure}[ht!]
    \centering
  \includegraphics[width=\columnwidth]{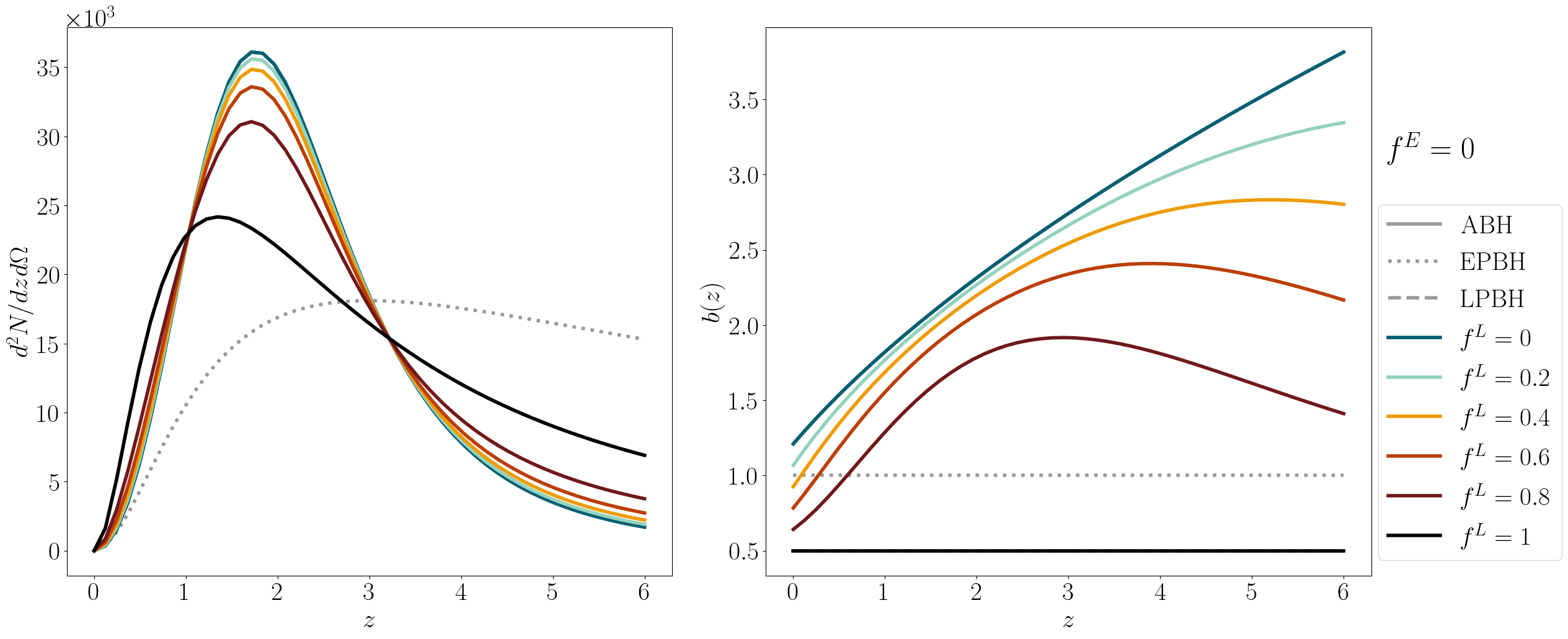}\vspace{-.5cm}
    \caption{Analogous to figure~\ref{fig:model_fid}, but in the case $f^E = 0$. \vspace{-.5cm}}
    \label{fig:models_0}
\end{figure}

\begin{figure}[ht!]
    \centering
  \includegraphics[width=\columnwidth]{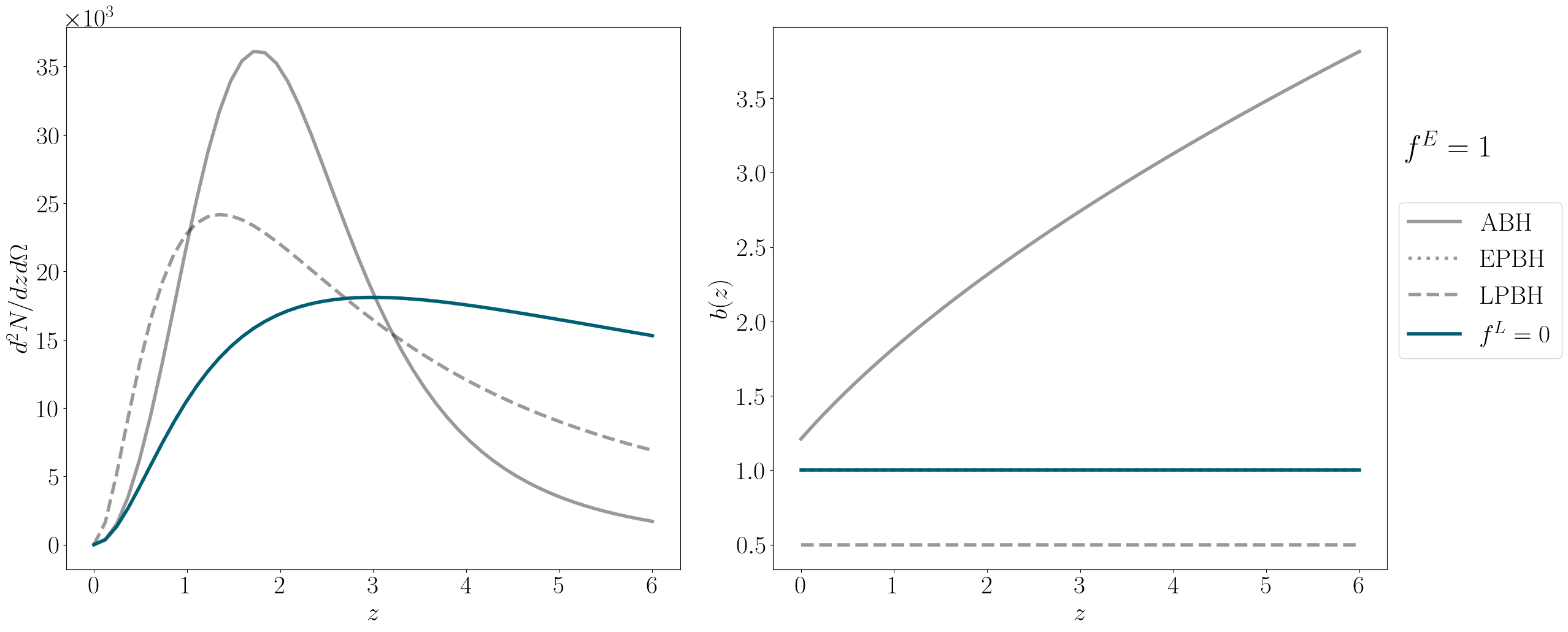}\vspace{-.3cm}
    \caption{Analogous to figure~\ref{fig:model_fid}, but in the case $f^E = 1$.\vspace{-.5cm}}
    \label{fig:models_1}
\end{figure}

We show how the observed number distribution $d^2N^{\rm GW}/dzd\Omega$ from eq.~\eqref{eq:numberdistr} and the effective bias $b(z)$ from eq.~\eqref{eq:bias} change accordingly to variations in the EPBH fraction $f^E$. All the plots in this appendix are analogous to figure~\ref{fig:model_fid}; the condition $f^E+f^L\geq 1$ reduces the cases of $f^L$ available when $f^E$ grows. The plots also show the degenerate cases $\{f^E,f^L\} = \{0,0\}$, only ABH (blue line in figure~\ref{fig:models_0}), $\{f^E,f^L\} = \{0,1\}$, only LPBH (black line in figure~\ref{fig:models_0}), $\{f^E,f^L\} = \{1,0\} = $ only EPBH (blue line in figure~\ref{fig:models_1}).

We here consider the detector setup ET2CE, in which the fiducial case is $\mathcal{R}^{\rm tot}_0 = 27\gpcyr$, $N^{\rm GW} = 1.1\times 10^5$ and $T_{\rm obs} = 10\,{\rm yr}$.
In the single ET case, instead, the overall observed number of events each year decreases to $N^{\rm GW} = 8\times 10^4$, therefore the overall normalization in the $d^2N^{\rm GW}/dzd\Omega$ plots is lower. The other modifications we introduce when dealing with this detector (namely, the different average beam size and the fraction of observed events described in figure~\ref{sec:analyses}) do affect neither the fiducial $d^2N^{\rm GW}/dzd\Omega$ nor $b(z)$. 
\vspace{-.3cm}
\begin{figure}[ht!]
    \centering
  \includegraphics[width=\columnwidth]{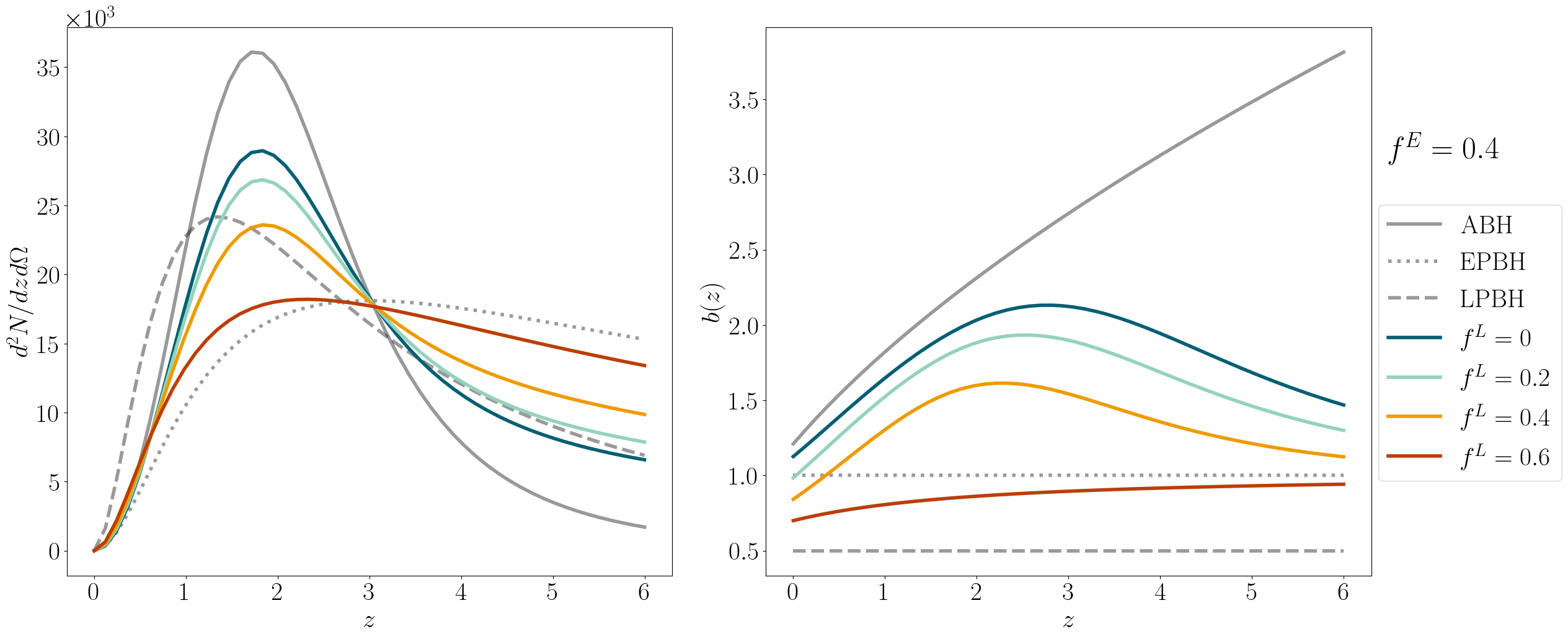}\vspace{-.5cm}
    \caption{Analogous to figure~\ref{fig:model_fid}, but in the case $f^E = 0.4$. \vspace{-.5cm}}
    \label{fig:models_02}
\end{figure}
\begin{figure}[ht!]
    \centering
  \includegraphics[width=\columnwidth]{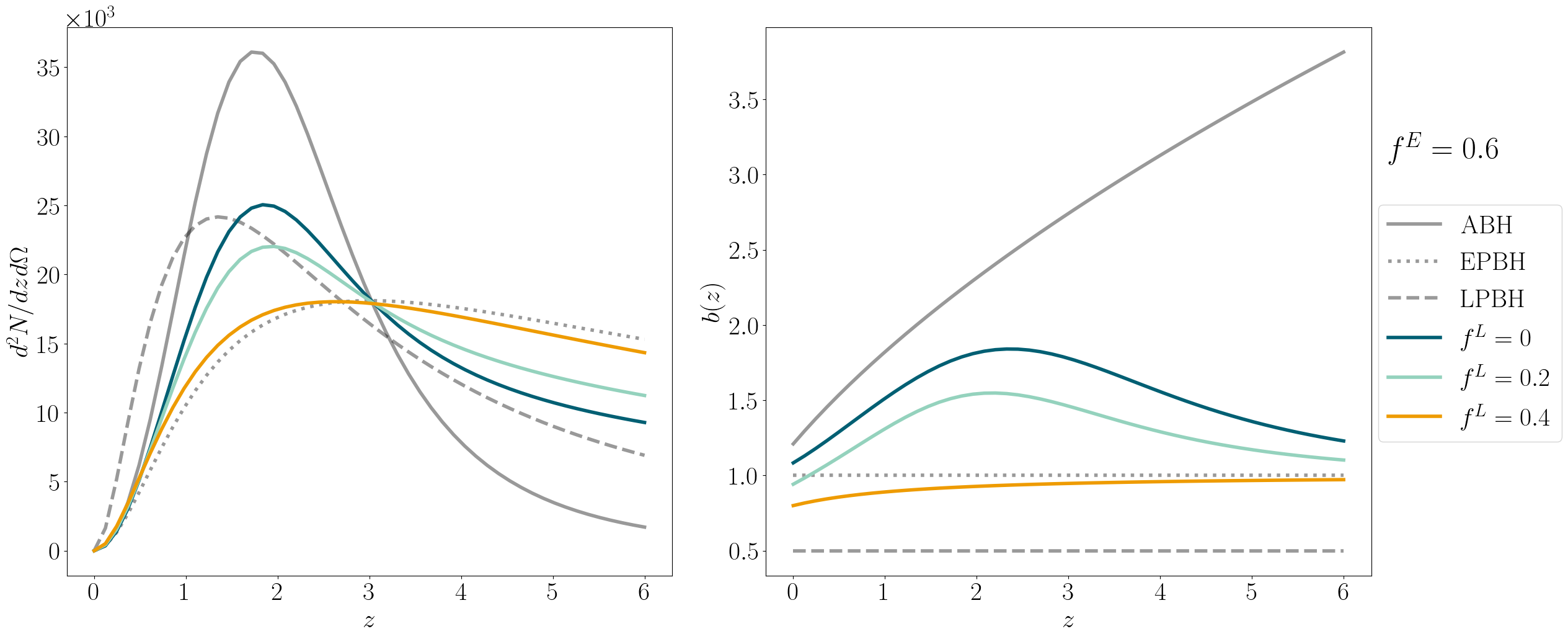}\vspace{-.5cm}
    \caption{Analogous to figure~\ref{fig:model_fid}, but in the case $f^E = 0.6$.\vspace{-.5cm}}
    \label{fig:models_06}
\end{figure}

\begin{figure}[ht!]
    \centering
  \includegraphics[width=\columnwidth]{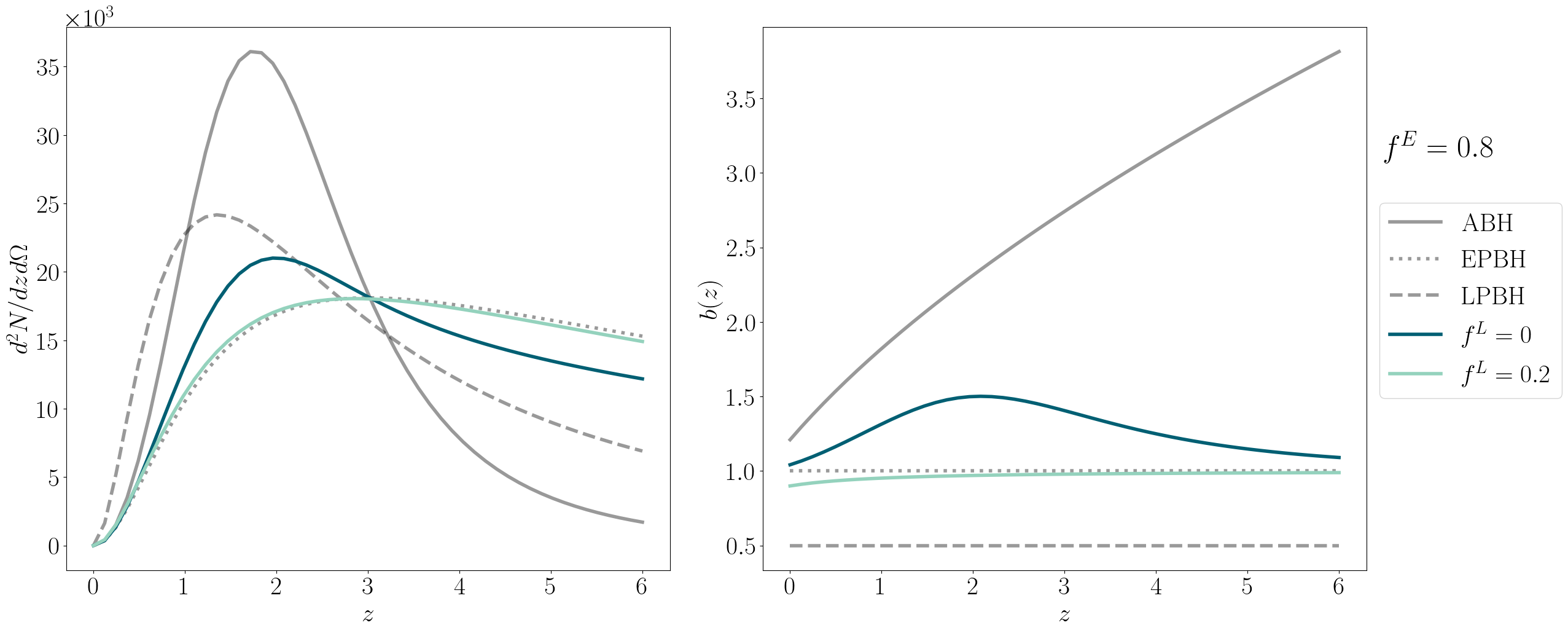}\vspace{-.3cm}
    \caption{Analogous to figure~\ref{fig:model_fid}, but in the case $f^E = 0.8$.\vspace{-.5cm}}
    \label{fig:models_08}
\end{figure}

\section{Angular power spectrum formalism} \label{app:cl}

Throughout this work, our observable is the angular power spectrum 
\begin{align}\label{eq:cl}
    &\tilde{C}_\ell(z_i,z_j) = B_\ell^2 C_\ell = \exp\biggl[- \frac{\ell(\ell+1)}{2}\frac{\Delta\tilde{\Omega}_{[\rm std]}}{8\,\log(2)}\biggr]C_\ell(z_i,z_j)\,,\\
    &C_\ell(z_i,z_j) = \frac{2}{\pi}\int dk \ k^2 P(k) \Delta_{\ell}(z_i,k)\Delta_{\ell}(z_j,k)\ \ ,
\end{align}
where $B_\ell$ is the smoothing factor due to the Gaussian beam from eq.~\eqref{eq:bl}, that exponentially decreases the power, and $C_\ell(z_i,z_j)$ is the standard angular power spectrum; $P(k)$ is the matter power spectrum, while the observed transfer functions $\Delta_{\ell}(z_{i,j},k)$ depend on the source number distribution, the window function and the theoretical transfer functions, which contain information on the density perturbations and the redshift space distortions (see e.g.,~\cite{Challinor2011} for the full expression and derivation). As we describe in section~\ref{sec:survey}, we compute the angular power spectrum in 5 tomographic bins (including both $i=j$ and $i \neq j$ components); to do so, we use and modify the public code \texttt{Multi\_Class}~\cite{Bellomo_2020, Bernal_2020},\footnote{\url{https://github.com/nbellomo/Multi_CLASS}} taking into account the ABH and PBH number distributions described in section~\ref{sec:ABH} and~\ref{sec:PBH}.

\begin{figure}[ht!]
    \centering
  \includegraphics[width=.7\columnwidth]{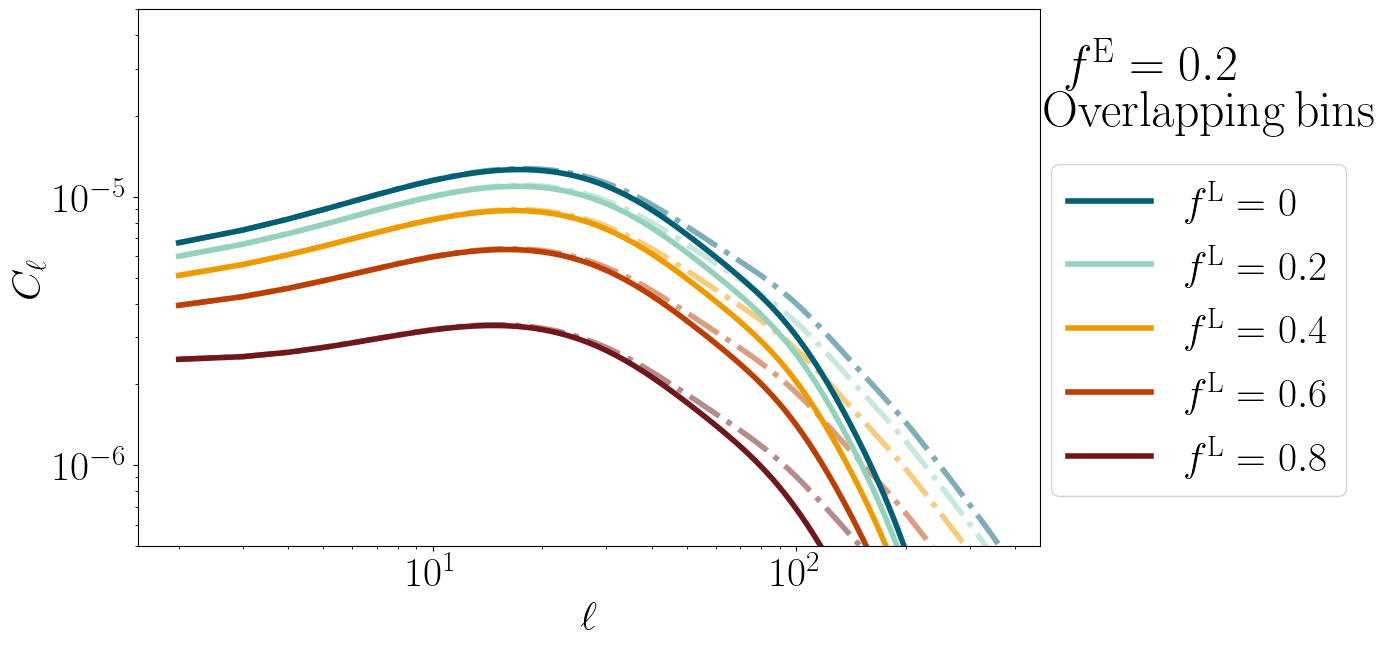}
  \includegraphics[width=.7\columnwidth]{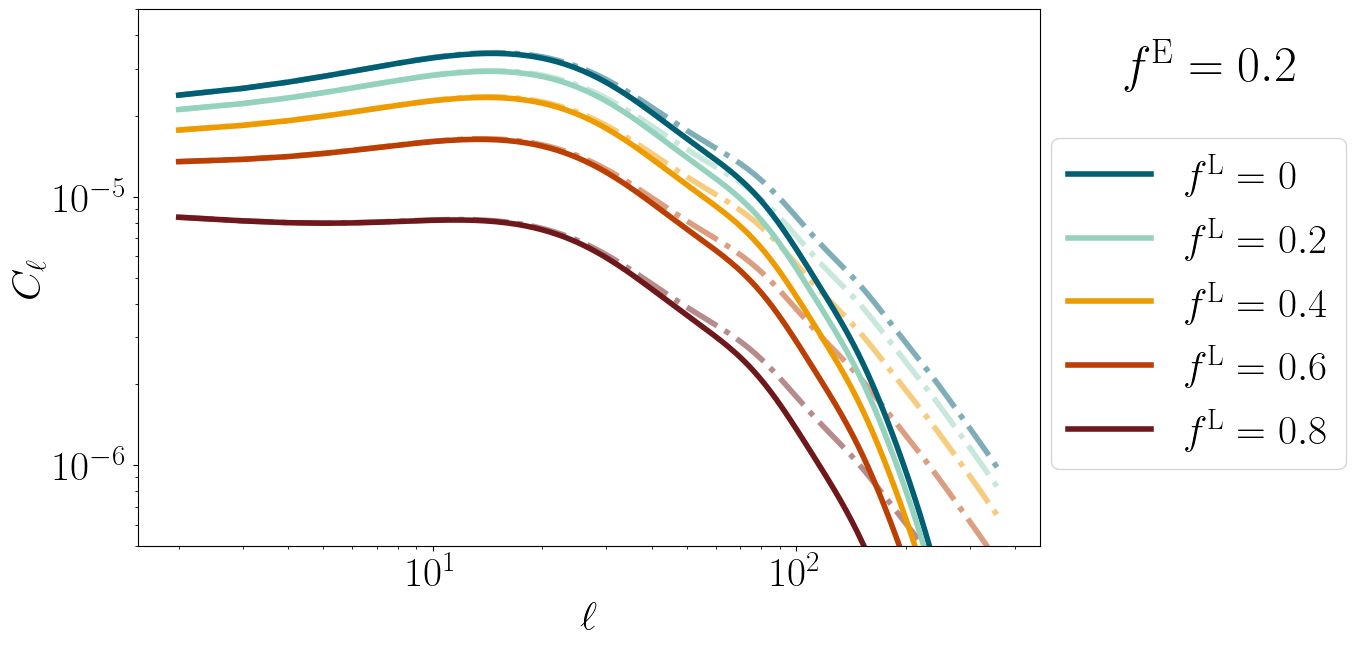}
    \caption{Angular power spectra in the $z_i = z_j = 0.4$ bin for $f^E = 0.2$, ET2CE case, using overlapping (top) and non-overlapping (bottom) redshift bins. Dot-dashed lines show $C_\ell(z_i,z_j)$, while continuous lines are $\tilde{C}_\ell(z_i,z_j)$ from eq.~\eqref{eq:cl}. }
    \label{fig:cl}
\end{figure}

Figure~\ref{fig:cl} compares the theoretical $C_\ell$ with the smoothed $\tilde{C}_\ell$ that we use as observable.

Eq.~\eqref{eq:cl} is used to compute the angular power spectrum for a single tracer, being this GW or galaxies. For cross-angular power spectrum, we use the general formulation
\begin{equation}\label{eq:cl_cross}
    \tilde{C}_\ell^{\rm GWg}(z_i,z_j) = \frac{2}{\pi}\int dk\,k^2P(k)\left[B_\ell\int_{z_i-\Delta z}^{z_i+\Delta z}b(z)\frac{dN^{\rm GW}}{dz}\tilde{\Delta}_\ell^{i}\right]\left[\int_{z_j-\Delta z}^{z_j+\Delta z}b_g(z)\frac{dN^{\rm g}}{dz}\tilde{\Delta}_\ell^{j}\right]\,,
\end{equation}
where $b(z),\,b_g(z)$ are respectively the GW effective bias and the galaxy bias, $dN^{\rm GW}/dz,\,dN^{\rm g}/dz$ the GW and galaxy number density per redshift bin and the transfer functions $\Delta^{i,j}$ include the density and redshift space distortion contributions.

\begin{figure}[ht!]
    \centering
  \includegraphics[width=.9\columnwidth]{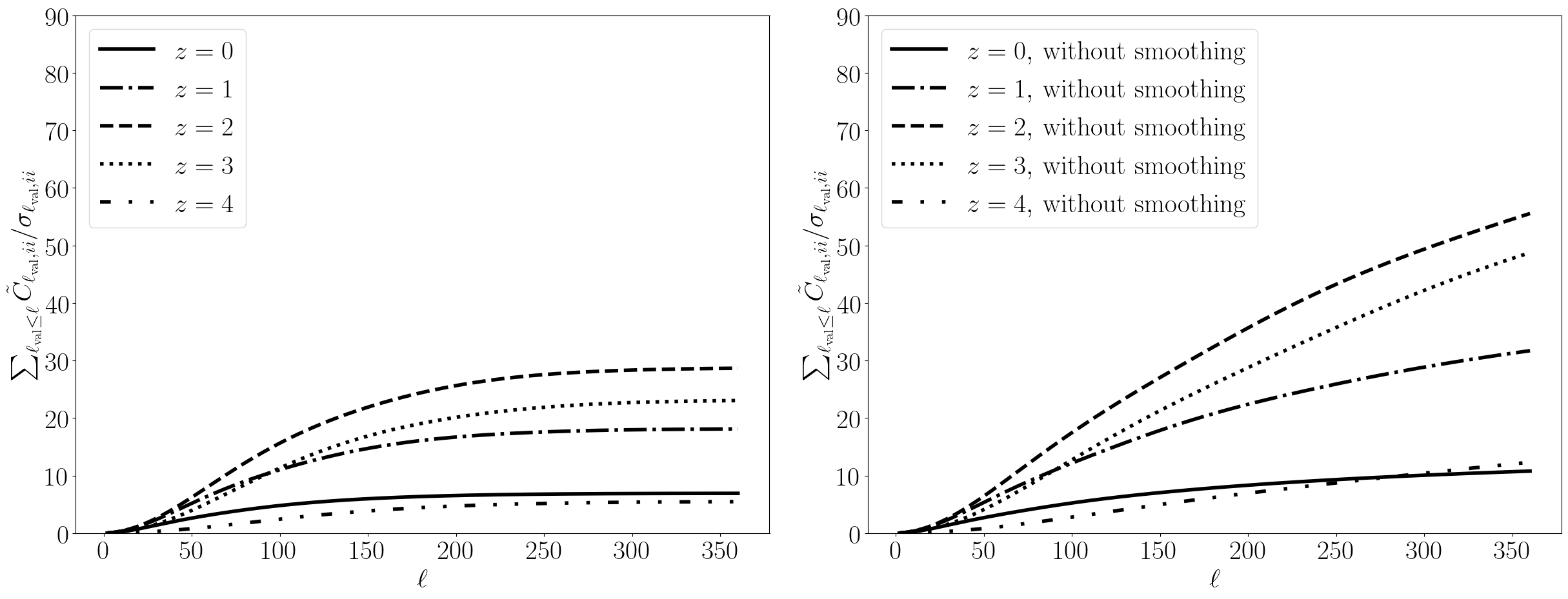}\\
  \includegraphics[width=.9\columnwidth]{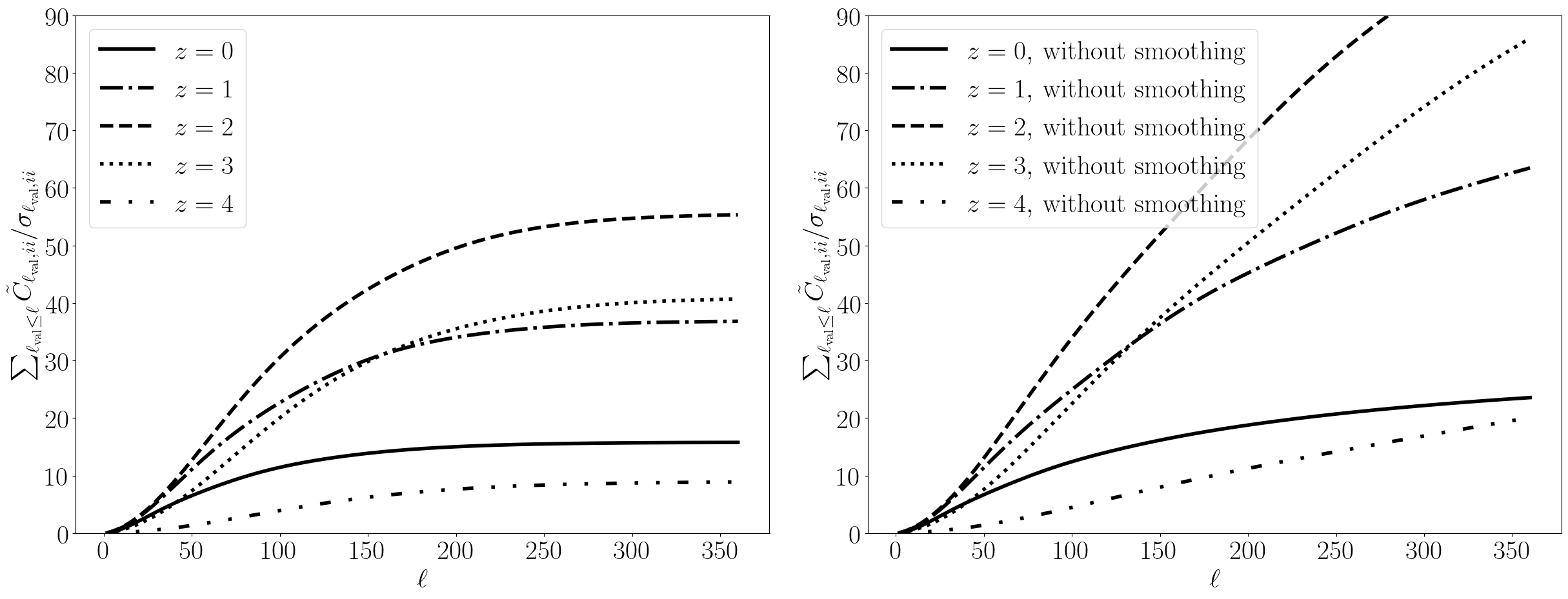}\\
    \caption{Cumulative signal-to-noise ratio of the ABH power spectrum in the $i = j$ bins of our analysis, ET2CE case (left panel). The saturation is due to the smoothing factor $B_\ell$ in  eq.~\eqref{eq:bl},~\eqref{eq:cl}; if this was not taken into account, the SNR would largely increase (right panel). Top row refers to the overlapping bins case, while in the bottom row non-overlapping bins are used. }
    \label{fig:snr_study}
\end{figure}

Before running the full analysis, we checked the SNR of our theoretical only-ABH assumption, in the case of auto-power spectra $(ij) = (pq)$. Figure~\ref{fig:snr_study} shows the $\ell$-cumulative SNR for ET2CE.
Note that the bulk of the information comes from intermediate redshifts $z\sim 2$, where the ABH number distribution peaks. Moreover, the $\tilde{C}_\ell$ SNR saturates only because of the smoothing factor on small scales adopted in eqs.~\eqref{eq:bl},~\eqref{eq:cl}: this intuitively suggests that our results will be quite conservative and the clustering study could provide more stringent constraints on the PBH merger contribution if the sky localization was improved.

\section{Full set of results}\label{app:all_plots_snr}

\subsection{SNR: detector dependence}

Different values of $\{f^E,f^L\}$ determine different $\rm SNR$ in eq.~\eqref{eq:snr}. Its measurement in each scenario depends also on the detector: tables~\ref{tab:snr} and~\ref{tab:snr_et} collects our results respectively for ET2CE and ET, alone or in cross-correlation with a wide or deep galaxy survey. 

\begin{table}[ht!]
\renewcommand{\arraystretch}{1.2}
\setlength{\tabcolsep}{3.9pt}
  \begin{tabular}{|cc|cccccc|cccccc|}
     \hline
      \multicolumn{2}{|c|}{\multirow{3}{*}{ET2CE}} & \multicolumn{12}{c|}{\qquad$f^E$} \\
   && \multicolumn{6}{c|}{Overlapping bins} &  \multicolumn{6}{c|}{Non-overlapping bins}\\
    & & 0  & 0.2 & 0.4 & 0.6 & 0.8 & 1 & 0  & 0.2 & 0.4 & 0.6 & 0.8 & 1  \\
    \hline
    \multirow{6}{*}{\rotatebox[origin=c]{90}{$f^L$}} & 0 & 0. & 0.43 & 0.82  & 1.25 & 1.77 &  2.39 & 0. & 0.60 & 1.23  & 2.00 & 2.94 &  4.09  \\ 
    & 0.2 & 0.19 & 0.66 & 1.16 & 1.74 & 2.44 &  - & 0.32 & 1.02 & 1.85 & 2.90 & 4.21 &  -\\ 
    & 0.4 & 0.44 & 1.03 &  1.70 & 2.51 & - & - & 0.77 & 1.68 &  2.84 & 4.36 & - & - \\ 
    & 0.6 & 0.83 & 1.63 &  2.61 & - & - & - & 1.45 & 2.76 &  4.54 & - & - & - \\ 
    & 0.8& 1.51 & 2.74 &  - & - & - & - & 2.63 & 4.80 &  - & - & - & - \\ 
    & 1 & 2.92 & - & -  & - & - &  - & 5.15 & - & -  & - & - &  -\\ 
    \hline
      \multicolumn{2}{|c|}{$\times$Wide } &  & \multicolumn{11}{c|}{ } \\
      \hline
    \multirow{6}{*}{\rotatebox[origin=c]{90}{$f^L$}} & 0 & 0.& 1.14 &2.47 & 4.18 & 6.57& 10.24 & 0. & 1.59 & 3.52  & 6.02 & 9.47 & 14.68 \\ 
    & 0.2 & 0.66 & 2.04&3.87 & 6.47 & 10.67 &- & 1.21 & 3.00 & 5.65 & 9.40 & 15.35 &  - \\ 
    & 0.4 & 1.61& 3.52& 6.36& 11.26 & -&- & 2.94 & 5.24 & 9.33  & 16.25 & - &  - \\ 
    & 0.6 & 3.13& 6.23 & 12.08& -& -&- & 5.65 & 9.24 & 17.49 & - & - & -  \\ 
    & 0.8 & 6.06& 13.31 &-& -&- & -& 10.74 & 19.31 &  - & - & - &  - \\ 
    & 1 & 15.40 &- &- & -&-&- &  26.63 & - & -  &-  & - &  - \\ 
     \hline
      \multicolumn{2}{|c|}{$\times$Deep } &  & \multicolumn{11}{c|}{ }  \\
    \hline
    \multirow{6}{*}{\rotatebox[origin=c]{90}{$f^L$}} & 0 &0. &1.13& 2.23& 3.55& 5.32 & 7.94 &  0. & 1.31  & 2.79  & 4.68 & 7.30 & 11.27 \\ 
    & 0.2 & 0.45& 1.75 &3.21 & 5.16& 8.19& -&  0.70 & 2.28 & 4.30  & 7.16& 11.71 &  - \\ 
    & 0.4 &1.10& 2.77 & 4.96& 8.53 &-& -&  1.71 & 3.85 & 6.98 & 12.30 & - &  - \\ 
    & 0.6 &2.16& 4.68 & 9.02 & -& -&- &-  3.32 & 6.75 & 13.14 & - & - &  - \\ 
    & 0.8 & 4.24& 9.78 &- & -& -&-&  6.43 & 14.42 &  - & - &  -&  - \\ 
    & 1 & 11.22& -&- &- & -& - &16.72 & - & -  & - & - & -  \\ 
    \hline
     \end{tabular}   
     \caption{$\rm SNR$ values in eq.~\eqref{eq:snr} for the fiducial ET2CE case and the different models taken into account. 
     Single tracer results provide above $2\sigma$ detection whenever PBH constitute $\gtrsim 60\%$\,-\,$80\%$ of the totality ($f^E+f^L\gtrsim 0.6-0.8$), the specific value depending on the relative abundance of EPBH and LPBH. When cross-correlations are considered, above $2$\,-\,$3\sigma$ detection is reached with PBH $\gtrsim 40\%$.}
    \label{tab:snr}
\end{table}

\begin{table}[ht!]
\centering
\renewcommand{\arraystretch}{1.3}
\setlength{\tabcolsep}{3.9pt}
  \begin{tabular}{|cc|cccccc|cccccc|}
     \hline
      \multicolumn{2}{|c|}{\multirow{3}{*}{ET}} & \multicolumn{12}{c|}{\qquad$f^E$} \\
   && \multicolumn{6}{c|}{Overlapping bins} &  \multicolumn{6}{c|}{Non-overlapping bins}\\
    & & 0  & 0.2 & 0.4 & 0.6 & 0.8 & 1 & 0  & 0.2 & 0.4 & 0.6 & 0.8 & 1  \\
    \hline
    \multirow{6}{*}{\rotatebox[origin=c]{90}{$f^L$}} & 0 & 0. & 0.01  & 0.02 & 0.03 & 0.04 & 0.06  & 0. & 0.02 &  0.03 & 0.06 & 0.09 & 0.12  \\ 
    & 0.2 &  0.01 & 0.02  & 0.03 & 0.04 & 0.06 &  - & 0.01 & 0.03 & 0.06  & 0.09 & 0.13 & -\\ 
    & 0.4 & 0.01 & 0.03 &  0.04 & 0.07  & - & - & 0.03 & 0.05 & 0.09 & 0.14 & - &  - \\ 
    & 0.6 & 0.02 & 0.04 & 0.07 & - & - & -  & 0.05 & 0.09 &  0.14 & - & - & - \\ 
    & 0.8& 0.04 & 0.07 &  - & - & - & - & 0.80 & 0.16 & - & - & - & - \\ 
    & 1 & 0.08 & - & -  & - & - &  - & 0.17 & - &  - & - & - & - \\ 
    \hline
      \multicolumn{2}{|c|}{$\times$Wide } &  & \multicolumn{11}{c|}{ } \\
    \hline
    \multirow{6}{*}{\rotatebox[origin=c]{90}{$f^L$}} & 0. & 0. & 0.08 & 0.17 & 0.29 & 0.47& 0.75& 0. & 0.13 & 0.30 & 0.53 & 0.84 & 1.34 \\ 
    & 0.2 & 0.05 & 0.15 & 0.28 & 0.47 & 0.78 & -& 0.12 & 0.28 & 0.52 & 0.86 & 1.42 &  - \\ 
    & 0.4 & 0.13 & 0.26 & 0.47 & 0.84 &- &- & 0.28 & 0.52 & 0.88  & 1.52 & - &  - \\ 
    & 0.6 & 0.25 & 0.47 & 0.91 &- & -& -&0.52 & 0.91 & 1.67 & - & - & -  \\ 
    & 0.8 & 0.47 & 1.02 &-&- & -& -&0.96  & 1.87 &  - & - & - &  - \\ 
    & 1 & 1.19 & - &- &- &-& -& 2.18  & - & -  &-  & - &  - \\ 
     \hline
      \multicolumn{2}{|c|}{$\times$Deep } &  & \multicolumn{11}{c|}{ }  \\
    \hline
    \multirow{6}{*}{\rotatebox[origin=c]{90}{$f^L$}} & 0 & 0. & 0.07 & 0.14 & 0.23 & 0.35 & 0.54 &  0. & 0.09  & 0.19 & 0.33 & 0.52 & 0.81 \\ 
    & 0.2 & 0.03& 0.11 &0.21 & 0.35 & 0.56 & -& 0.06  & 0.17 & 0.32 &0.52  & 0.85 &  - \\ 
    & 0.4 & 0.08 & 0.19 & 0.34 & 0.59  &- & -& 0.15  & 0.30 & 0.52 & 0.90 & - &  - \\ 
    & 0.6 & 0.16 & 0.32 & 0.63 & -&- &- & 0.29  & 0.53 & 0.98 & - & - &  - \\ 
    & 0.8 & 0.31 & 0.69  & -& -& - & -& 0.53  & 1.09 &  - & - &  -&  - \\ 
    & 1 & 0.80 & - &-  &- &- & - & 1.27 & - & -  & - & - & -  \\ 
    \hline
     \end{tabular}   
     \label{tab:snr_et}
\caption{Analogous to table~\ref{tab:snr}, considering single ET as GW detector.}
\end{table}

\subsection{Fisher matrix: model dependence}

\begin{figure}[ht!]
\begin{minipage}{0.49\linewidth}
  \includegraphics[width=\columnwidth]{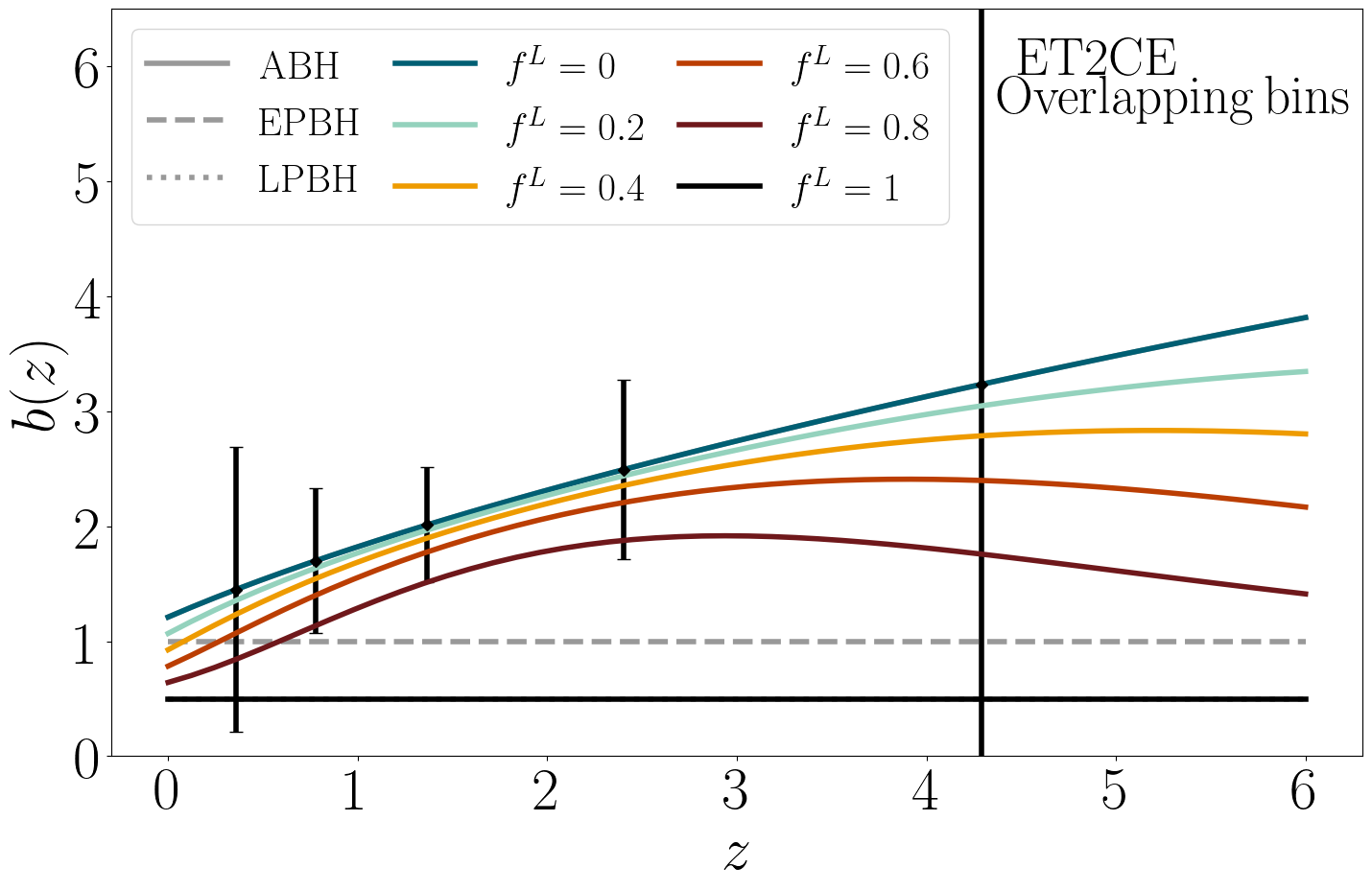}
  \includegraphics[width=\columnwidth]{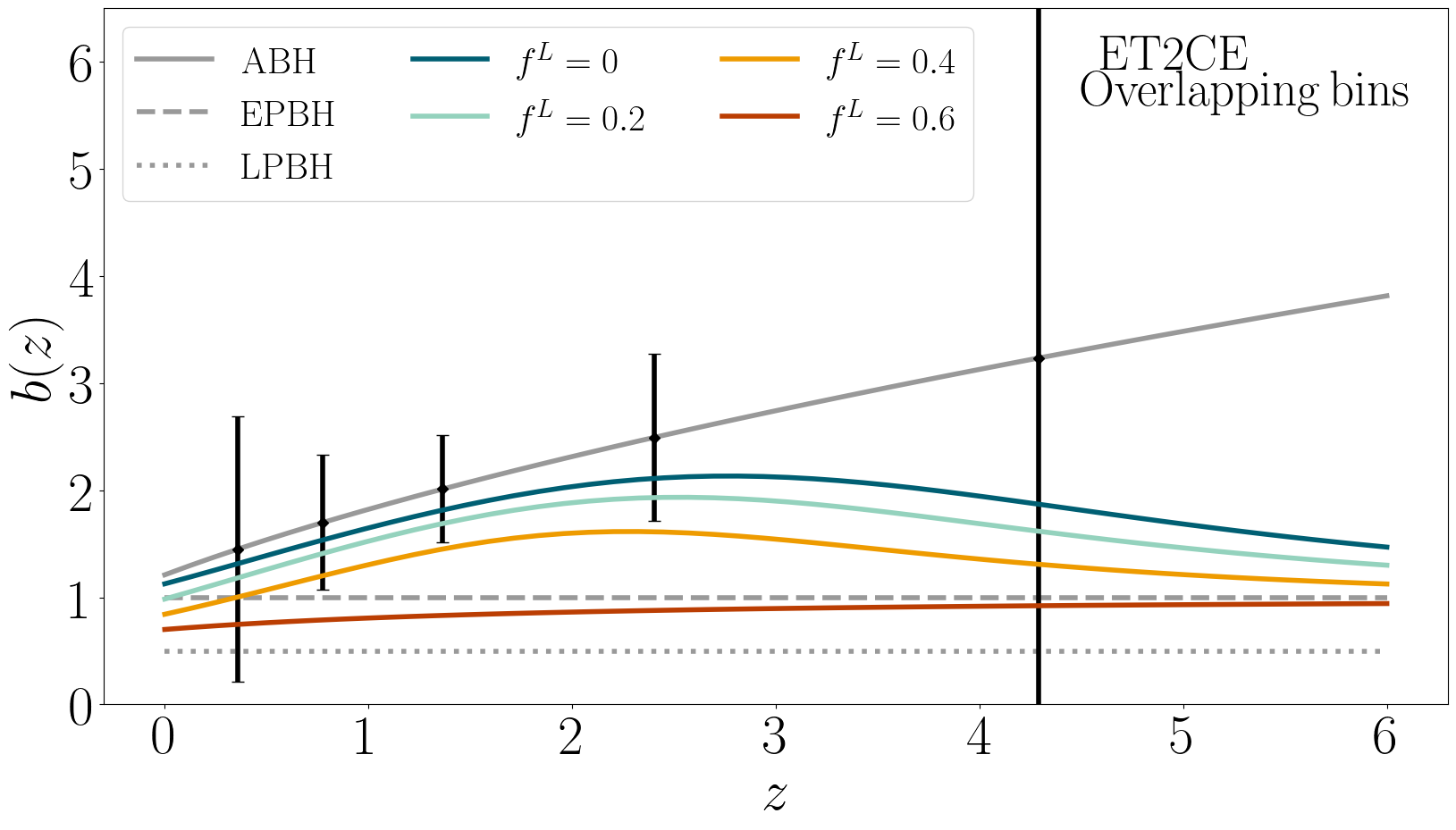}\\
  \includegraphics[width=\columnwidth]{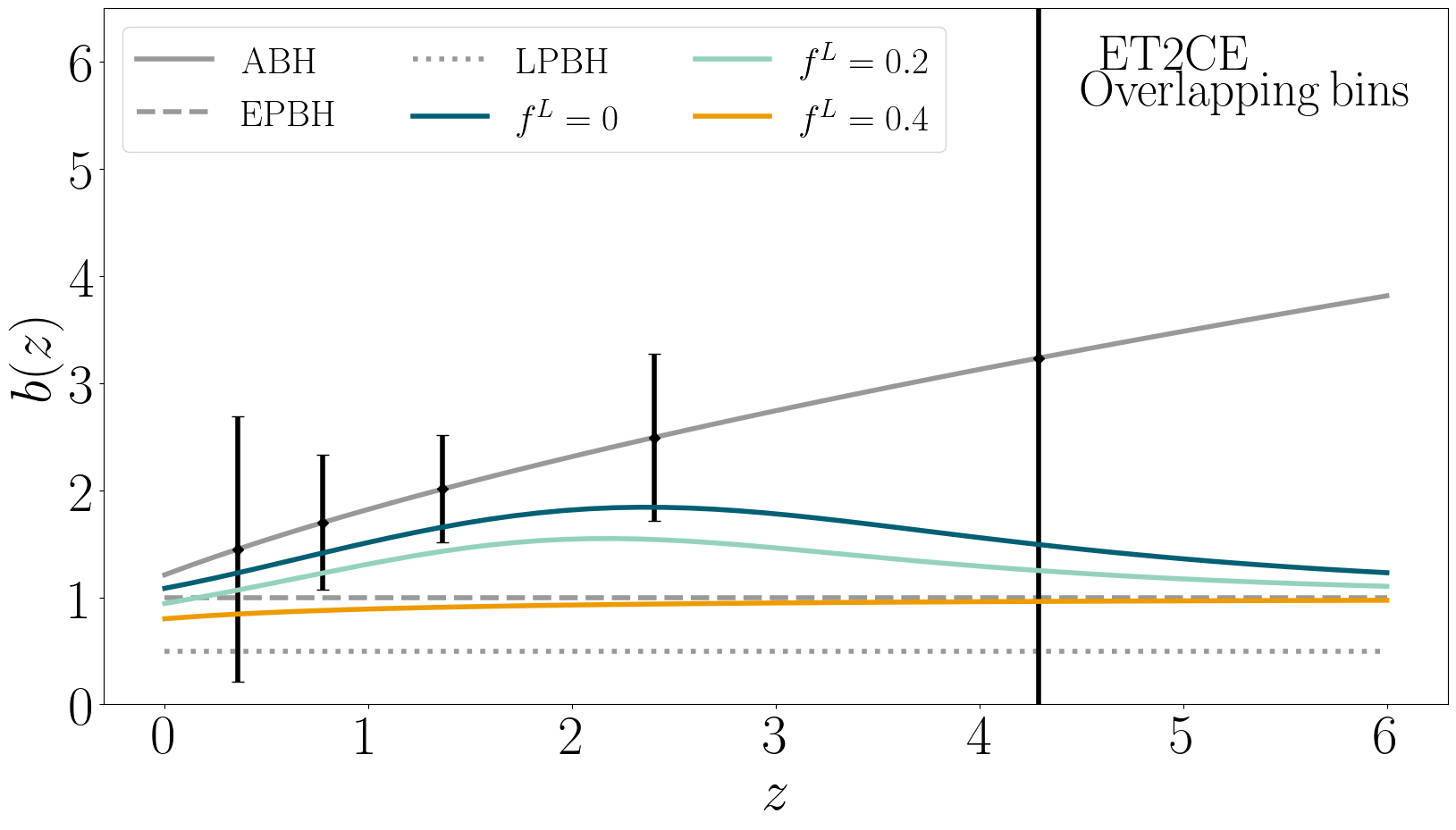}
  \includegraphics[width=\columnwidth]{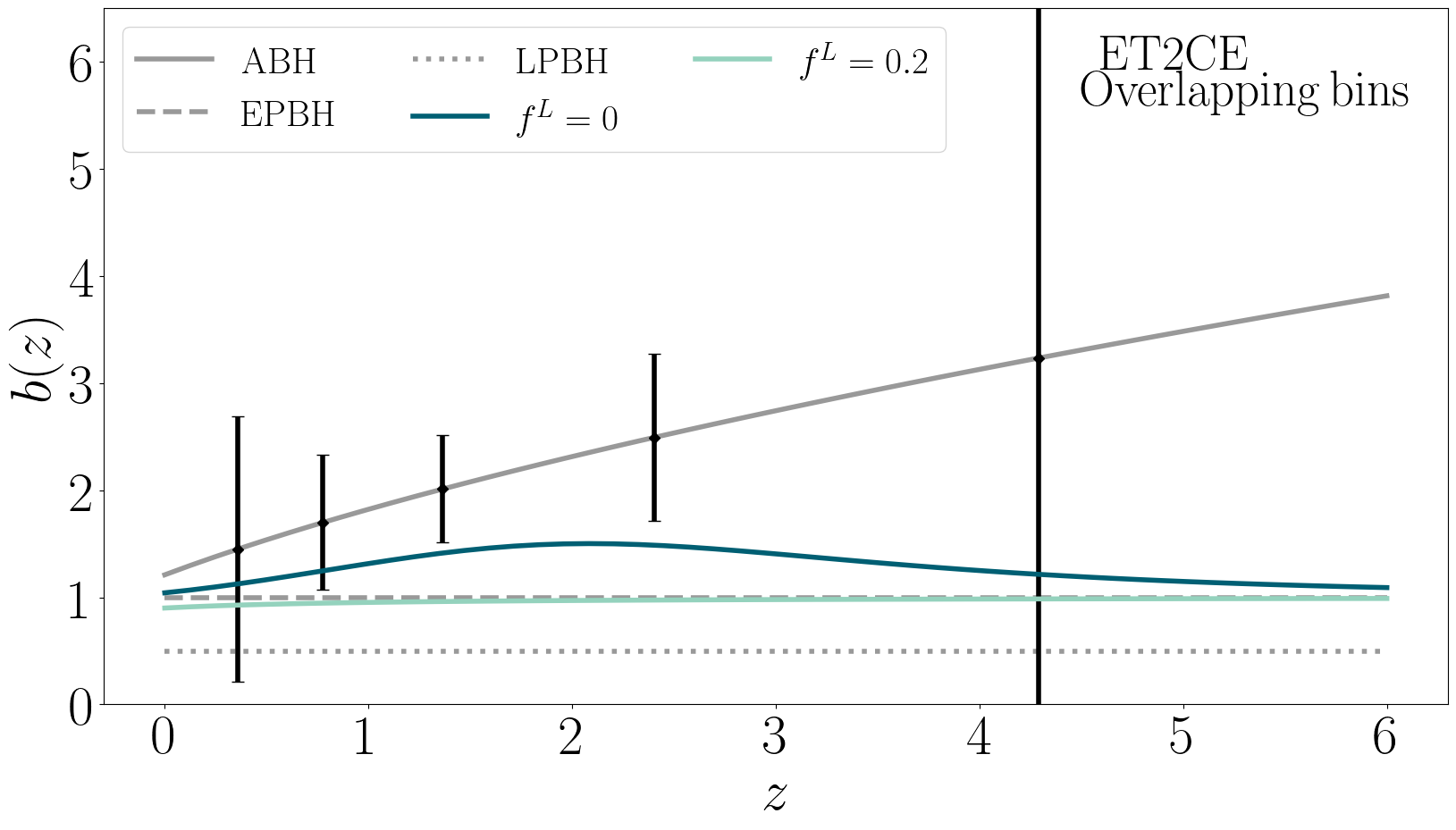}\\
  \includegraphics[width=\columnwidth]{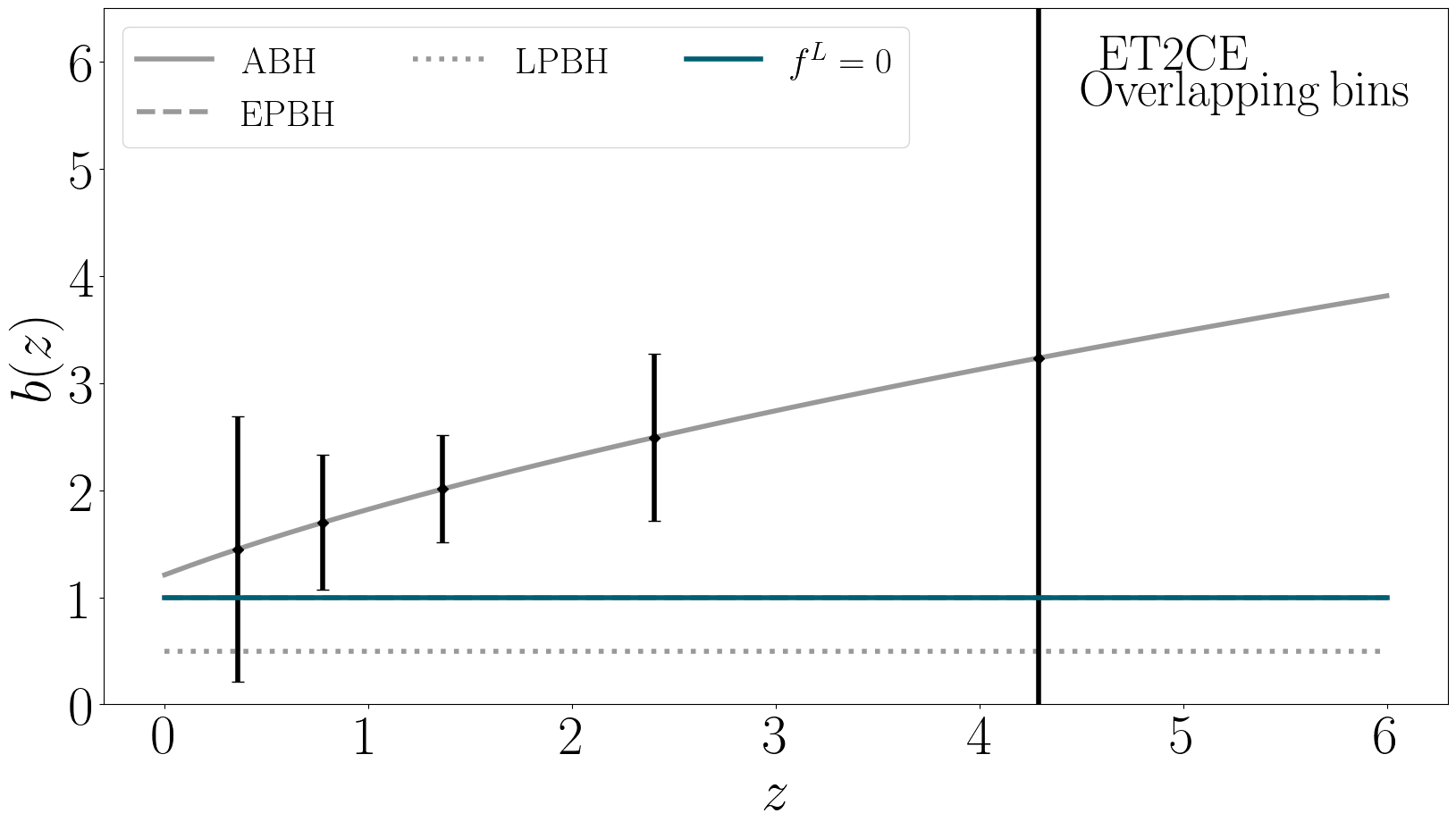}
\end{minipage}
\begin{minipage}{0.49\linewidth}
  \includegraphics[width=\columnwidth,height=.65\columnwidth]{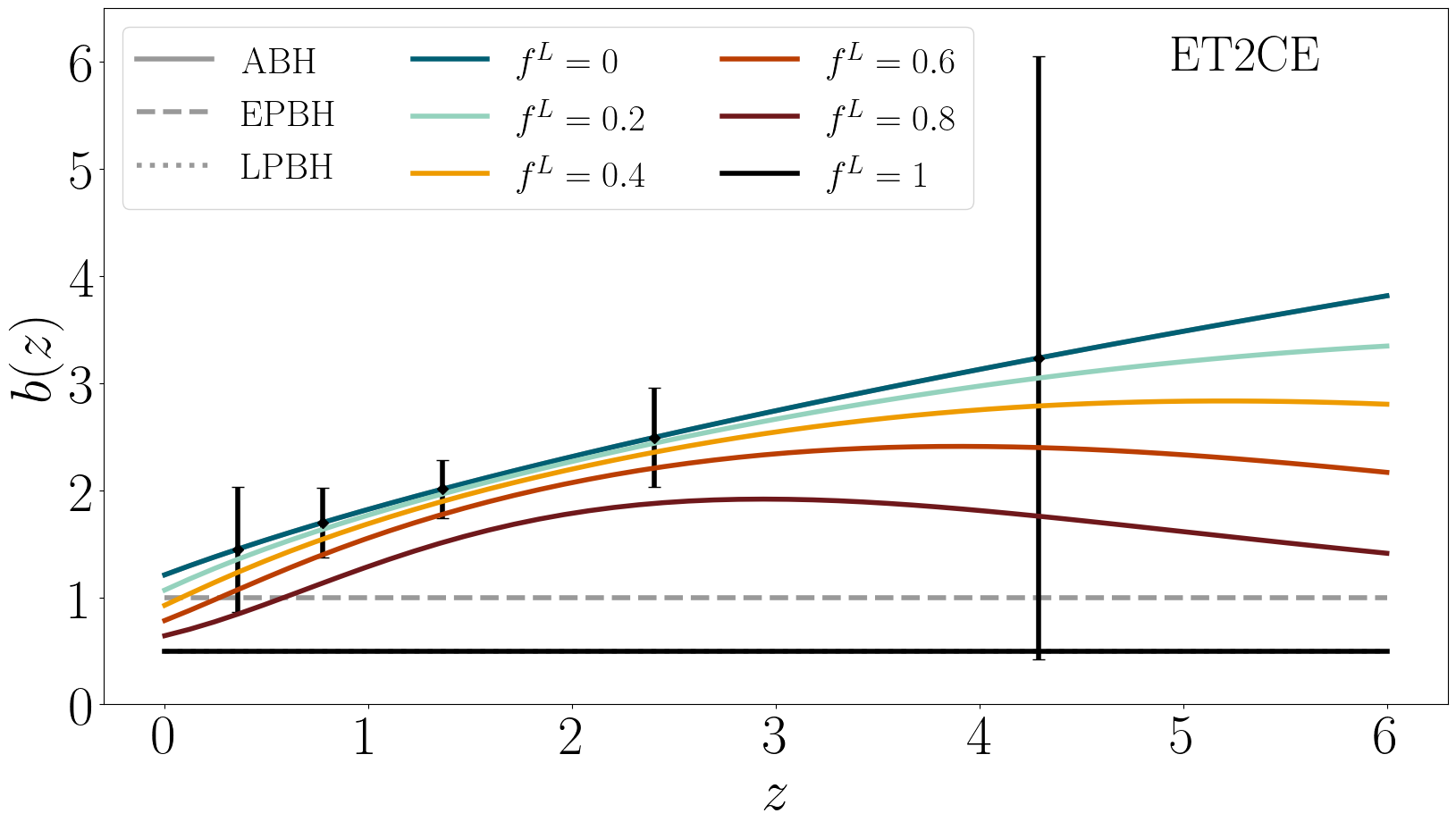}
  \includegraphics[width=\columnwidth]{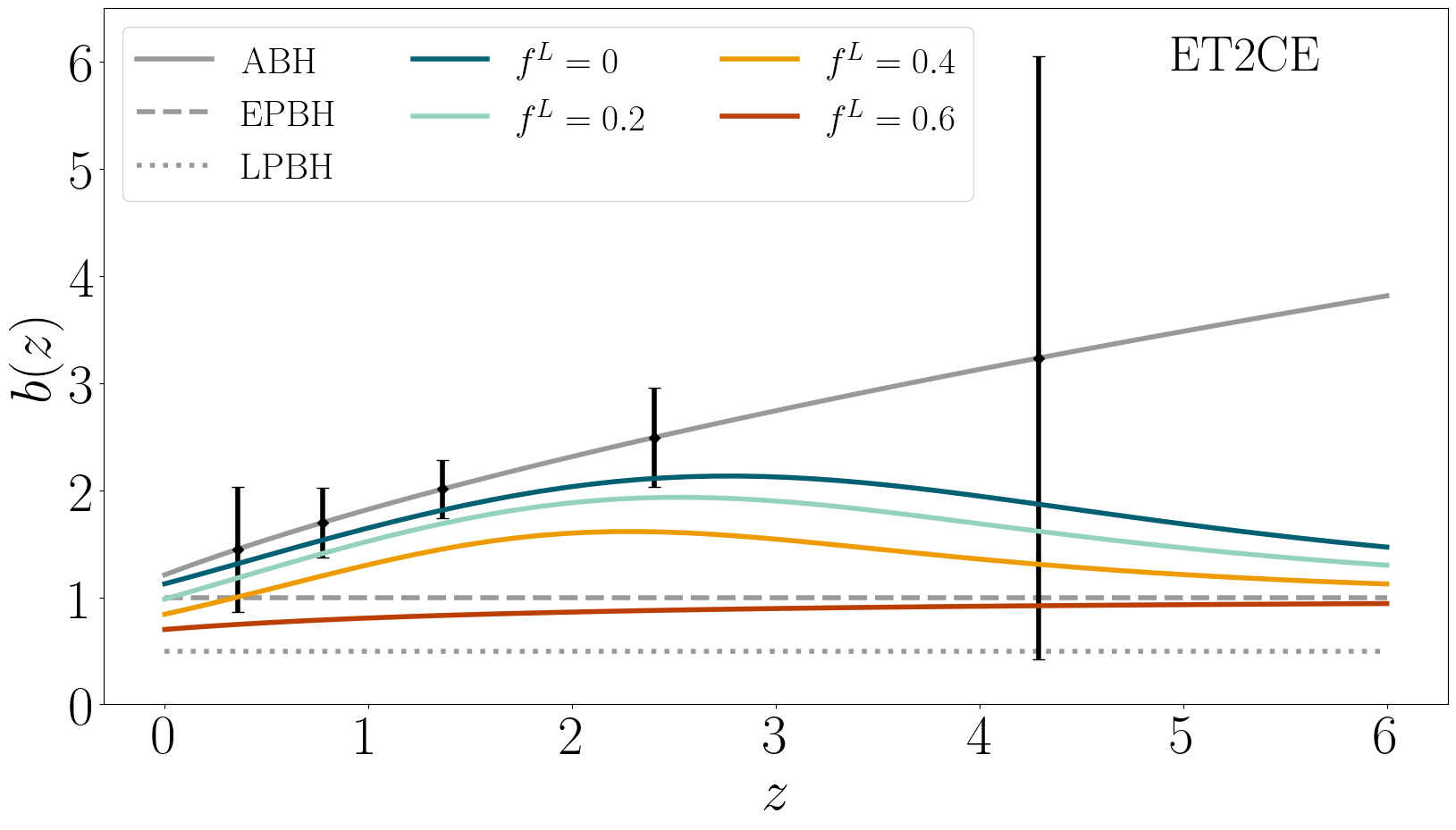}\\
  \includegraphics[width=\columnwidth]{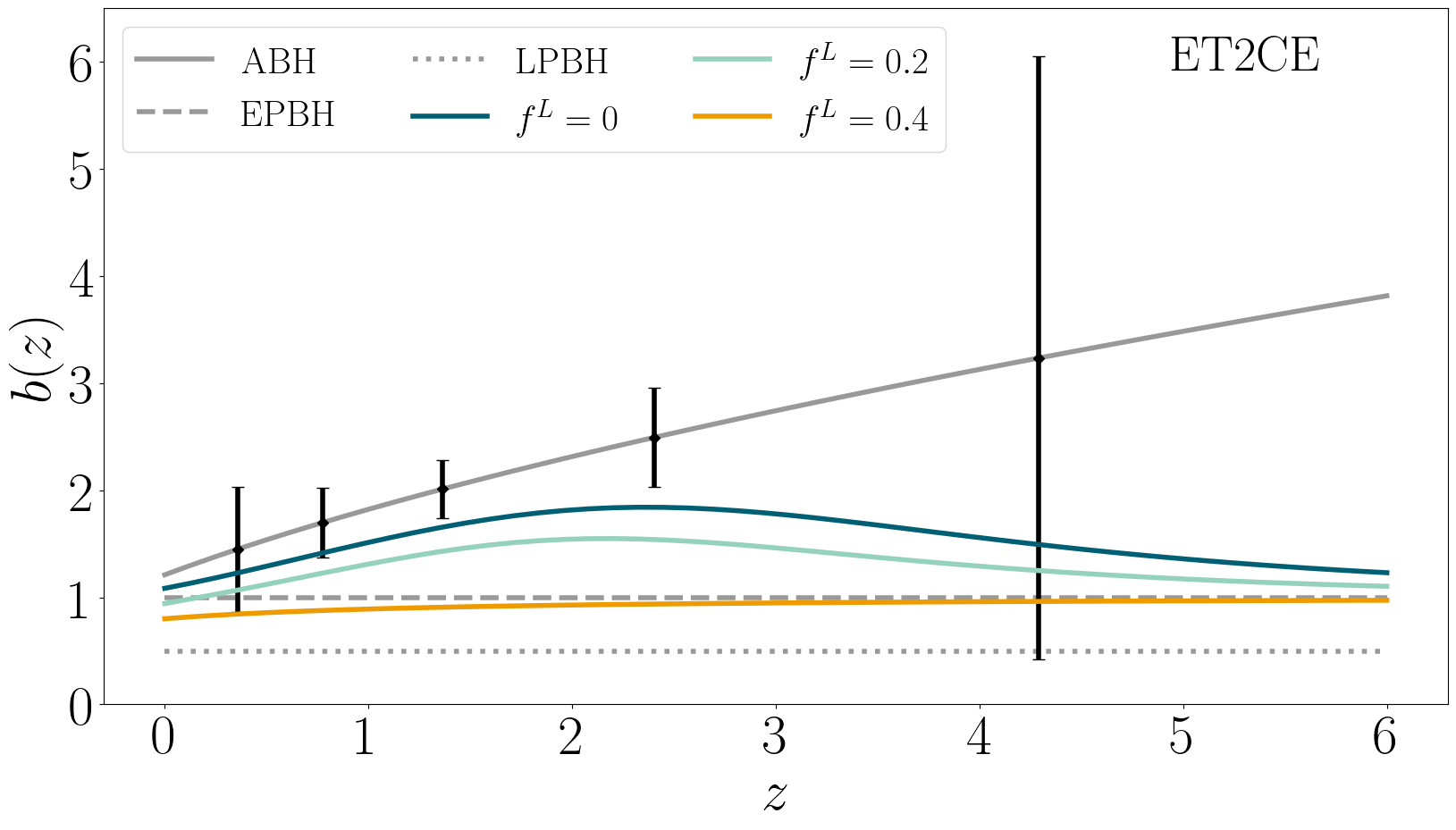}
  \includegraphics[width=\columnwidth]{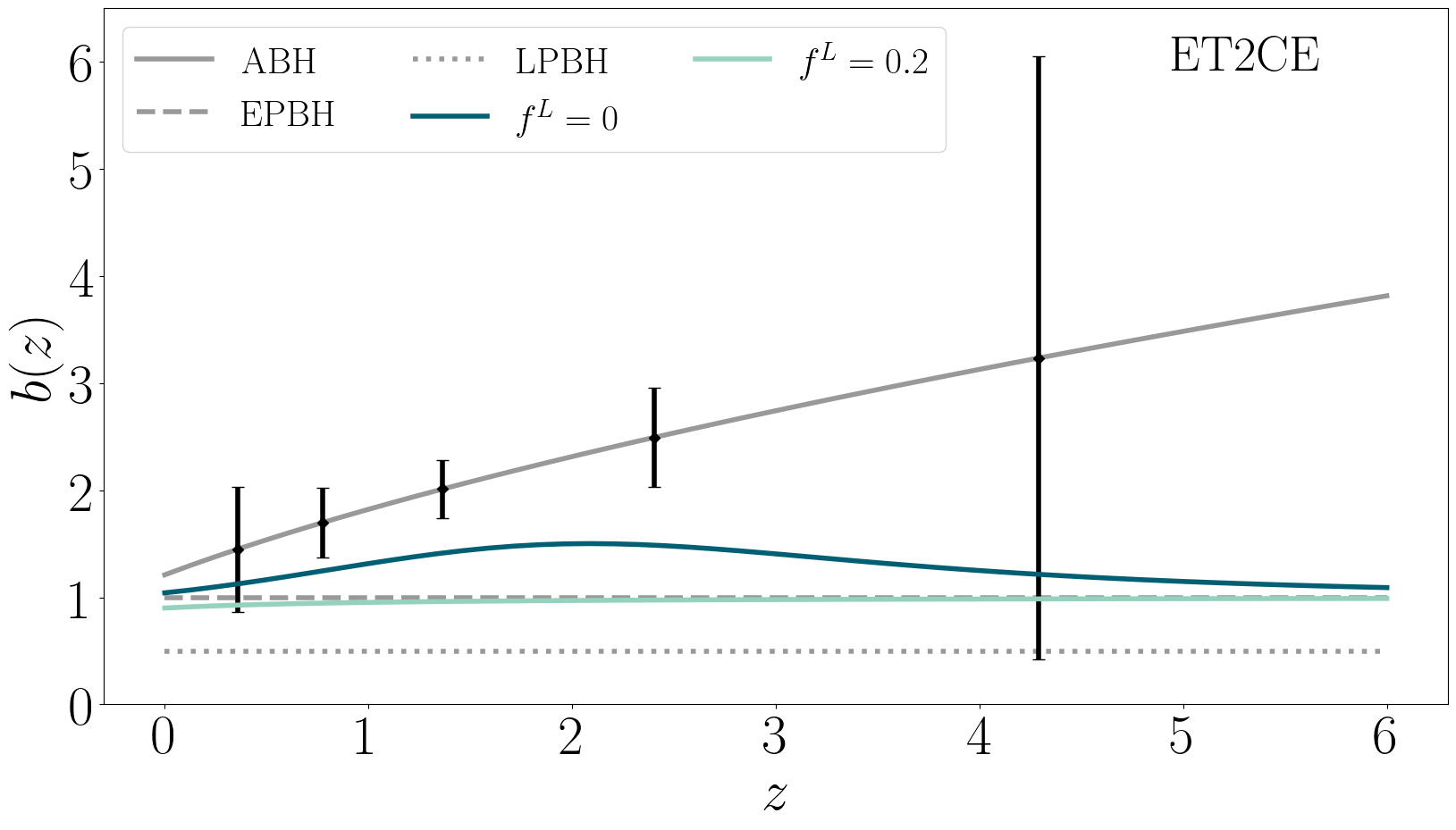}\\
  \includegraphics[width=\columnwidth]{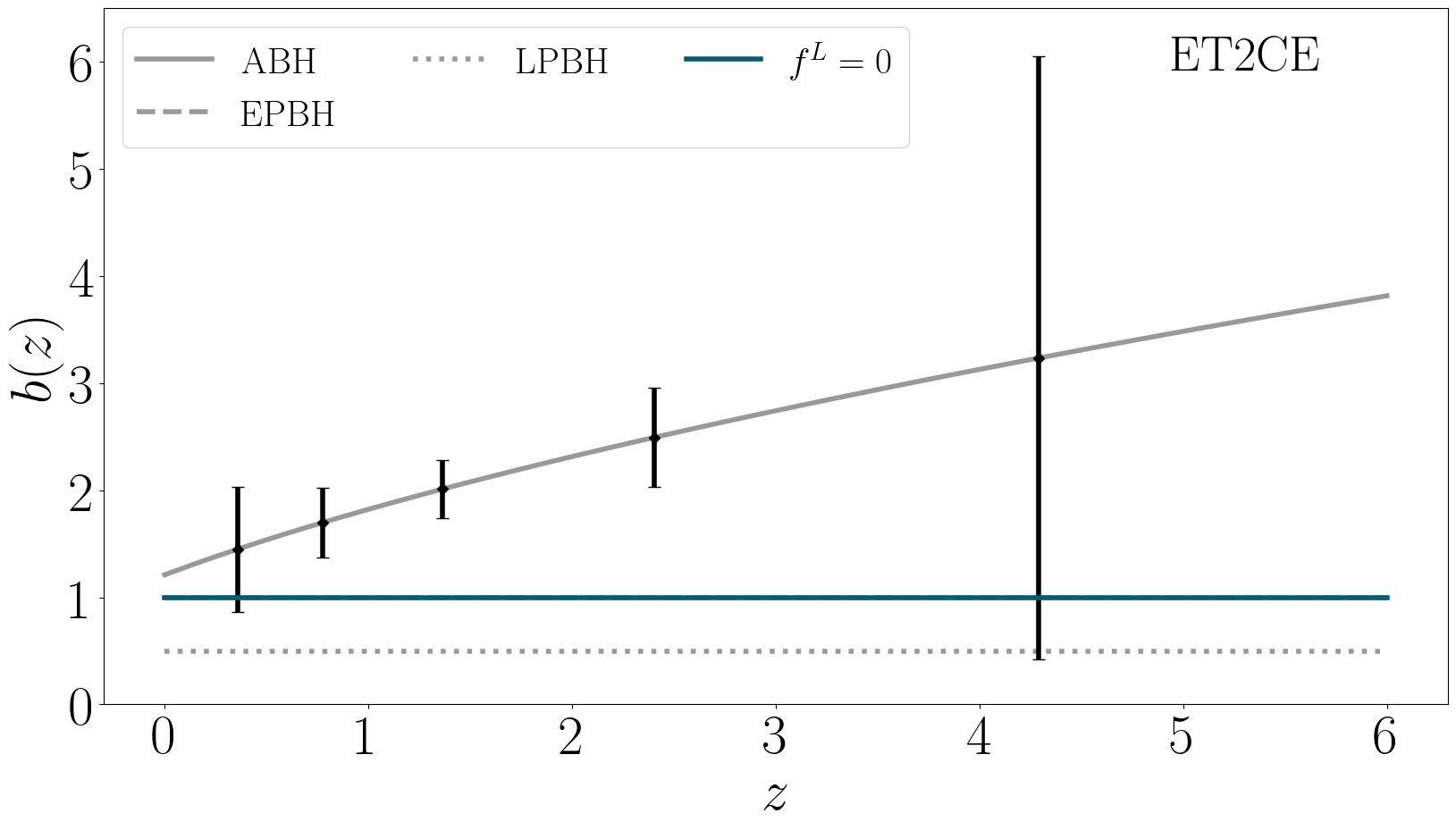}
\end{minipage}
    \caption{Analogous to the upper panel of figure~\ref{fig:bias_forecast}, but in the cases $f^E = 0,0.4, 0.6, 0.8, 1$. Left panels refer to overlapping bins, right panels to non-overlapping bins.}
    \label{fig:bias_all}
\end{figure}

In section~\ref{sec:bias_forecast} we computed forecasts for marginalized errors on ABH bias parameters and we showed how they can be used to distinguish the only-ABH scenario from the case in which both ABH and PBH contribute to the merger rate. Figure~\ref{fig:bias_forecast} compared ABH bias errorbars with the effective bias in the cases $f^E = 0.2$; panels in figure from~\ref{fig:bias_all} show the comparison in the cases $f^E = \{0,0.4,0.6,0.8,1\}$ for the ET2CE scenario. Finally, table~\ref{tab:bias_forecasts_cross_et} collects results for the ET scenario, alone or in cross-correlation with galaxy surveys (similar plots to the previous ones can be produced for ET, ET$\times$Wide survey and ET$\times$Deep survey as well).

\subsection{Fisher matrix: detector dependence}

The predicted constraining power on bias parameters depends on the detector properties. The following table is analogous to~\ref{tab:bias_forecasts_cross} when we adopt ET instead of ET2CE. The only-GW survey has no constraining power with uninformative prior, while it reduces to the amplitude of the prior itself when the Gaussian $50\%\sigma_b$ is considered. Cross-correlations improve the results; in particular, the wide survey leads to $\lesssim 100\%$ error at low-$z$ in the case of uninformative prior.

Note that this result could be improved considering a larger local merger rate or a less conservative cut/smoothing on the small scales: for instance, in~\cite{libanore2020} marginalized errors $\lesssim 100\%$ were found up to $z\sim 3$ by using a sharp cutoff (i.e.,~no exponential smoothing from eq.~\eqref{eq:bl},~\eqref{eq:cl}) associated with $\Delta\Omega = 100\,{\rm deg^2}$.

\begin{table}[ht!]
\setlength{\tabcolsep}{3.9pt}
\renewcommand{\arraystretch}{1.2}
    \begin{tabular}{|c|ccccc|ccccc|}
    \hline
    & \multicolumn{5}{c|}{Overlapping bins}
    & \multicolumn{5}{c|}{Non-overlapping bins}\\
         & $b_1$ & $b_2$ & $b_3$ & $b_4$ & $b_5$& $b_1$ & $b_2$ & $b_3$ & $b_4$ & $b_5$ \\
    \hline
        ET & 1693\% & 1054\% & 988\% & 1634\% & 7266\% & 667\% & 494\% & 476\% & 880\% & 4997\% \\
        ET, ${b_i}^p$ prior & 50\% & 50\% & 50\% & 50\% & 50\% & 50\% & 50\% & 50\% & 50\% & 50\% \\
        \hline
        $\times$Wide survey & 54\% & 35\% & 39\% & 120\% & 4302\%& 67\% & 53\% & 55\% & 131\% & 4309\% \\
        \hline
        $\times$Deep survey  & 236\% & 160\% & 150\% & 210\% & 1000\%& 140\% & 107\% & 104\% & 149\% & 936\% \\
    \hline
    \end{tabular}
    \caption{Analogous to table~\ref{tab:bias_forecasts_cross}, considering single ET as GW detector. Note that when we use a prior on the bias parameters, results are completely dominated by its value. }
    \label{tab:bias_forecasts_cross_et}
\end{table}

\clearpage
\bibliography{biblio}

\end{document}